\documentclass[a4paper,fleqn,usenatbib]{mnras}

\usepackage[T1]{fontenc}
\usepackage{ae,aecompl}

\usepackage{graphicx}	
\usepackage{amsmath}	
\usepackage{amssymb}	
\usepackage{natbib}

\usepackage{txfonts}

\usepackage{subfig}
\usepackage{rotfloat}
\restylefloat{figure}

\title[MOCCA -- VI. Bimodal spatial distribution of blue stragglers]{MOCCA code for star cluster
simulations -- VI. Bimodal spatial distribution of blue stragglers}

\author[Arkadiusz Hypki and Mirek Giersz]{
Arkadiusz Hypki,$^{1,2}$\thanks{E-mail: ahypki@strw.leidenuniv.nl} and  
Mirek Giersz$^{2}$\\
\\
$^{1}$Leiden Observatory, Leiden University, PO Box 9513, NL-2300 RA
Leiden, the Netherlands\\
$^{2}$Nicolaus Copernicus Astronomical Center, Bartycka 18, 00--716 Warsaw,
Poland
}

\date{Accepted XXX. Received YYY; in original form ZZZ}

\pubyear{2016}

\begin{document}
\label{firstpage}
\pagerange{\pageref{firstpage}--\pageref{lastpage}}
\maketitle

\begin{abstract}

The paper presents an analysis of formation mechanism and properties of spatial
distributions of blue stragglers in evolving globular clusters, based on
numerical simulations done with the \textsc{mocca} code.
First, there are presented N-body and \textsc{mocca} simulations which try to
reproduce the simulations presented by \citet{Ferraro2012Natur.492..393F}. Then,
the agreement between N-body and the \textsc{mocca} code is shown. Finally, we
discuss the formation process of the bimodal distribution. We report that so-called
 bimodal spatial distribution of blue stragglers  is a very transient
feature. It is formed for one snapshot in time and it can easily vanish in the
next one. Moreover, we show that the radius of avoidance proposed by
\citet{Ferraro2012Natur.492..393F} goes out of sync with the apparent minimum of
the bimodal distribution after about two half-mass relaxation times.
This finding creates a real challenge for the dynamical clock, which uses this
radius to determine the dynamical age of globular clusters. Additionally, the paper
discusses a few important problems concerning the apparent visibilities of the
bimodal distributions which have to be taken into account while studying the
spatial distributions of blue stragglers.

\end{abstract}




\section{Introduction}
\label{sec:Intro}

Blue straggler stars (BSs) are important members of globular clusters (GCs). They are
promising tools to study the complex interplay between the
dynamical evolution of globular clusters and stellar evolution. BSs are defined
as stars which are brighter than the main-sequence (MS) turn-off point (also their
mass is larger than the stars on the turn-off point). They lie on the extension
of the main-sequence in color-magnitude diagram (CMD). Thus, they had to acquire somehow an
additional mass to stay on the main-sequence longer. Two possible channels of their formation
involve a physical collision with another star or some mass transfer.
BSs were first discovered by \citet{1953AJ.....58...61S} in M3, later they were
observed in essentially all clusters \citep{2004ApJ...604L.109P}. BSs were
discovered also in open clusters, e.g. \citet{Mathieu2009Natur.462.1032M} and
dwarf galaxies e.g. \citet{Mateo1995AJ....110.2166M},
\citet{Mapelli2007MNRAS.380.1127M} or \citet{Monelli2012ApJ...744..157M}.

In order to study radial positions of BSs in GCs one needs to compare their
number as a function of radius with other types of stars which are assumed that
they trace the radial density distribution of stars in a cluster. In the
literature the best two, from among such candidate populations, are HB stars and RGB stars. Their
number is sufficiently high, they are well visible on the CMD, and they are
present essentially on any radial distance from the center throughout the
entire cluster.
Additionally, their lifetimes are rather short, and thus, they are rather unchanged by stellar dynamics
significantly during their lifetimes (like more massive BSs). Thus, HB and RGB
radial positions trace clusters densities very well.


For a quantitative analysis, GC is divided into a number of concentric annuli
around the cluster center and a specific frequency definition is introduced
\citep{Ferraro1993AJ....106.2324F}. The specific frequency, called also the double
normalized ratio, is defined by the equation:
\begin{equation} \label{eq:BSsRelativeFreq}
R_{pop} = \frac{N_{pop} / N^{tot}_{pop}}{L_{samp} / L^{tot}_{samp}}
\end{equation}
where $pop$ described BSs, HB or RGB stars, $N_{pop}$ is the number of stars of
a given population (e.g. $N_{BSS}$), and $N^{tot}_{pop}$ is the total number of stars from
a given population. $L_{samp}$ denotes the sampled luminosity measured in
each annulus and $N^{tot}_{samp}$ is the total luminosity for all annuli. The
luminosity in each annulus is calculated by integrating the single-mass King
model that best fits the observed surface density profile (see e.g.
\citet{Lanzoni2007ApJ...668L.139L}). The reddening, distance to the cluster, and
possible incompleteness in spatial coverage of the outermost annuli should be
taken into account as well. The specific frequency, $R_{pop}$, is a very useful
quantity which allows to show whether the number of stars from a given population shows signs of an increased number
in some parts of a GC. The luminosity of GCs is very smooth through out the
entire cluster and very well represents ,,underlying'' mass of GC. 
By scaling the number of BSs, or another population, with
luminosities, one can compare GCs of various sizes with various dynamical
statuses. For example, if the number of BSs was significantly larger in the
center of GC, the specific frequency would reveal it with values larger than
1.0. The bimodal, unimodal and flat radial
distributions of BSs are nicely visible with the specific frequency (see below).

\citet{2003ApJ...588..464F} defined the BSs specific frequency as the number of
BSs, normalized to the number of the horizontal branch stars.
They examined six GCs and found that the BSs' specific frequency varies from 0.07 to 0.92, and it does
not depend on the central density, total mass, and velocity dispersion. What is
surprising, they found the largest BSs specific frequencies for clusters with
the lowest central density (NGC~288) and the highest central density (M80).
\citet{2003ApJ...588..464F} claim that these two kinds of BSs formation
processes, i.e., mass transfer and mergers, can have comparable efficiency in
producing BSs in their respective typical environments.

\citet{2008A&A...481..701S} found a strong correlation between the BSs specific
frequency and the linear combination of the binary fraction ($\xi_{bin}$) and velocity
dispersion ($\sigma_{v}$): $\xi_{bin} + \alpha \sigma_{v}$, where $\alpha =
-4.62$. This indicates that, for a given binary fraction, the BSs specific frequency
decreases with increasing velocity dispersion. Small cluster velocity dispersion
corresponds to a lower binding energy limit between soft and hard binaries (to a
larger fraction of hard binaries). Since the natural evolution of hard binaries
leads to an increase of their binding energy \citep{Heggie1975MNRAS.173..729H}, low
velocity dispersion GCs should host a larger fraction of hard binaries, which
are able to survive both, stellar encounters and activate mass-transfer,
and/or merging processes between the companions \citep{2008A&A...481..701S}.
Therefore, more BSs formed by the evolution of primordial binaries is expected
to form in lower velocity dispersion GCs.

The radial distribution of BSs in many clusters is bimodal. First discoveries
of bimodal distributions were made for M3 by \citet{Ferraro1993AJ....106.2324F,
1997A&A...324..915F} and by \citet{Zaggia1997A&A...327.1004Z} for M55. The BSs
radial distribution for the M3 cluster clearly shows a maximum at the center
of the cluster, a clear-cut dip in the intermediate region, and again a rise of BSs
in the outer region of the cluster (but lower than the central value). The
bimodal distribution of BSs was later shown by other authors for other clusters
like 47~Tuc \citep{2004ApJ...603..127F}, NGC~6752 \citep{2004ApJ...617.1296S},
M55 \citep{2007ApJ...670.1065L}, M5 \citep{2006ApJ...646..881W,
Lanzoni2007ApJ...663..267L} and others. \citet{1994ApJ...431L.115S} suggested
that BSs were formed by direct collisions in the center of the cluster and then
ejected to the outer part of the system as a result of a dynamical interaction.
Ejected BSs were afterwards moved back to the center of the cluster because of
the mass segregation, which leads to the increase of the number of BSs in the
center of the system. If dynamical interactions were energetic enough, then
BSs would stay outside of the cluster for longer time and this could be the
reason why there is a higher rate of formation of BSs in the outer part of the
cluster, i.e., the second peak of the BSs in the bimodal distribution. Later, the
bimodal distribution of BSs in the cluster was explained differently by
\citet{1997A&A...324..915F}. They showed it as a result of different processes
forming BSs in different parts of the cluster -- mass transfer for the outer
BSs and stellar collisions leading to mergers for BSs in the center of the
cluster. Furthermore, \citet{2004ApJ...605L..29M, 2006MNRAS.373..361M} and
\citet{Lanzoni2007ApJ...663..267L}, by performing numerical simulations, showed
that the bimodal distribution in the cluster cannot be explained only by a
collisional scenario in which BSs are created in the center of the cluster and
some of them are ejected to the outer part of the system. This process is
believed to be not efficient enough, and $\sim 20-40\%$ of BSs have to be
created in the peripherals in order to get the required number of BSs for the
cluster. It is believed that in the outer part of a star cluster, binaries can
start mass transfer in isolation without suffering from energetic dynamical
interactions with field stars. Even if one can observe a bimodal distribution
of BSs for many clusters, one can not generalize this feature. There are known
clusters for which radial distributions are even flat, like for NGC~2419
\citep{2008ApJ...677.1069D,Contreras2012ApJ...748...91C}.


\index{radius of avoidance}
For GCs, which have signs of the bimodal spatial distribution, the radius at
the minimum of the specific frequency is called the radius of avoidance ($r_{avoid}$).
See \citet[Fig.~12]{Lanzoni2007ApJ...663..267L} for a few examples of $r_{avoid}$ in GCs with respect to the GCs core radii. The area surrounding the radius of
avoidance is called ,,zone of avoidance'' \citep{Mapelli2004ApJ...605L..29M}.
The radius of avoidance is a quantitative value which describes the radius below
which all heavier objects are expected to have enough time to already sink to the center of GC. BSs are example of such objects. They have masses larger than $m_{turn-off}$, some of them can have masses even close to $2 \times m_{turn-off}$ (collisional BSs) -- they are significantly larger than the average mas of stars in a GC. Some BSs appear in binaries, which lowers the mass segregation time for them even more. 
The radius of avoidance is a value which divides the GC essentially into two
regions. The structure below the $r_{avoid}$ is expected to be mass-segregated and
thus it should not reflect the initial GC properties. More massive objects, like
BSs, are expected to be found already in the deep center of the GC ($< r_c$).
However, the stars at the distances larger than $r_{avoid}$ should 
more-less resemble the initial GC properties, because at such large radii dynamical
processes in GC should not change significantly the structure of the GC. Thus, it
is expected e.g. for mass-transfer BSs (EMT), created as a result of
unperturbed evolution, to stay more or less on their initial orbits. The radius of
avoidance can be calculated from the equation for the dynamical friction time
$t_{df}$ by \citep{Binney1987gady.book.....B}:
\begin{equation} \label{eq:RAvoid}
t_{df} = \frac{3}{4~ln \Lambda~G^2~(2\Pi)^{0.5}} \frac{\sigma(r)^3}{m_{BSS}~
\rho(r)}
\end{equation}
where $ln \Lambda$ is the Coulomb logarithm, \textit{G} is gravitational
constant and $\sigma(r)$, $\rho(r)$ are cluster velocity dispersion and density
at the radius $r$. Masses of the
BSs are usually assumed to be $m_{BSS} = 1.2 M_{\odot}$.

NGC~6388 is another example of a very well studied globular cluster in terms of the
BSs population and radial properties, see \citet{Dalessandro2008ApJ...677.1069D}.
They show this bimodal spatial
distribution of BSs in NGC~6388 together with a flat distribution of HB
(which are used here to trace the entire GC mass distribution). In NGC~6388
 several groups of HB stars were recognized. HB referred here is the red-HB
(RGH) subgroup -- the most numerous group of HB stars characteristic for
metal-reach stellar populations (for more details about HB subgroups see
\citet{Dalessandro2008ApJ...677.1069D}). The bimodal distribution is clearly
visible with its central peak in the cluster center, then  a clear fall around $4-5
r/r_c$, and again an increase of the number of BSs in the outer parts of GCs ($> 5
r/r_c$).

The bimodal radial distribution is also visible in the cumulative plot which
very well represents the internal radial structure of the GC
\citep{Dalessandro2008ApJ...677.1069D}.
It represents the cumulative number of BSs ($\Phi$) and the corresponding cumulative number of HB stars starting from the deep center ($\Phi = 0$) of the cluster up to its outer
regions ($\Phi = 1$). The fraction of BSs is clearly larger in the
center ($\Phi \leq 0.5$), which corresponds to the most central $\sim 30''$ of
the GC. Then, the fraction of HB stars outnumbers the BSs population. The
cumulative number of BSs raises again near $100''$ of the center of GC. It is
worth to notice how smooth the cumulative number of HB stars is, which makes them
a very good reference population for the radial positions of BSs.


\citet{Dalessandro2008ApJ...677.1069D} claim that the radial distribution of BSs
in NGC~6388 is peculiar. Although it is clearly bimodal (see
\citet[Fig.~11]{Dalessandro2008ApJ...677.1069D}), BSs occupy regions which already should be cleared from them due to dynamical friction.
The position of the radius of avoidance ($r_{avoid}$) for NGC~6388 does not
correspond to the dip in the intermediate region, where the number of BSs is
the lowest \citep{Dalessandro2008ApJ...677.1069D}. The radius of
avoidance, $r_{avoid} \simeq 15 r_c$, was calculated using Eq.~\ref{eq:RAvoid},
GC age was assumed to be $t_{age} = 12$~Gyr, velocity dispersion $\sigma_0 =
18.9 km/s$, and stellar density of the order of $10^6 stars/pc^3$
\citep{Pryor1993ASPC...50..357P}. The position of $r_{avoid}$ is rather unexpected.
It is about 3 times larger than the dip in the bimodal spatial distribution ($5
r_c$). This could suggest that the dynamical friction, responsible for creating
so-called ,,zone of avoidance'', is not as efficient as it was previously thought.
NGC~6388 GC seems to be ,,dynamically younger'' than assumed because
BSs segregated only up to $5 r_c$, rather than to the expected $15 r_c$.
\citet{Dalessandro2008ApJ...677.1069D} discuss the possibility of the existence
of IMBH in the center of NGC~6388 but it is not clear how it affects GC at radii
of about $5-15 r_c$. Moreover, if IMBH existed in the center of NGC~6388 and
BSs were influenced by strong interactions with central objects, then BSs
would be ejected to the outer regions but across all radii. The ejected BSs
would fill the dip around $5 r_c$ rather than concentrating in the
outskirts of GC. It is especially important from the point of view of this
paper, which also suggests that dynamical friction is itself a very interesting
explanation, but it does not explain all features observed in the \textsc{mocca}
simulations. The picture of formation of bimodal spatial distribution for GCs
seems to be more complex (see Section.~\ref{chap:Bimodality}).

Dynamical simulations of \citet{Mapelli2004ApJ...605L..29M,
Mapelli2006MNRAS.373..361M} and \citet{Lanzoni2007ApJ...663..267L} showed that
rather a large number of collisional BSs ($\lesssim 50\%$) is needed to
reproduce the number of BSs in the central peak. Additionally, about 20-40\% of
mass-transfer BSs are needed to reproduce the number of BSs in the outer regions
of GCs.

A detail study of BSs radial distribution was performed by
\citet{Lanzoni2007ApJ...663.1040L} for NGC~1904 (M79). They used wide-field
ground based ESO-WFI and space GALEX observations to collect a multiwavelength
photometric data (from far UV to near infrared). What is important, this
extensive work covered entire cluster extension from the very central regions up
to the tidal radius. In total, position of 39 BSs were analyzed and they have
been found to be highly segregated in the cluster core. No other increase of
the number of BSs were found in the outskirts of GC.
No evidence for the bimodal spatial distribution in this case is consistent with the formation mechanism
based on the dynamical friction which drives BSs into the center of the GC. In the
Harris catalogue \citep{Harris1996AJ....112.1487H} NGC~1904 is denoted as
core-collapsed GC and thus stars in this GC are most probably already fully segregated
 and there is no expectation to observe the bimodal spatial
distribution.
Radius of avoidance for NGC~1904 is $r_{avoid} \sim 30 r_c$ (see
Eq.~\ref{eq:RAvoid}) which supports the
conclusions of the fully mass segregation of this GC
\citep{Lanzoni2007ApJ...663.1040L}.


For the GC $\omega$~Cen the radial distribution of BSs is flat
\citep{Ferraro2006ApJ...638..433F}. This GC is simply too large and mass
segregation processes did not have enough time to alter the BSs positions. There is no increase in the number of BSs in the center
and there is no evidence of the second peak in the outskirts of GC.


Blue stragglers are luminous stars in GCs and thus they are
useful to probe the dynamical evolution of stellar systems.
\citet{Ferraro2012Natur.492..393F} used them to build a \textit{dynamical clock} of
GCs. Using their location inside GCs, they tried to
infer the dynamical stage reached by GCs. They suggest that the
only physical mechanism which lies behind such a clock is the dynamical friction.

At the end of the introductory section it is worth to mention that in the
literature, terms \textit{collision} and \textit{merger} are used differently by
different authors. In this paper the term \textit{collision} is defined as a
physical collision between at least two stars during a dynamical interaction,
while the term \textit{merger} is defined as a coalescence between stars from
one binary as a result of stellar evolution.

This paper is organized as follows. The Sect.~\ref{sec:NumericalSimulations}
briefly describes the properties of various models computed for the needs of
this work. In the Sect.~\ref{chap:Bimodality} there is detailed analysis of
the formation and evolution of the bimodal spatial distribution of BSs. First,
we described an attempt to reproduce available N-body simulations, then we present
comparison between our N-body simulations and \textsc{mocca} code simulations.
Finally, in this section we discuss in details the formation mechanism of the
bimodal spatial distribution for real-size GCs. In the
Sect.\ref{sec:Bim:DynamicalClock} we summarize or work and discuss its influence
on the observations of GCs.

\section{Numerical simulations}
\label{sec:NumericalSimulations}

The numerical simulations were performed with the
\textsc{mocca}\footnote{\url{http://moccacode.net}} code
\citep{Hypki2013MNRAS.429.1221H}. 

The \textsc{mocca} code is one of the most advanced codes 
to simulate real size globular clusters and provide full information on the
stellar and dynamical evolution of all stars in such system. It originates
from the Monte Carlo code for star clusters simulations developed by
\citet{Henon1971Ap&SS..14..151H}, then improved by
\citet{Stodolkiewicz1986AcA....36...19S} and later heavily developed by
Giersz and collaborators
\citep{Giersz2013MNRAS.431.2184G,Giersz2015MNRAS.454.3150G}. The stellar evolution
is done for both single and binary stars using SSE and BSE codes
\citep{Hurley2000MNRAS.315..543H, Hurley2002MNRAS.329..897H}. Recently, the Monte
Carlo code was integrated with \textsc{fewbody} code of \citet{Fregeau2007ApJ...658.1047F} to deal with strong
interactions like in N-body codes. The code got a new name -- \textsc{mocca}.
This addition allows the \textsc{mocca} code to follow the formation and
evolution of exotic objects, for instance blue stragglers
\citep{Hypki2013MNRAS.429.1221H}.


\subsection{Initial parameters for the \textsc{mocca} code simulations}
\label{sec:InitialParameters}

The \textsc{mocca} code was used to compute many models of globular clusters
with various initial conditions. The models and their properties are summarized
in Tab.~\ref{tab:Sim:Ini1} and Tab.~\ref{tab:Sim:Ini2}. 

The models from Tab.~\ref{tab:Sim:Ini1} and Tab.~\ref{tab:Sim:Ini2} are
identical to the models presented in another paper in the series about BSs
(Hypki et al.
2016, \textit{MOCCA code for star cluster
simulations -- V. Initial globular cluster conditions influence on blue
stragglers}, submitted). This other paper shows how various initial conditions
of globular clusters and various initial properties of binaries influence the
population of BSs. The copy of Tab.~\ref{tab:Sim:Ini1} and
Tab.~\ref{tab:Sim:Ini2} is however
left in this paper for the sake of completeness.

\begin{table*}
    \begin{tabular}{c|c|c|c|c|c|c|c|c|c|c|c}
\multicolumn{12}{c}{\textsc{initial mass function of the \textsc{mocca} simulations (Part~I)}}\\
Name            &$N$ &$f_b$&IM&$IMF_s$&$IMF_b$&q&a&e&z&$r_{tid}$&$r_h$\\
\hline
\textsc{mocca-1}&300k&0.1&P&K93&K91&U&UL&T&0.001&69&6.9\\
\textsc{mocca-2}&300k&0.2&P&K93&K91&U&UL&T&0.001&15&1.5\\
\textsc{mocca-3}&300k&0.2&P&K93&K91&U&UL&T&0.001&25&2.5\\
\textsc{mocca-4}&300k&0.2&P&K93&K91&U&UL&T&0.001&35&3.5\\
\textsc{mocca-5}&300k&0.2&P&K93&K91&U&UL&T&0.001&45&4.5\\
\textsc{mocca-6}&300k&0.2&P&K93&K91&U&UL&T&0.001&69&1.2\\ 
\textsc{mocca-7}&300k&0.2&P&K93&K91&U&UL&T&0.001&69&1.7\\ 
\textsc{mocca-8}&300k&0.2&P&K93&K91&U&UL&T&0.001&69&2.3\\ 
\textsc{mocca-9}&300k&0.2&P&K93&K91&U&UL&T&0.001&69&2.8\\ 
\textsc{mocca-10}&300k&0.2&P&K93&K91&U&UL&T&0.001&69&3.5\\ 
\textsc{mocca-11}&300k&0.2&P&K93&K91&U&UL&T&0.001&69&4.6\\ 
\textsc{mocca-12}&300k&0.2&P&K93&K91&U&UL&T&0.001&69&6.9\\ 
\textsc{mocca-13}&300k&0.2&P&K93&K91&U&UL&T&0.001&69&9.9\\ 
\textsc{mocca-14}&300k&0.2&P&K93&K91&U&UL&T&0.001&69&17.3\\ 
\textsc{mocca-15}&300k&0.2&P&K93&K91&U&UL&T&0.001&85&8.5\\
\textsc{mocca-16}&300k&0.2&P&K93&K91&U&UL&T&0.001&135&13.5\\
\textsc{mocca-17}&300k&0.2&P&K93&K91&U&UL&T&0.001&235&23.5\\
\textsc{mocca-18}&300k&0.2&P&K93&K91&U&UL&T&0.001&335&33.5\\
\textsc{mocca-19}&300k&0.3&P&K93&K91&U&UL&T&0.001&69&9.6\\
\textsc{mocca-20}&300k&0.5&P&K93&K91&U&UL&T&0.001&69&9.6\\

\textsc{mocca-21}&600k&0.05&P&K93&K91&U&UL&T&0.001&100&10.0\\ 
\textsc{mocca-22}&600k&0.1&P&K93&K91&U&UL&T&0.001&100&10.0\\ 
\textsc{mocca-23}&600k&0.2&P&K93&K91&U&UL&T&0.001&25&2.5\\ 
\textsc{mocca-24}&600k&0.2&P&K93&K91&U&UL&T&0.001&35&0.9\\ 
\textsc{mocca-25}&600k&0.2&P&K93&K91&U&UL&T&0.001&35&1.2\\ 
\textsc{mocca-26}&600k&0.2&P&K93&K91&U&UL&T&0.001&35&1.8\\ 
\textsc{mocca-27}&600k&0.2&P&K93&K91&U&UL&T&0.001&35&3.5\\ 
\textsc{mocca-28}&600k&0.2&P&K93&K91&U&UL&T&0.001&55&1.4\\ 
\textsc{mocca-29}&600k&0.2&P&K93&K91&U&UL&T&0.001&55&1.8\\ 
\textsc{mocca-30}&600k&0.2&P&K93&K91&U&UL&T&0.001&55&2.8\\ 
\textsc{mocca-31}&600k&0.2&P&K93&K91&U&UL&T&0.001&55&5.5\\ 
\textsc{mocca-32}&600k&0.2&P&K93&K91&U&UL&T&0.001&100&1.7\\ 
\textsc{mocca-33}&600k&0.2&P&K93&K91&U&UL&T&0.001&100&2.5\\ 
\textsc{mocca-34}&600k&0.2&P&K93&K91&U&UL&T&0.001&100&5.0\\ 
\textsc{mocca-35}&600k&0.2&P&K93&K91&U&UL&T&0.001&100&10.0\\ 
\textsc{mocca-36}&600k&0.2&P&K93&K91&U&UL&T&0.001&100&20.0\\ 
\textsc{mocca-37}&600k&0.2&P&K93&K91&U&UL&T&0.001&180&18.0\\ 
\textsc{mocca-38}&600k&0.2&P&K93&K91&U&UL&T&0.001&130&13.0\\ 
\textsc{mocca-39}&600k&0.2&P&K93&K91&U&UL&T&0.001&230&23.0\\ 
\textsc{mocca-40}&600k&0.2&P&K93&K91&U&UL&T&0.001&300&30.0\\ 
\textsc{mocca-41}&600k&0.2&P&K93&K91&U&UL&T&0.001&400&40.0\\ 
\textsc{mocca-42}&600k&0.4&P&K93&K91&U&UL&T&0.001&100&10.0\\ 
\textsc{mocca-43}&600k&0.5&P&K93&K91&U&UL&T&0.001&100&10.0\\ 

    \end{tabular}
\caption[Initial conditions of \textsc{mocca} simulations done for the purpose
of this paper -- Part I]{Initial conditions of \textsc{mocca} simulations done
for the purpose of this paper. Symbols have the following meaning: N -- initial
number of objects (single + binary stars), $f_b$ -- initial binary fraction,
$f_b = N_b / N$ ($N_b$ -- number of binaries), $IM$ -- initial model, P --
Plummer model, $IMF_s$ -- Initial Mass Function for single stars, K93 --
\citet{Kroupa1993MNRAS.262..545K} in the range $[0.1; 100] \mathrm{M_{\odot}}$,
$IMF_b$ -- Initial Mass Function for binary stars, K91 --
\citet[eq.~1]{Kroupa1991MNRAS.251..293K}, binary masses from 0.2 to
100~$\mathrm{M_{\odot}}$, q -- distribution of mass ratios between stars in
binaries, U -- uniform distribution of mass ratios, R -- random pairing of
masses for binary components, a -- semi-major axes distribution, UL -- uniform
distribution of semi-major axes in the logarithmic scale from $2(R_1+R_2)$ to
100~AU, L -- lognormal distribution of semi-major axes from $2(R_1+R_2)$ to
100~AU, K95 -- binary period distribution from
\citet{Kroupa1995aMNRAS.277.1491K}, K95E -- distribution of semi-major axes with
eigenevolution and feeding algorithm \citep{Kroupa1995aMNRAS.277.1491K}, K13 --
new eigenevolution and feeding algorithm \citep{Kroupa2013pss5.book..115K}, e --
eccentricity distribution, T -- thermal eccentricity distribution, TE -- thermal
eccentricity distribution with eigenevolution, z -- mettalicity (e.g.~0.001 =
1/20 of the solar metallicity 0.02), $r_{tid}$ -- tidal radius in pc, $r_h$ --
half-mass radius in pc. The main difference in the simulation from
\textsc{mocca-1} to \textsc{mocca-43} is in the dynamical timescales of GCs'
evolution. 
}
	\label{tab:Sim:Ini1}
\end{table*}

\begin{table*}
    \begin{tabular}{c|c|c|c|c|c|c|c|c|c|c|c}
\multicolumn{12}{c}{\textsc{initial mass function of the \textsc{mocca} simulations (Part~I)}}\\
Name            &$N$ &$f_b$&IM&$IMF_s$&$IMF_b$&q&a&e&z&$r_{tid}$&$r_h$\\
\hline
\textsc{mocca-44}&300k&0.2&P&K93&K91&R&UL  &T&0.001&69&6.9\\
\textsc{mocca-45}&300k&0.2&P&K93&K91&R&L   &T&0.001&69&6.9\\
\textsc{mocca-46}&300k&0.2&P&K93&K91&R&K95 &T&0.001&69&6.9\\
\textsc{mocca-47}&300k&0.2&P&K93&K91&R&K95E&T&0.001&69&6.9\\
\textsc{mocca-48}&300k&0.2&P&K93&K91&R&K13 &T&0.001&69&6.9\\

\textsc{mocca-49}&300k&0.2&P&K93&K91&R&UL  &TE&0.001&69&6.9\\
\textsc{mocca-50}&300k&0.2&P&K93&K91&R&L   &TE&0.001&69&6.9\\
\textsc{mocca-51}&300k&0.2&P&K93&K91&R&K95 &TE&0.001&69&6.9\\
\textsc{mocca-52}&300k&0.2&P&K93&K91&R&K95E&TE&0.001&69&6.9\\
\textsc{mocca-53}&300k&0.2&P&K93&K91&R&K13 &TE&0.001&69&6.9\\

\textsc{mocca-54}&300k&0.2&P&K93&K91&U&L   &T&0.001&69&6.9\\
\textsc{mocca-55}&300k&0.2&P&K93&K91&U&K95 &T&0.001&69&6.9\\
\textsc{mocca-56}&300k&0.2&P&K93&K91&U&K95E&T&0.001&69&6.9\\
\textsc{mocca-57}&300k&0.2&P&K93&K91&U&K13 &T&0.001&69&6.9\\

\textsc{mocca-58}&300k&0.2&P&K93&K91&U&UL  &TE&0.001&69&6.9\\
\textsc{mocca-59}&300k&0.2&P&K93&K91&U&L   &TE&0.001&69&6.9\\
\textsc{mocca-60}&300k&0.2&P&K93&K91&U&K95 &TE&0.001&69&6.9\\
\textsc{mocca-61}&300k&0.2&P&K93&K91&U&K95E&TE&0.001&69&6.9\\
\textsc{mocca-62}&300k&0.2&P&K93&K91&U&K13 &TE&0.001&69&6.9\\

\textsc{mocca-63}&600k&0.2&P&K93&K91&U&K13&TE&0.001&55&5.5\\ 

    \end{tabular}
\caption[Initial conditions of \textsc{mocca} simulations done for the purpose
of this paper -- Part II]{For description see Tab.~\ref{tab:Sim:Ini1}}
	\label{tab:Sim:Ini2}
\end{table*}

The models from \textsc{mocca-1} up to \textsc{mocca-43}
(Tab.~\ref{tab:Sim:Ini1}) differ mainly in the dynamical scales of their
evolution. These are models which have mainly different tidal radii (distance
from the host galaxy) and different concentrations. These are the parameters which are
expected to have the biggest influence on the spatial evolution of BSs. They
cover the models from slowly evolving ones to the models which evolve very fast
and dissolve very quickly (even in several Gyr).
Whereas, the models with identifiers larger than 43 (Tab.~\ref{tab:Sim:Ini2})
differ mainly in the initial parameters of binaries (e.g. different initial
semi-major axes distributions, eccentricity distributions). They were used
mostly in the already mentioned other paper about BSs. Nevertheless, they
were also used in this paper to check whether the different initial binaries
properties could have some additional influence on the formation of the bimodal
spatial distribution of blue stragglers.

All models from Tab.~\ref{tab:Sim:Ini1} and Tab.~\ref{tab:Sim:Ini2} were
used to study the evolution of the spatial distribution of BSs in
evolving globular clusters. However, eventually we have carefully chosen only three
models to support the conclusions of this work (see
Sect.~\ref{sec:Bim:RAvoidDrift}).

The core radius ($r_c$) referred in this work is computed with
\citet{1985ApJ...298...80C}, the relaxation time ($t_{rh}$) is actually the
half-mass relaxation time (unless noted otherwise) and a star is considered as
the blue straggler in \textsc{mocca} simulations if it exceeds the turn-off
mass by at leas 2\% (for details see \citet{Hypki2013MNRAS.429.1221H}).

\section{Bimodal spatial distribution of blue stragglers}
\label{chap:Bimodality}

This section presents a study of the spatial distribution of blue stragglers in
GCs. For a few of them the bimodal distribution is observed (see
Sect.~\ref{sec:Intro}). In the first subsection we present a comparison between
N-body and \textsc{mocca} simulations in terms of the spatial distribution of
blue stragglers. We show that the \textsc{mocca} code is a suitable tool to
track positions of BSs in stellar systems.
The next subsection shows the formation and evolution of the bimodal spatial
distribution for the real-size star clusters. This subsection points out also its important implications on the
determination of the dynamical ages of GCs.

\subsection{N-body and MOCCA simulations of simplified models}
\label{sec:Bim:Simulations}

The origin and evolution of the bimodal spatial distribution in the
simulations done with the \textsc{mocca} code for real-size star clusters was very challenging from
the beginning. It was very difficult to obtain clear signs of the bimodality of blue stragglers. Moreover, the bimodal distribution
appeared to be very transient.
It was present for some snapshots in time and then after one or several next 
snapshots ($\sim$~hundreds of Myr) the signs of bimodality disappeared.
It was very difficult to recreate observations or results from the N-body simulations presented
already in the literature \citep{Ferraro2012Natur.492..393F}. Thus, we decided to perform a series of sanity
checks and comparisons with N-body models. The results of these tests are
presented here. They should give a proper confidence to the results obtained by
the \textsc{mocca} code for simulations of real-size star clusters presented and
discussed in Sect.~\ref{sec:Bim:RAvoidDrift}.

Simulations which show the formation of bimodal spatial distribution were
performed by \citet{Ferraro2012Natur.492..393F} with \textsc{nbody6} code
\citep{Aarseth2003gnbs.book.....A, Nitadori2012MNRAS.424..545N}. The results of
the simulations are shown in \citet[Supplementary
Fig.~2]{Ferraro2012Natur.492..393F}. They performed 8~N-body simulations with 16k particles. For each simulation 3 projections of stars' positions to the plane of the sky were made along three main axes to have a
better statistics.  The initial conditions for simulations were simplified. Each
simulation consisted initially only of stars with three different masses: 1\% of
heavy BSs with masses $1.2 M_{\odot}$, 10\% of RGB stars with masses $0.8
M_{\odot}$ and 89\% of MS stars with masses $0.4 M_{\odot}$. The initial model
was a King model with $W_0 = 6$, which corresponds to the concentration $c =
r_{tid} / r_c \cong 18$. The stellar evolution was switched off, there were no
primordial binaries and there was no initial mass segregation present in the
system.
The simplified initial conditions were chosen to have a very simple physics in the
system but with all needed processes essential to form the bimodal spatial
distribution of BSs. The values on Y axis are the relative frequency of BSs
normalized to the number of RGB stars ($R_{BSS}(r) = [N_{BSS}(r)/N_{BSS,tot}] /
[N_{RGB}(r)/N_{RGB,tot}]$). The values on X axis are distances from the center
of star cluster in the units of current cluster core radius.
The gray strip around unity is drawn for the reference and by definition it
represents regions of star cluster in which stars were not  yet fully affected by
the mass segregation -- thus, the relative number between BSs and RGB stars is $\sim
1$.
The error bars are calculated based on Poisson counting statistics. The N-body
simulations done by \citet{Ferraro2012Natur.492..393F} are referred in this
paper as \textsc{ff} simulations.

The \textsc{ff} simulations show the drift
of $r_{min}$, denoted with arrows, from the very center to the outskirts of the
star cluster with increasing relaxation time. The radius $r_{min}$ is roughly
the position around which the bimodal distribution is visible (roughly the lowest value of $R_{BSS}$). The width of the dip around $r_{min}$ is also increasing with time. Note that the minima $r_{min}$ for all snapshots except $t
= 0$ reach very low values $\sim 0.1$~$R_{BSS}$. The purpose of these
simulations was to show the ongoing evolution of $r_{min}$ due to the mass
segregation in the system. \citet{Ferraro2012Natur.492..393F} used these
simulations to show that the physical process behind the formation
of the bimodal spatial distribution of BSs is the mass segregation. These
simulations were not intended to be compared with any observational data.

Our goal was
to try to reproduce \textsc{ff} simulations done by
\citet{Ferraro2012Natur.492..393F} and to verify the mass segregation as the formation process of the bimodal distributions in GCs.
First, a number of tests were performed with the \textsc{nbody6} and the \textsc{mocca} 
codes to show that there is an agreement between them.

\begin{figure}
  \includegraphics[width=\columnwidth]{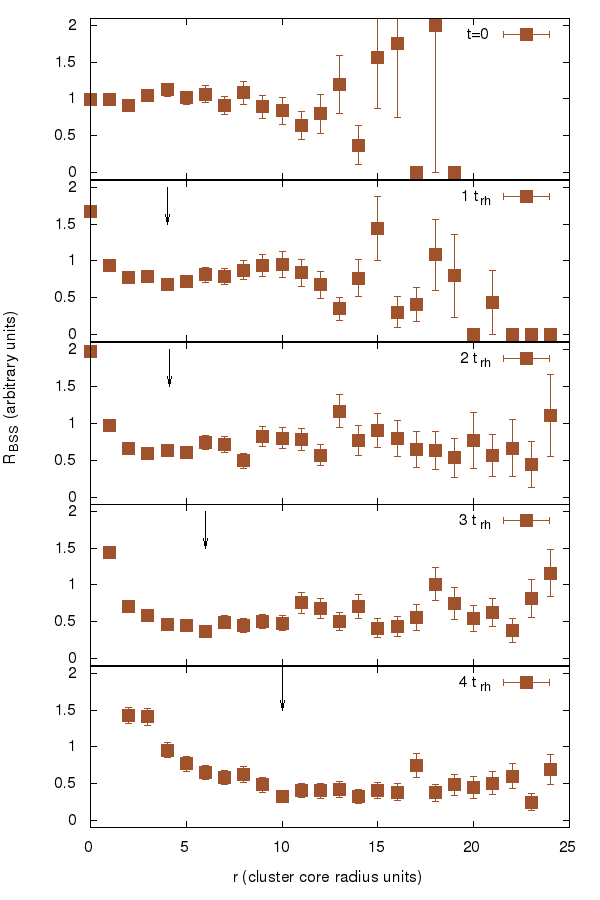}
  \caption[The relative frequency of BSs obtained from direct N-body
simulations as an attempt to reproduce results from
\citet{Ferraro2012Natur.492..393F}]{The relative frequency of BSs obtained from direct N-body
simulations as an attempt to reproduce results from
\citet{Ferraro2012Natur.492..393F}. These are simulations which are referred to as
\textsc{dh} simulations in the text.
The initial conditions for N-body models are the same as for the simulation showed
in \citet{Ferraro2012Natur.492..393F}. The positions of $r_{min}$, chosen by eye,
are denoted with arrows. For later times ($> 2 t_{rh}$) an estimation of the
position of $r_{min}$ is rather debatable. However, the overall
drift of $r_{min}$ with time should be well visible with the arrows.}
  \label{fig:Bim:DouglasNbody}
\end{figure}

The N-body simulations, performed for the purpose of this paper, were done
with the same initial conditions as for N-body simulation done by
\citet{Ferraro2012Natur.492..393F} and with the same numerical code, i.e., \textsc{nbody6}. In total,
there were made 8 N-body simulations (with 3 projections). They are referred here as \textsc{dh}
simulations. They were calculated until the core collapse only ($\sim 5 t_{rh}$). The snapshots with the drift or $r_{min}$
value for a few dynamical times ($1, 2, 3, 4
t_{rh}$) are presented in Fig.~\ref{fig:Bim:DouglasNbody}. Values for
$R_{BSS}$ were calculated in the same way as for \textsc{ff} simulations. The
error bars are calculated based on Poisson counting statistics ($R_{BSS} /
\sqrt{n_{BSS}}$).
The drift is denoted with the arrows and their places were chosen by eye
(taking into account error bars).
The $r_{min}$ is moving to the outer regions of star cluster with time.
This is in agreement with the results of \textsc{ff} simulations.

However, \textsc{dh} simulations were not able to reproduce fully the results
obtained by \textsc{ff} simulations \citep{Ferraro2012Natur.492..393F}.
In \textsc{ff} simulations the minimum around $r_{min}$ reaches values of $R_{BSS} \sim 0.2$.
These results were unable to reproduce such a large dip
around $r_{min}$ for any of the times ($1, 2, 3, 4 t_{rh}$, see
Fig.~\ref{fig:Bim:DouglasNbody}). 

The large differences between \textsc{ff} and \textsc{dh} simulations concern different binning. In \citet{Ferraro2012Natur.492..393F} the widths of bins are different for different distances from the center of the star
cluster. In the \textsc{dh} simulations the width of bins were the same across
all distances (in cluster core units). The bins have the same size in order to
have a clear picture of the formation and evolution of the $r_{min}$ radius, since
variable sizes for bins would introduce only artificial changes of the radial
distribution of BSs. The careful selection of the sizes of bins can have a large
impact of the visible appearance of the bimodal spatial distributions (see
discussion in Sect.~\ref{sec:Bim:Stochasticity}). 
Interestingly, for the
first bin for time $4~t_{rh}$ the value for $R_{BSS}$ runs out of the scale already.


Noticeable differences between \textsc{ff} and \textsc{dh} are also in terms of
the noise (error bars).
\textsc{ff} simulations do not show all bins -- e.g. it is not possible that for
$t = 0$ there are no BSs stars outside $\sim 6 r_c$ (see
\citet[Supplementary
Fig.~2]{Ferraro2012Natur.492..393F}). Thus, it is difficult to realize what  the
noise in the outer regions of star clusters is (after $r_{min}$). However, based on
the \textsc{dh} simulations one can see that the noise is very large. The error
bars are largest for the bins which are outside $r_{min}$ due to a small number of BSs there.

For the \textsc{dh} simulations there are plotted all 25 first bins, regardless of how many BSs are present in the bin.
We decided to have the same sizes for all bins to avoid adding any
artificial changes to the signs of bimodal spatial distribution. Together with
large errors, the values for $R_{BSS}$ scatter a lot too. Even for time $t =
0.0$ many values are not around 1.0 -- this is the expected value for
non-segregated regions of the star cluster. In the case of \textsc{ff} simulations such errors are not discussed.

However, the overall drift of $r_{min}$ was reproduced in \textsc{dh} simulations but without such a clear minimum around $r_{min}$, like in
\textsc{ff} simulations. In the \textsc{ff} model for time $t = 2 t_{rh}$ it
reaches values $R_{BSS} \sim 0.2$, and after $r_{min}$ it raises to values $\sim
1.0$. In the \textsc{dh} model the values are $\sim 0.5$ and in general do not
raise to value $\sim 1.0$ after $r_{min}$. This is also the reason why the
proper selection of the $r_{min}$ was so difficult.

The positions of $r_{min}$ in the units of the core radii were not reproduced.
The radii $r_{min}$ in \citet{Ferraro2012Natur.492..393F} are significantly
further away from the center than in \textsc{dh} simulations. For the time $2
t_{rh}$ \textsc{ff} simulation has $r_{min} \sim 10 [r / r_c]$, whereas in
\textsc{dh} simulations $r_{min}$ is about two times smaller ($r_{min} \sim 5 [r
/ r_c]$).

\begin{figure}
  \includegraphics[width=\columnwidth]{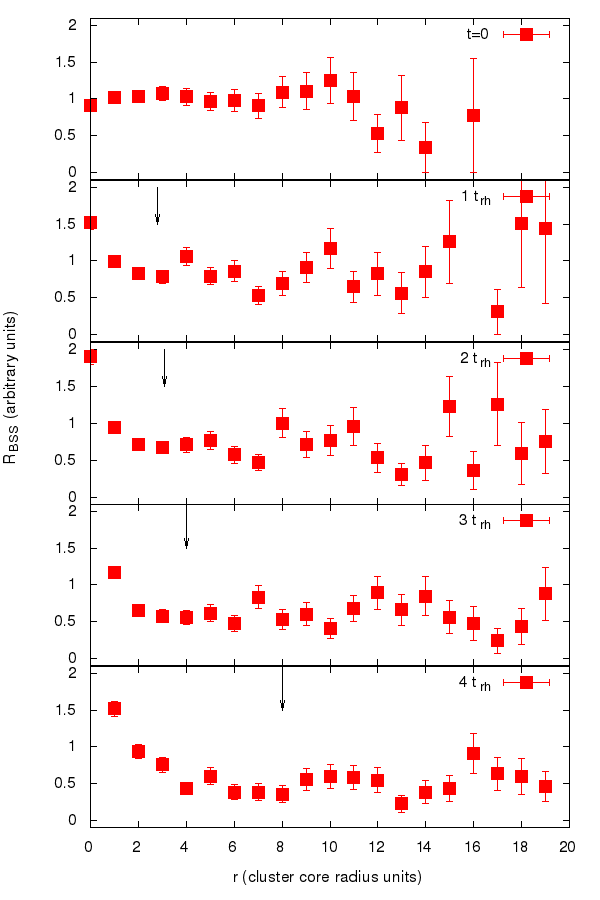}
  \caption[MOCCA simulation showing a drift of $r_{min}$ values with time]{BSs
  radial distributions obtained from the \textsc{mocca} simulation as an attempt to reproduce the
  $r_{min}$ drift showed by \citet{Ferraro2012Natur.492..393F} --
 this simulation is referred in the text as \textsc{ah} simulation.
Its initial conditions are the same as for
\textsc{ff} and \textsc{dh} simulations (see \citet[Supplementary
Fig.~2]{Ferraro2012Natur.492..393F}
and Fig.~\ref{fig:Bim:DouglasNbody}) with the exception that there was made only
one \textsc{mocca} simulation but with 100k initial stars (see details in the text). The
positions $r_{min}$, chosen by eye, are denoted with arrows. For later times ($>
2 t_{rh}$) the proper estimation on the location of $r_{min}$ is very
debatable. However, the overall drift of $r_{min}$ with time should be well
visible with the arrows.} 
  \label{fig:Bim:MoccaFerraroModel}
\end{figure}

The \textsc{mocca} simulation showing the drift of $r_{min}$ is presented
in Fig.~\ref{fig:Bim:MoccaFerraroModel} and it is called \textsc{ah} simulation.
The model is essentially the same as for \textsc{ff} and \textsc{dh} simulations
with the only difference that for \textsc{ah} simulation the initial number was
100k stars. Thus, the total number of BSs, RGB and MS stars is
essentially the same as for  \textsc{ff} and \textsc{dh} simulations (8
simulations $\times$ 16k $\sim$ 128k).
All the other parameters for \textsc{ah} simulation are the same and 
the overall evolution between models is very similar as well (e.g. the core collapse
for \textsc{dh} and \textsc{ah} models in the units of $t_{rh}$ is the same). The positions of $r_{min}$ are chosen by eye too.

The drift of $r_{min}$ is reproduced in the \textsc{ah} simulation too.
Additionally, the dip around $r_{min}$ is also not as significant as in
\textsc{ff} simulations -- the results are more consistent actually with
\textsc{dh} models. The noise in the outer regions is as large as in
\textsc{dh} simulations and also the values of $r_{min}$ are different than in \textsc{ff} simulations. These values are, in turn, more consistent with \textsc{dh} simulations.

\begin{figure}
  \includegraphics[width=\columnwidth]{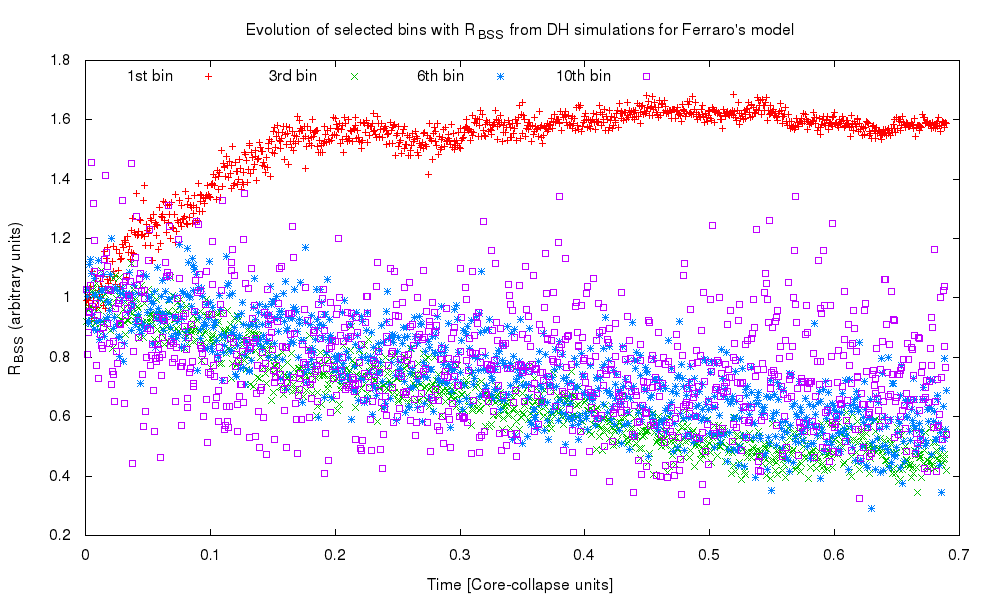}
  \includegraphics[width=\columnwidth]{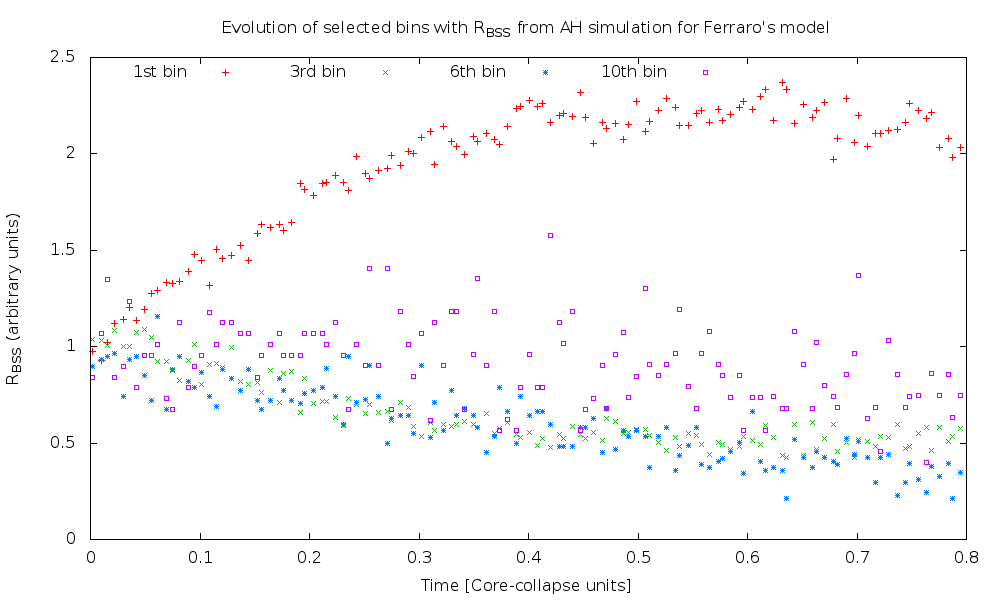}
  \caption[The values of normalized number of BSs for all snapshots for
  N-body and \textsc{mocca} simulation for simplified models]{The values of normalized
  number of BSs ($R_{BSS}$) for the 1st, 3rd, 6th and 10th bins for all
  snapshots for \textsc{dh} (top) and \textsc{ah} simulations (bottom). Initial
  parameters are discussed in the text. The time of the simulation on the X axis is
  scaled to the core collapse units. The snapshots for \textsc{ah} simulation are performed less
  frequently, thus, there are much less points for the bottom plot. The first bin for
  \textsc{ah} simulation has a slightly larger value than for \textsc{dh}
  simulation. However, the overall trend agrees and is well visible for both simulations.
The values for $R_{BSS}$ scatter for both simulations significantly (except the 1st bin).}
  \label{fig:Bim:AllTimesteps}
\end{figure}

Large errors for bins outside the $r_{min}$ were the reason to check carefully
how the noise really looks like for all snapshots throughout the whole
simulation.
Fig.~\ref{fig:Bim:AllTimesteps} shows the radial distribution
of normalized number of BSs ($R_{BSS}$) for the 1st, 3rd, 6th and 10th bin for all snapshots. The top panel is for 
\textsc{dh} simulations, and
the bottom one for \textsc{ah} simulation. In other words,
Fig.~\ref{fig:Bim:AllTimesteps} is like Fig.~\ref{fig:Bim:DouglasNbody}, only for
selected bins and for all snapshots.
Initial parameters for both types of simulations are discussed in the
text.
The time of the simulation on the X axis is scaled to the core collapse units -- in
this way it is easier to compare the simulations. The snapshots for \textsc{ah}
simulation are performed less frequently, thus, there is less points
for it. The first bin for \textsc{ah} simulation has slightly larger values than for
\textsc{dh} simulation. It means that for the \textsc{mocca} code there is a slightly larger mass segregation rate. It is a consequence of the larger number of stars for \textsc{ah} simulations (core collapses more significantly for GCs with higher numbers of stars). 
However, this is not important in the context of the bimodal distribution formation.

The overall trend in Fig.~\ref{fig:Bim:AllTimesteps} is well visible for both types of
simulations.
The most central bin constantly increases up to the value $\sim 2.0$. It is
caused by the increasing number of BSs in the central region. The 3rd and 6th
bins constantly decrease, which corresponds to the continuous decrease of the
region around $r_{min}$. The 10th bin remains more or less constant but it has
a very large dispersion.
The values for $R_{BSS}$, for bins $> 1st$, scatter for both simulations
significantly. The values for 3rd and 6th bins are alternately smaller and
larger with respect to each other. The overall trend is well visible but
differences between consequent snapshots are huge. It covers up the signs
of bimodal spatial distribution between consequent snapshots. The purpose of
this figure is twofold. First of all, it shows that the agreement between N-body
\textsc{dh} simulations and \textsc{mocca} \textsc{ah} simulation is good. It shows
that the \textsc{mocca} code properly deals with the mass segregation process and thus it is a suitable tool to study radial distribution of
exotic objects like BSs.
The second goal was to show how greatly chaotic bins are for all snapshots in time, for bins outside the central 1st bin.

The signs of bimodal spatial distribution for the simplified model defined by
\citet{Ferraro2012Natur.492..393F} is very chaotic both for N-body (\textsc{dh})
and \textsc{mocca} (\textsc{ah}) simulations. Thus, we decided to check the
signs of the bimodal distribution for even a simpler model with only two test
masses: 99\% of MS stars with $0.4 M_{\odot}$ and 1\% of BSs stars with $1.2
M_{\odot}$. This model is called \textsc{ah2} simulation.
All other initial conditions for simulation \textsc{ah2} are the same as for the
\textsc{ah} simulation. Thus, the model \textsc{ah2} is the simplest possible,
for which the mass segregation works only for 2 different masses.
The goal was to check whether the signs of the bimodal spatial distributions
will be easier to see in comparison to \textsc{ah} simulation (with 3 different
masses). The scaling of the normalized number of BSs ($R_{BSS}$) for
\textsc{ah2} model is done with luminosities (there are no RGB stars there).

\begin{figure}
  \includegraphics[width=\columnwidth]{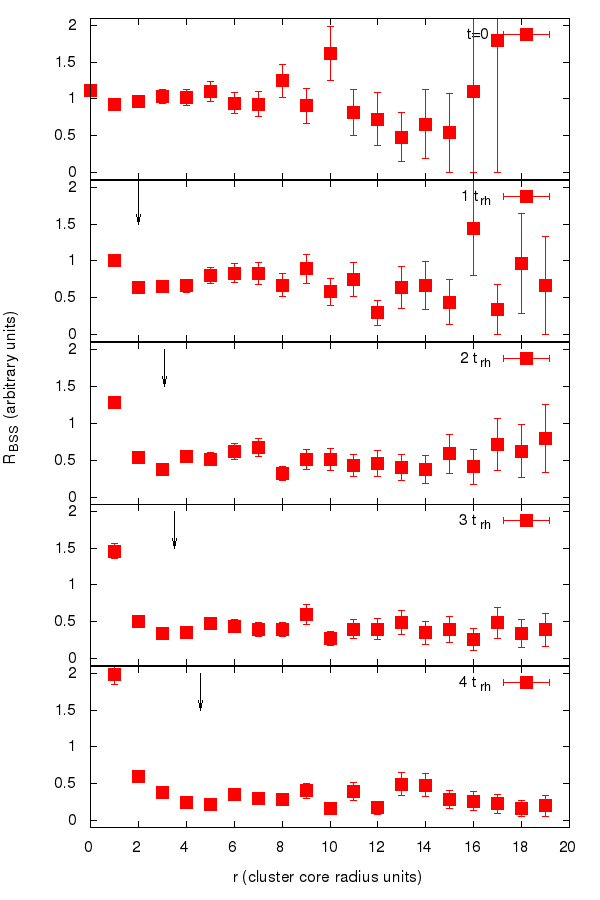}
  \caption[MOCCA simulation showing a drift of $r_{min}$ values with time for the
  simplified model consisting only of 2 different masses]{BSS radial
  distributions of BSs ($R_{BSS}$) obtained from the
  \textsc{mocca} simulation as an attempt to reproduce $r_{min}$ drift showed by
  \citet{Ferraro2012Natur.492..393F} but for the simplified model
  consisting of only 2 masses: 99\% of MS stars with $0.4 M_{\odot}$ and 1\% of
  BSs with $1.2 M_{\odot}$.}
  \label{fig:Bim:2mass}
\end{figure}

The results
for a few half-mass relaxation times for \textsc{ah2} simulation are presented
in Fig.~\ref{fig:Bim:2mass}. It shows the formation
and evolution of the dip around $r_{min}$ value. The minima turned out to be
hardly visible, even more difficult than for the \textsc{ah} simulation. The signs
of the bimodal spatial distribution are chaotic too. The differences between consequent
snapshots are significant -- the sign of bimodality can disappear between two
following snapshots in time. Even for the simplest possible model, the
signs of bimodal distribution are very hard to observe. Thus, for the real-size
star cluster the signs of bimodality are expected to be very chaotic, i.e., present
for some snapshot and vanishing within the next one or more snapshots.

Another significant difference between \textsc{ff} simulations and \textsc{dh},
\textsc{ah}, \textsc{ah2} simulations concerns the average value of $R_{BSS}$
 outside the $r_{min}$ values. It decreases with time. However, we
expected to see them more or less around the value 1.0. The $R_{BSS} \sim 1.0$
is the expected value for the regions which are not yet affected much by the
mass segregation (due to larger distances from the center of star cluster). This
feature was looked for in \textsc{ff} simulations in order to find the
$r_{min}$. However, the values of $R_{BSS}$ are constantly decreasing with time
for all of the \textsc{dh}, \textsc{ah}, \textsc{ah2} simulations (see
Fig.~\ref{fig:Bim:DouglasNbody}, Fig.~\ref{fig:Bim:MoccaFerraroModel},
Fig.~\ref{fig:Bim:2mass}). This constant decrease is likely to originate from the fact that some of BSs move to the cluster's center faster
than the local mass-segregation time. If it happens by a chance that a star (on
elongated orbit), while moving through the pericenter of its orbit, will have a
close two-body interaction, it may loose more energy and sink to the center more
quickly. Nevertheless, \textsc{dh}, \textsc{ah}, \textsc{ah2} simulations showed
that it is rather a wrong assumption to expect the $R_{BSS}$ values to stay at
$\sim 1.0$ outside the $r_{min}$.

The comparison between \textsc{dh} and \textsc{ah} simulations shows that the
\textsc{mocca} code can follow the mass segregation and thus the changes in positions of BSs as accurately as N-body
codes.
Thus, the \textsc{mocca} code is a proper tool to study the formation and evolution of the bimodal
spatial distribution for real-size stars clusters. 




\subsection{Bimodal distribution for real-size globular clusters}
\label{sec:Bim:RAvoidDrift}

\begin{figure*}
		\includegraphics[width=9cm]{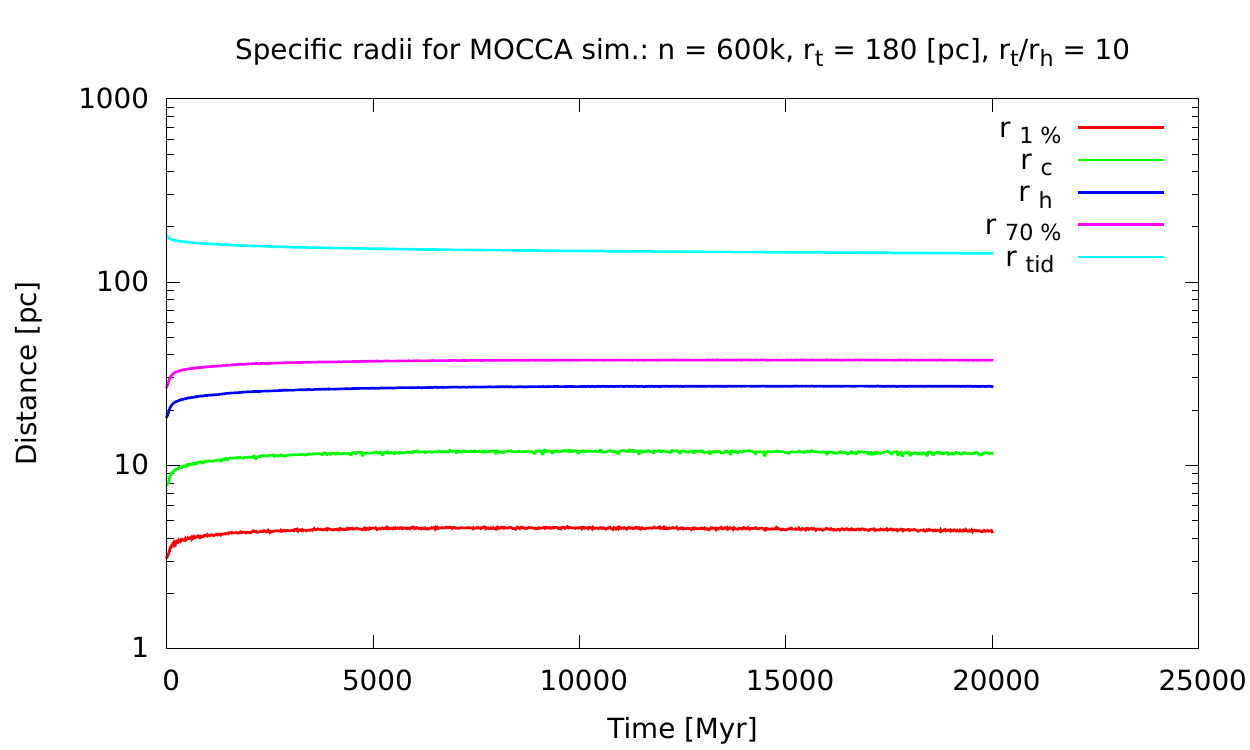}
		\includegraphics[width=9cm]{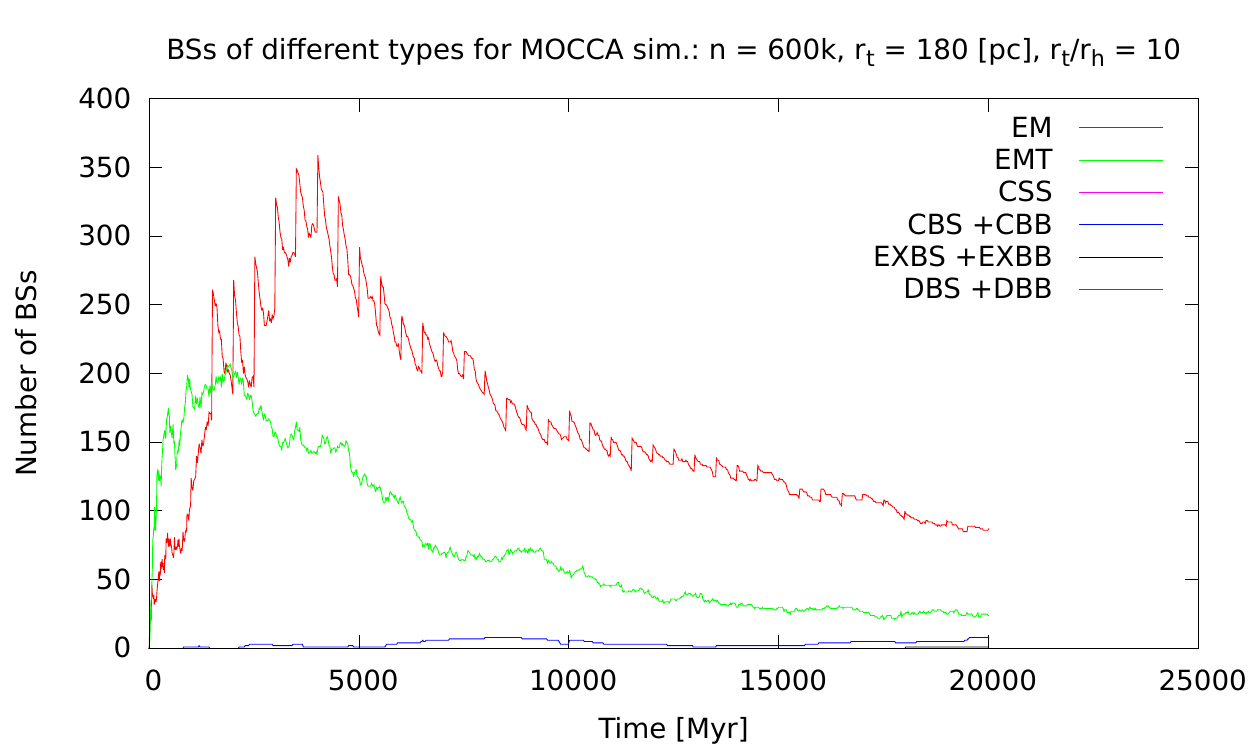}
		
		\includegraphics[width=9cm]{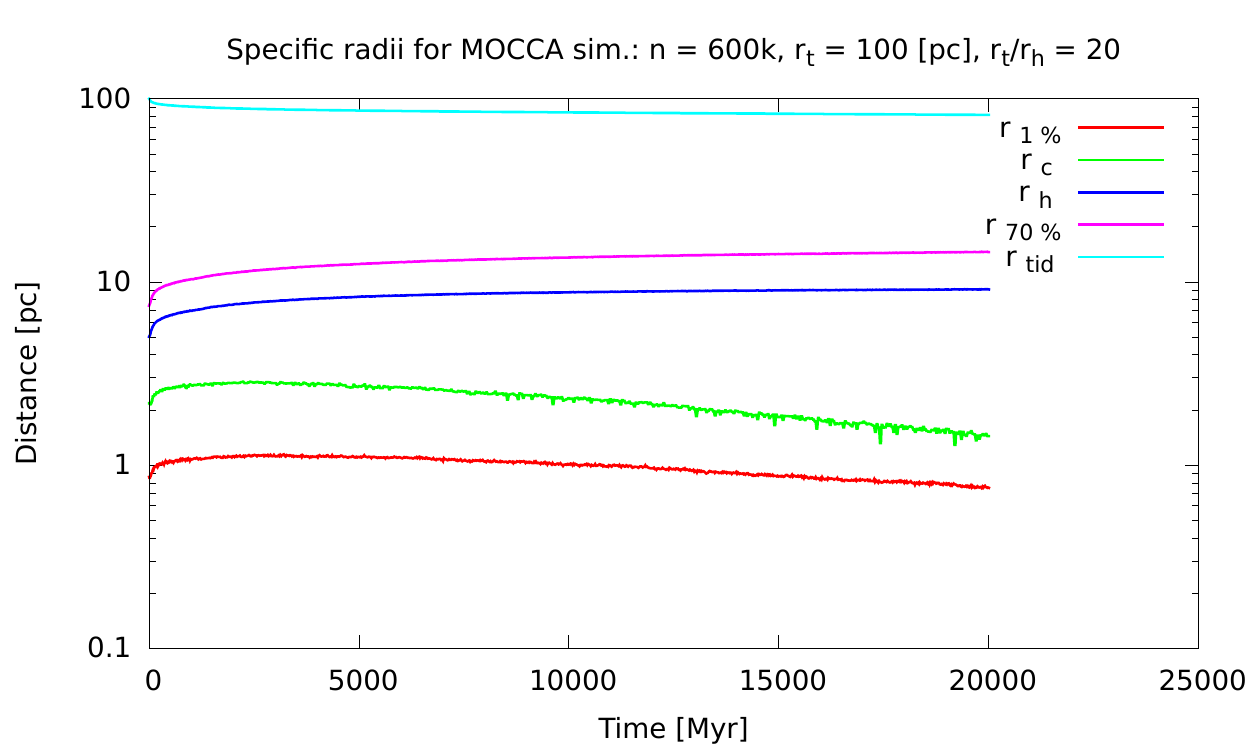}
		\includegraphics[width=9cm]{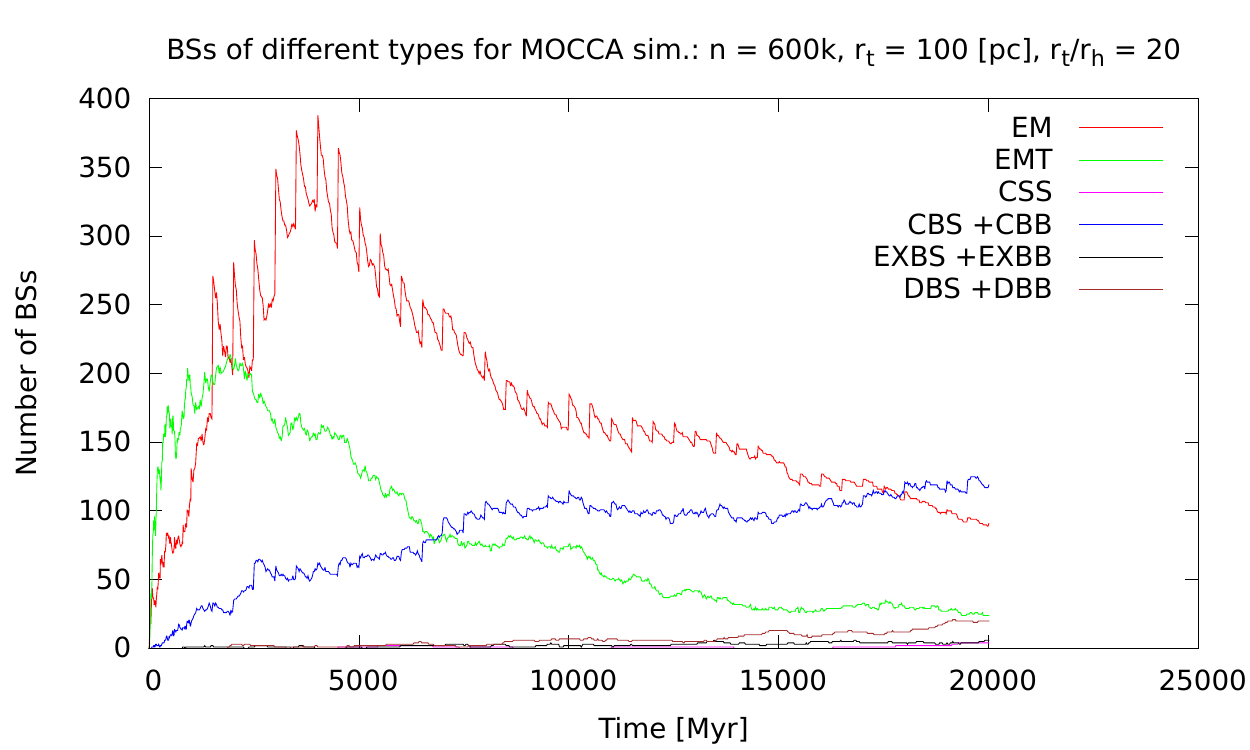}
		
		\includegraphics[width=9cm]{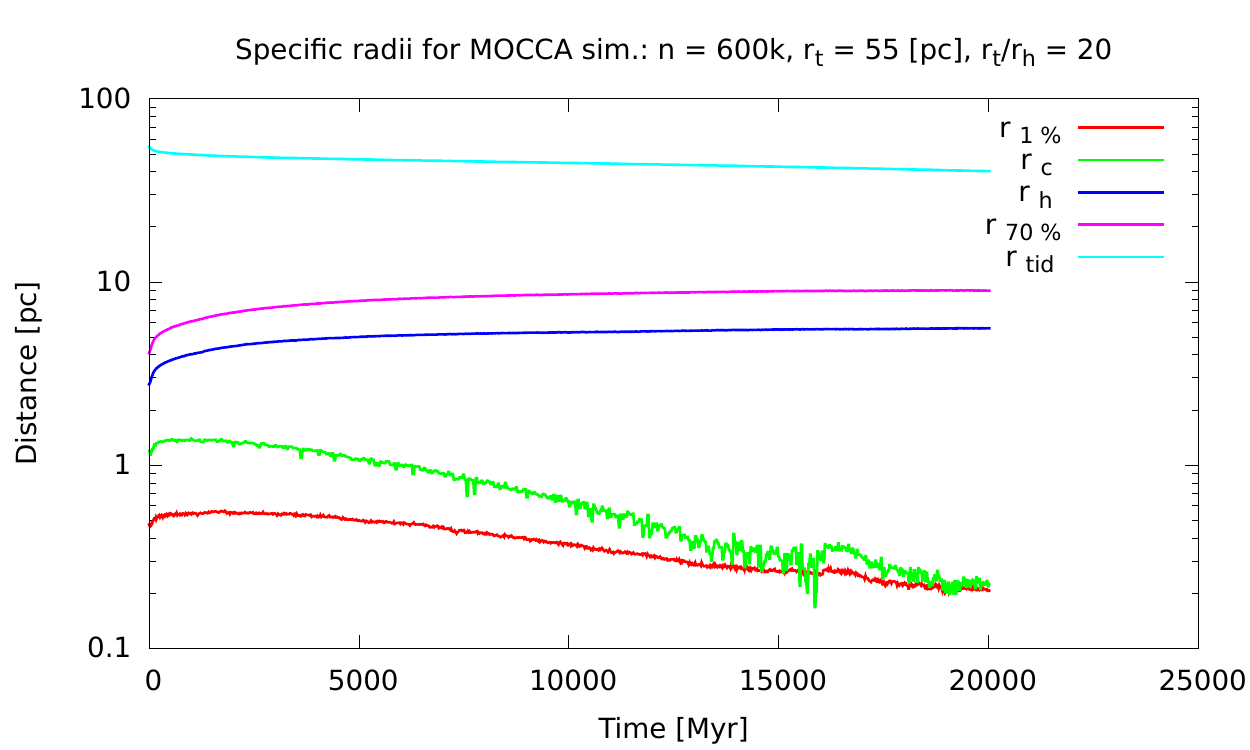}		
		\includegraphics[width=9cm]{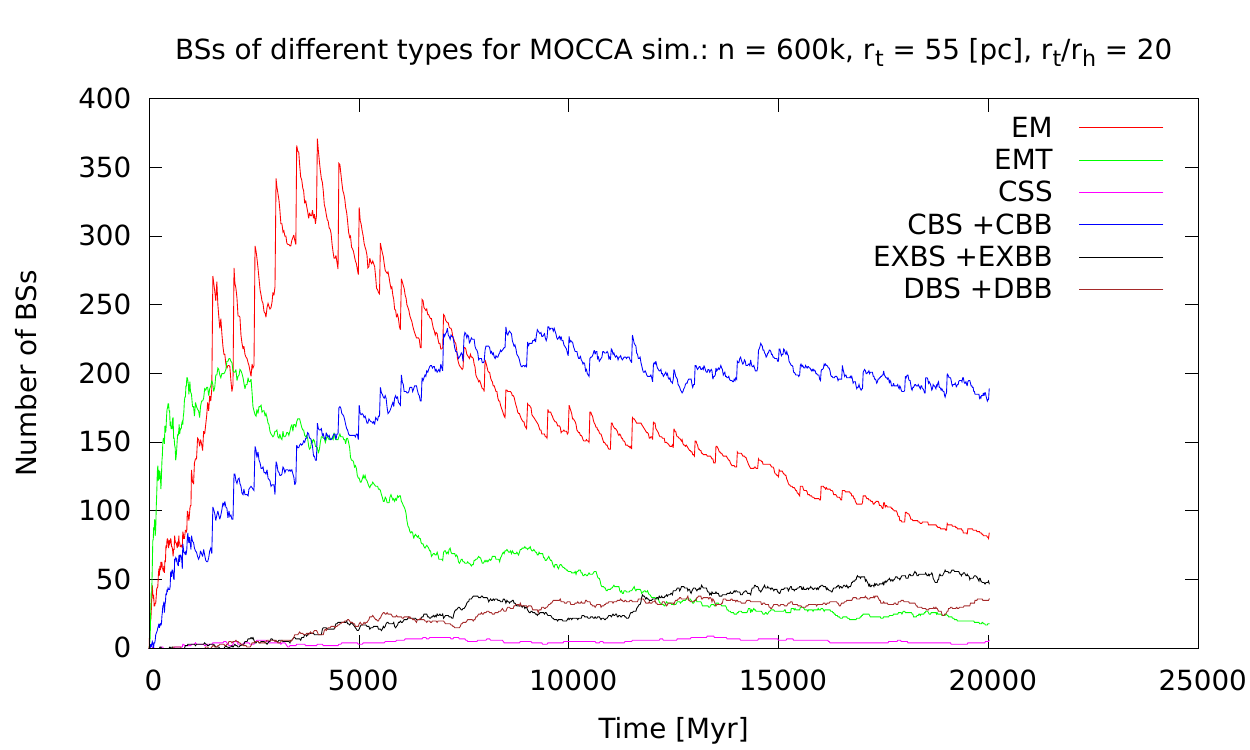}
\caption[Characteristic radii and number of BSs of different types for three
models with different dynamical scales as examples for different rates of the
formation of the bimodal spatial distribution]{Characteristic radii and number
of BSs of different types for \textsc{mocca-slow} (top row),
\textsc{mocca-medium} (middle row) and \textsc{mocca-fast} simulations
(bottom row).
The first column presents a few specific radii as a function of time, like
core radius ($r_c$), half-mass radius ($r_h$), tidal radius ($t_{tid}$), and some 
Lagrangian radii. The second column shows the number of BSs of
different types: EM -- evolutionary mergers, EMT -- evolution mass transfer,
CBS/CBB -- collisional binary-single, binary-binary, EXBS/EXBB -- exchanged
binary-single, binary-binary, and DBS/DBB -- dissolved binary-single,
binary-binary. For the initial conditions of the models see
Tab.~\ref{tab:Bim:InitParams} and for the discussion of the plots see
Sect.~\ref{sec:Bim:RAvoidDrift}.}
\label{fig:Bim:Radii}
\end{figure*}
\begin{figure*}
		\includegraphics[width=9cm]{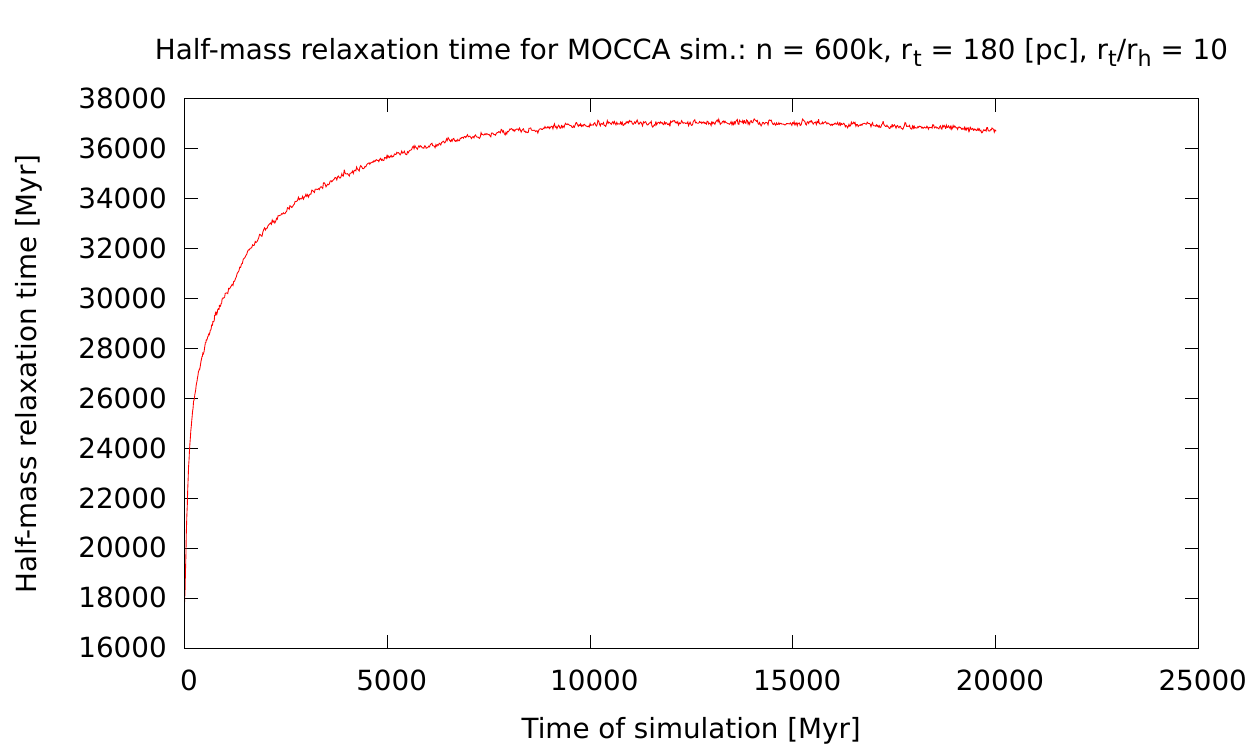}
		\includegraphics[width=9cm]{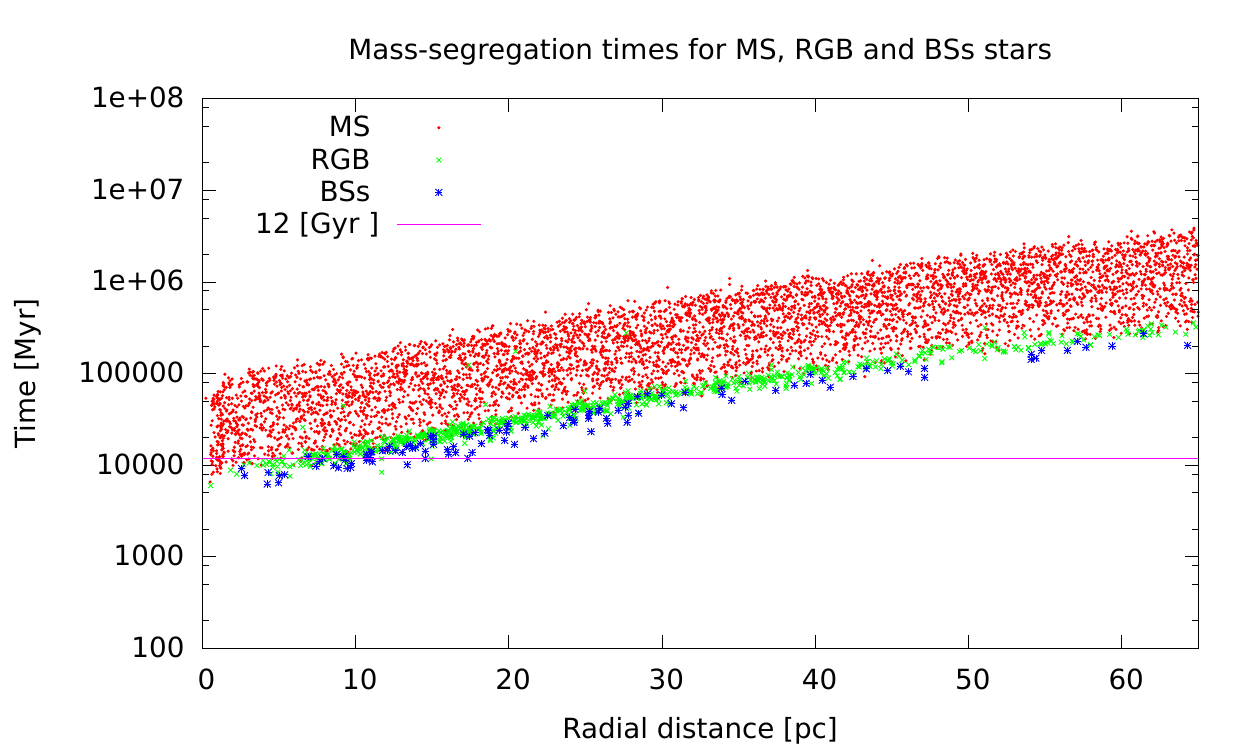}
		
		\includegraphics[width=9cm]{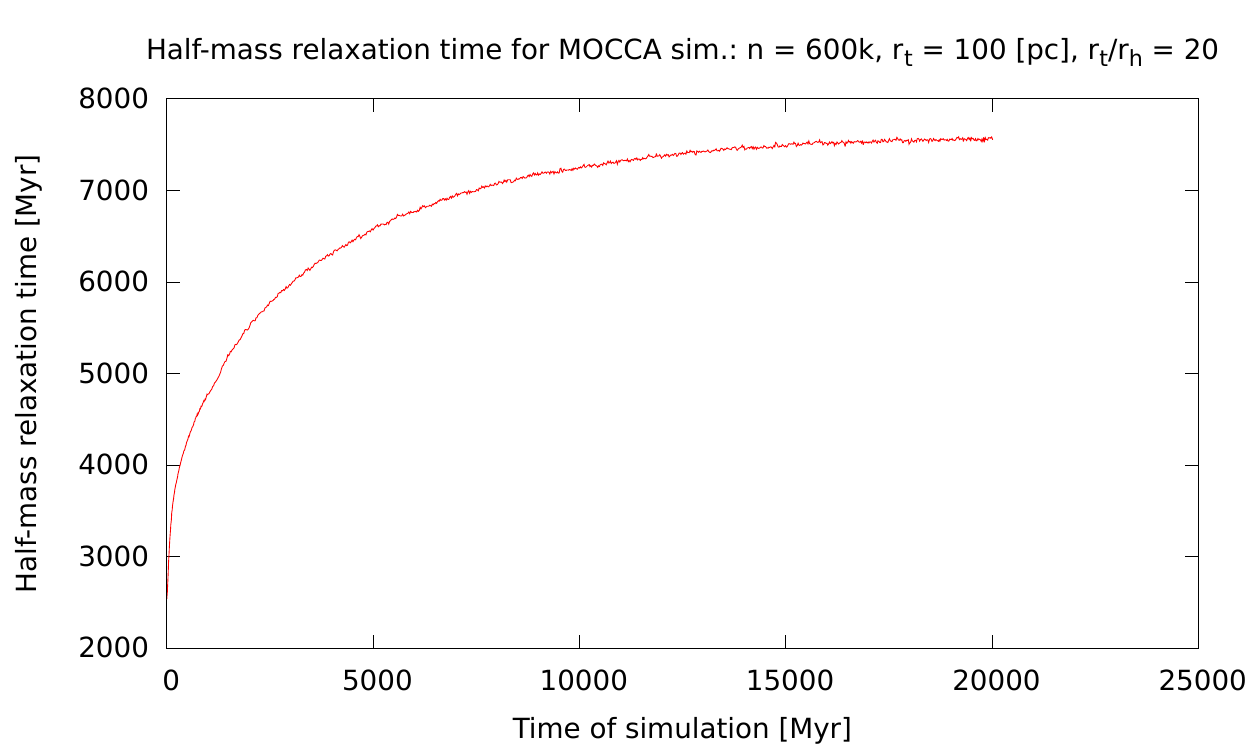}
		\includegraphics[width=9cm]{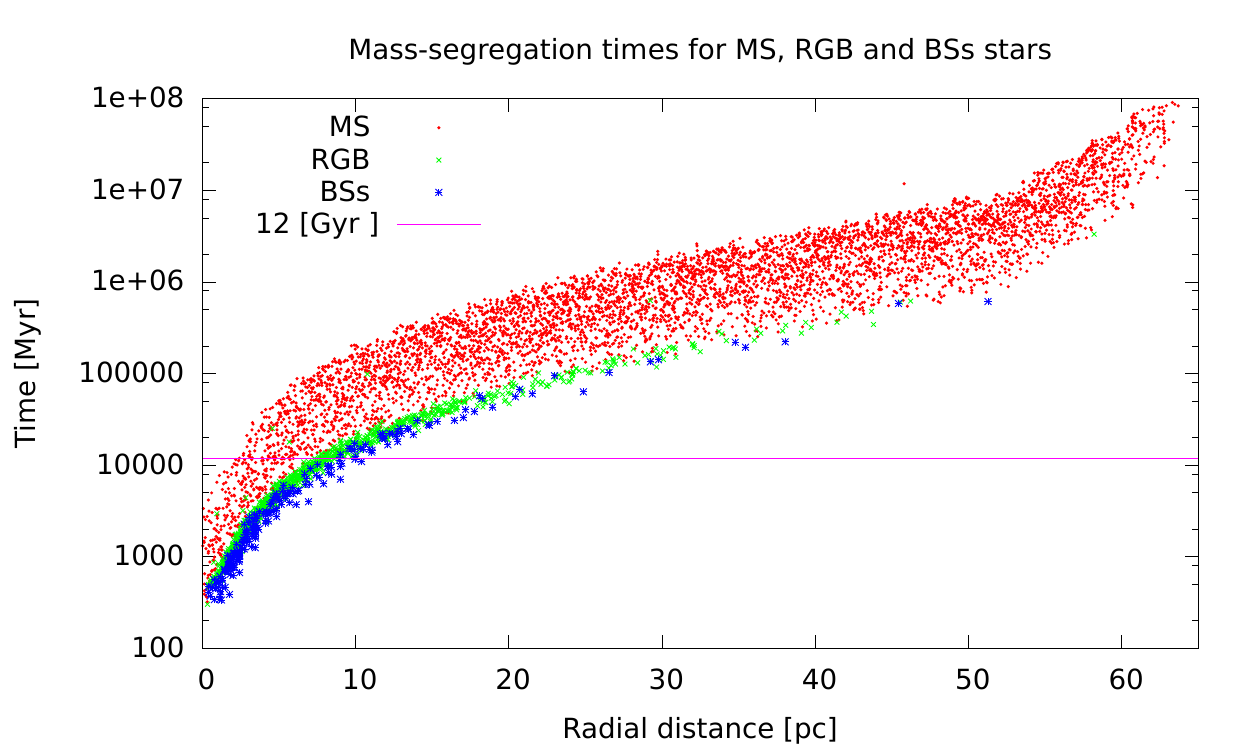}

		\includegraphics[width=9cm]{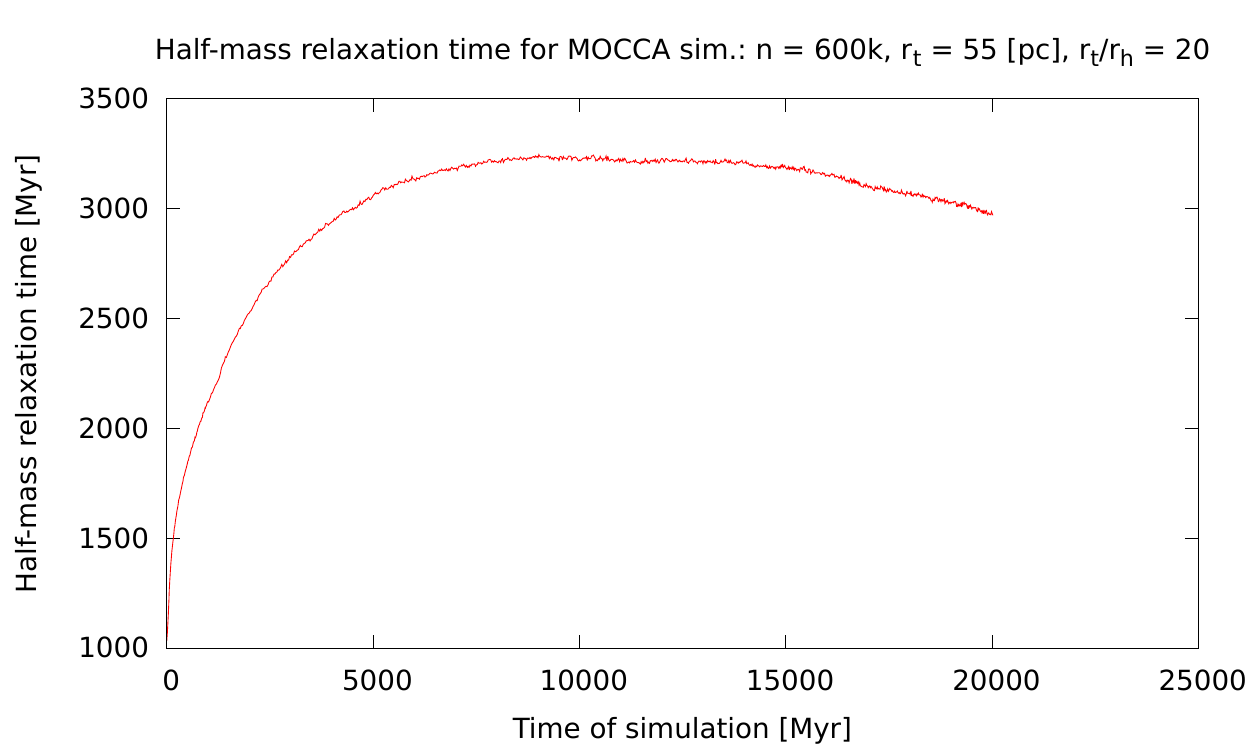}
		\includegraphics[width=9cm]{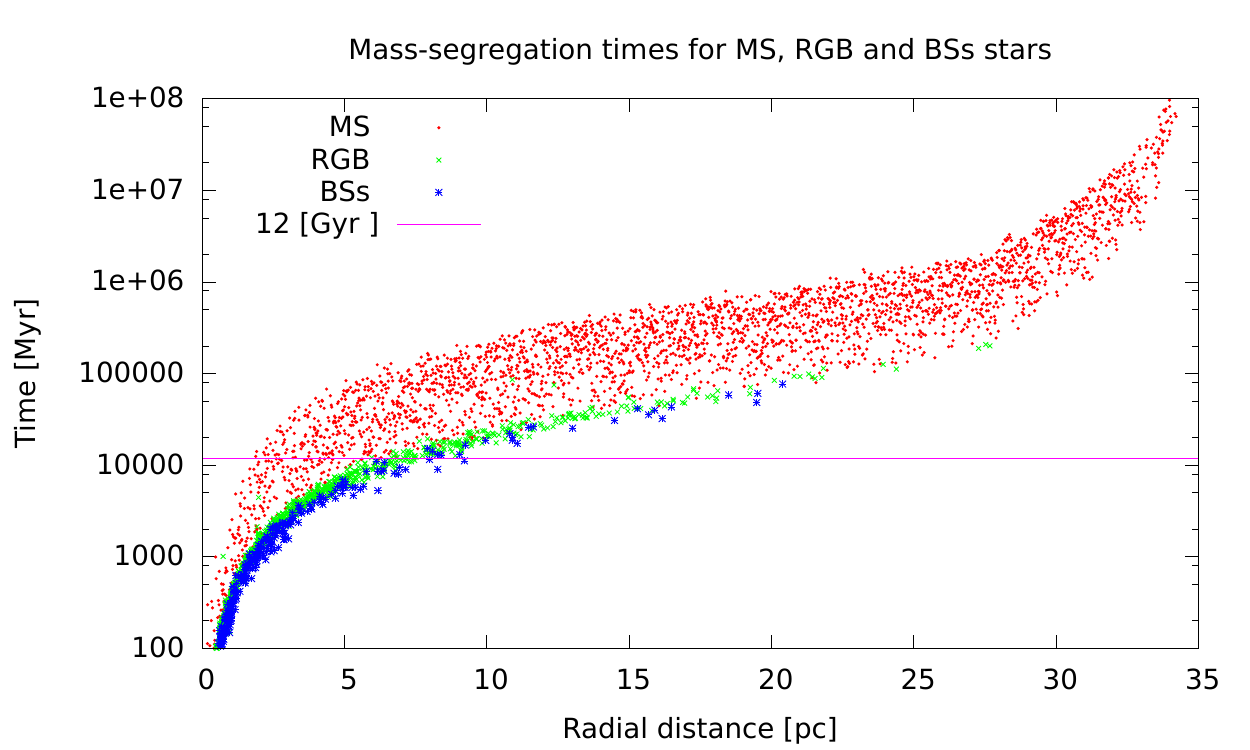}		
\caption[Half-mass relaxation and mass-segregation times for three models with different dynamical scales as examples for different rates of the formation of the bimodal spatial distribution]{Half-mass relaxation and mass-segregation times for
\textsc{mocca-slow} (top row), \textsc{mocca-medium} (middle row) and
\textsc{mocca-fast} simulations (bottom row). The first column shows the changes
with time of the half-mass relaxation time.
The second column contains plots with mass-segregation times for BSs, RGB
and MS stars at the time 12~Gyr. The violet line
represents the time 12~Gyr itself.
For the initial conditions
of the models see Tab.~\ref{tab:Bim:InitParams}, and for the discussion of the
plots see Sect.~\ref{sec:Bim:RAvoidDrift}.}
\label{fig:Bim:TimeScales}
\end{figure*}

Three selected models, showing the formation and evolution of the spatial
distribution of BSs, were chosen to have different relaxation times and thus
different rates of mass segregations.
Tab.~\ref{tab:Bim:InitParams} summarizes the initial parameters for them. The
only differences between the models are in the concentrations and the tidal
radii. Thus, their rate of the dynamical evolution varies. The slowest evolving
model, \textsc{mocca-slow}, has large tidal radius $r_{tid} = 180$~[pc], and
small concentration $c = r_{tid} / r_{h} = 10$.
The GC with slightly faster dynamical evolution, \textsc{mocca-medium}, has
smaller tidal radius $r_{tid} = 100$~[pc], and larger concentration $c = 20$.
The fastest evolving GC, \textsc{mocca-fast}, has even smaller tidal radius
$r_{tid} = 55$~[pc] with the same concentration $c = 20$. All simulations were
computed up to 20~Gyr. The implications of the different rate of evolution of
GCs on the population of BSs are presented in Fig.~\ref{fig:Bim:Radii} and
Fig.~\ref{fig:Bim:TimeScales}. The selected models are actually equivalent to
models presented in Sect.~\ref{sec:InitialParameters}, but they are renamed in
this section for the sake of clarity. Model \textsc{mocca-slow} is equal to
model \textsc{mocca-37}, \textsc{mocca-medium} to \textsc{mocca-34} and
\textsc{mocca-fast} to \textsc{mocca-30}.

\begin{table*}
	    \begin{tabular}{p{3cm} | p{3cm} | p{3cm} | p{3cm}}
	    Parameter  & \textsc{mocca-slow} & \textsc{mocca-medium} &
	    \textsc{mocca-fast}
	    \\
	    \hline
	    Single stars ($N_s$) & \multicolumn{3}{c}{480k}\\
	    \hline
	    Binary stars ($N_b$) & \multicolumn{3}{c}{120k}\\
	    \hline
 	    Binary fraction ($f_b = \frac{N_s}{N_b + N_s}$) &
 	    \multicolumn{3}{c}{0.2}\\
	    \hline
 	    Initial model & \multicolumn{3}{c}{Plummer}\\
	    \hline
 	    IMF of stars  & \multicolumn{3}{c}{\citet{Kroupa1993MNRAS.262..545K} in
 	    the range $[0.1; 100] \mathrm{M_{\odot}}$} \\
	    \hline
 	    IMF of binaries & \multicolumn{3}{c}{\citet[eq.
 	    1]{Kroupa1991MNRAS.251..293K}, binary masses from 0.2 to
 	    100~$\mathrm{M_{\odot}}$}\\
	    \hline
 	    Total mass (M(0)) & \multicolumn{3}{c}{$3.4 \times 10^{5}
 	    \mathrm{M_{\odot}}$}\\
	    \hline
 	    Binary mass ratios & \multicolumn{3}{c}{Uniform} \\
	    \hline
 	    Binary semi-major axes & \multicolumn{3}{c}{Uniform in the logarithmic
 	    scale from $2(R_1+R_2)$ to 100~AU}\\
	    \hline
 	    Binary eccentricities & \multicolumn{3}{c}{Thermal (modified by
 	    \citet[eq.~1]{Hurley2005MNRAS.363..293H})}\\
	    \hline
 	    Metallicity  & \multicolumn{3}{c}{0.001 (1/20 of the solar metallicity
 	    0.02)}
 	    \\
	    \hline
 	    Initial tidal radius ($r_{tid}$) & 180 pc & 100 pc & 55 pc \\
	    \hline
 	    Initial half-mass radius ($r_{h}$) & 18 pc & 5 pc & 2.75 pc\\
	    \hline
 	    Initial core radius ($r_{c}$) & 8.2 pc & 3.8 pc & 2.0 pc\\
	    \end{tabular}
	\caption[Initial conditions for three models with different dynamical scales as examples for different rates of the formation of the bimodal spatial distribution]{Initial conditions for \textsc{mocca-slow}, \textsc{mocca-medium} and
	\textsc{mocca-fast} simulations. They represent slowly, slightly faster and
	fast evolving GCs. Such models were
	chosen to have different rates of dynamical evolution 
	to study the different speeds of formation and evolution of the bimodal
	spatial distribution in the real-size GCs. See details in Sect.~\ref{sec:Bim:RAvoidDrift}.}
	\label{tab:Bim:InitParams}
\end{table*}

The first column in Fig.~\ref{fig:Bim:Radii} presents the evolution of
characteristic radii for the \textsc{mocca-slow}, \textsc{mocca-medium} and \textsc{mocca-fast}
simulations. Each plot contains core radius $r_c$, half-mass radius $r_h$, tidal
radius $r_{tid}$, and two additional Lagrangian radii, $r_{1\%}$ and $r_{70\%}$.
The changes in radii show the dynamical evolution of the models. For the
\textsc{mocca-slow} simulation there are actually no changes of the  inner most
radii which means that this GC is not affected by the dynamical interactions
much. Its evolution is driven mainly by stellar evolution.
The faster evolving model, \textsc{mocca-medium}, shows changes in the $r_{1\%}$
and $r_c$. The radii get smaller with time, which means that the density
increases in the core region of the GC. The fastest evolving
GC, \textsc{mocca-fast}, gets even denser and around 16~Gyr one can see the sign
of the core collapse in $r_c$ -- it gets dynamically old.

The second column in the Fig.~\ref{fig:Bim:Radii} shows the number of BSs of
different types as a function of time for the \textsc{mocca-slow},
\textsc{mocca-medium} and \textsc{mocca-fast} simulations. 
The number of EM and EMT changes similarly
for all three simulations. It is a consequence of the same initial parameters
for binaries for the three models. EM and EMT channels in general are not
affected by dynamical interactions, thus their numbers are essentially the same
for the models with different densities but with the same initial mass
functions. The differences in the GCs' densities have a great influence on the
dynamically created BSs (CBS, CBB).
Their number differ a lot between the models. The densest and fastest evolving
model, \textsc{mocca-fast}, has the largest number of them because of the more
frequent strong dynamical interactions between stars which lead to the BSs
creation. \textsc{mocca-fast} has also the highest number of BSs which had some
interactions with other binaries and changed their companions (EXBS, EXBB) or
were disrupted (DBS, DBB). For the details about the formation processes of BSs
see \citet{Hypki2013MNRAS.429.1221H}.

Fig.~\ref{fig:Bim:TimeScales} shows the timescales of the GCs evolution which
are relevant for the formation of the signs of the bimodal spatial
distributions. The first column in Fig.~\ref{fig:Bim:TimeScales} shows the
half-mass relaxation times ($t_{rh}$) for the \textsc{mocca-slow},
\textsc{mocca-medium} and \textsc{mocca-fast} models. The $t_{rh}$ is the
largest for slowest evolving model and is much larger than the age of the Universe.
Whereas the $t_{rh}$ increases for \textsc{mocca-fast} models to only about 3~Gyr
after the first few Gyr of the simulation, which is much less than the Hubble time.
The second column of Fig.~\ref{fig:Bim:TimeScales} presents the mass-segregation
times 
for 12~Gyr snapshot for BSs, RGB and MS stars. The
mass-segregation time gives an impression on how much time is needed for a star
at a distance r to sink to the center of GC.
For most massive stars, BSs, the times are lowest.
The RGB stars, with masses just slightly smaller than these of BSs, have the
mass-segregation times only slightly larger. The violet line which denotes
12~Gyr is plotted for a reference to show which portion of stars could be affected already by the mass-segregation process. For the \textsc{mocca-slow}
one can see that almost all BSs and RGB stars are above this 12~Gyr limit.
Whereas for \textsc{mocca-fast} almost all BSs are well below 12~Gyr.

For the models with different rates of the dynamical
evolution the formation of the signs of bimodal spatial distribution should be
different. For the \textsc{mocca-slow} we expected that the bimodality
will not be visible for a very long time. Whereas for
\textsc{mocca-fast} we expected to observe it earlier and with a larger dip around
$r_{avoid}$. Additionally, the drift of $r_{avoid}$ should be also different for
the models. For the faster evolving GCs its value in the units of e.g. the
core radius should be larger.

\begin{figure*}
		\includegraphics[width=5.8cm,height=3.3cm]{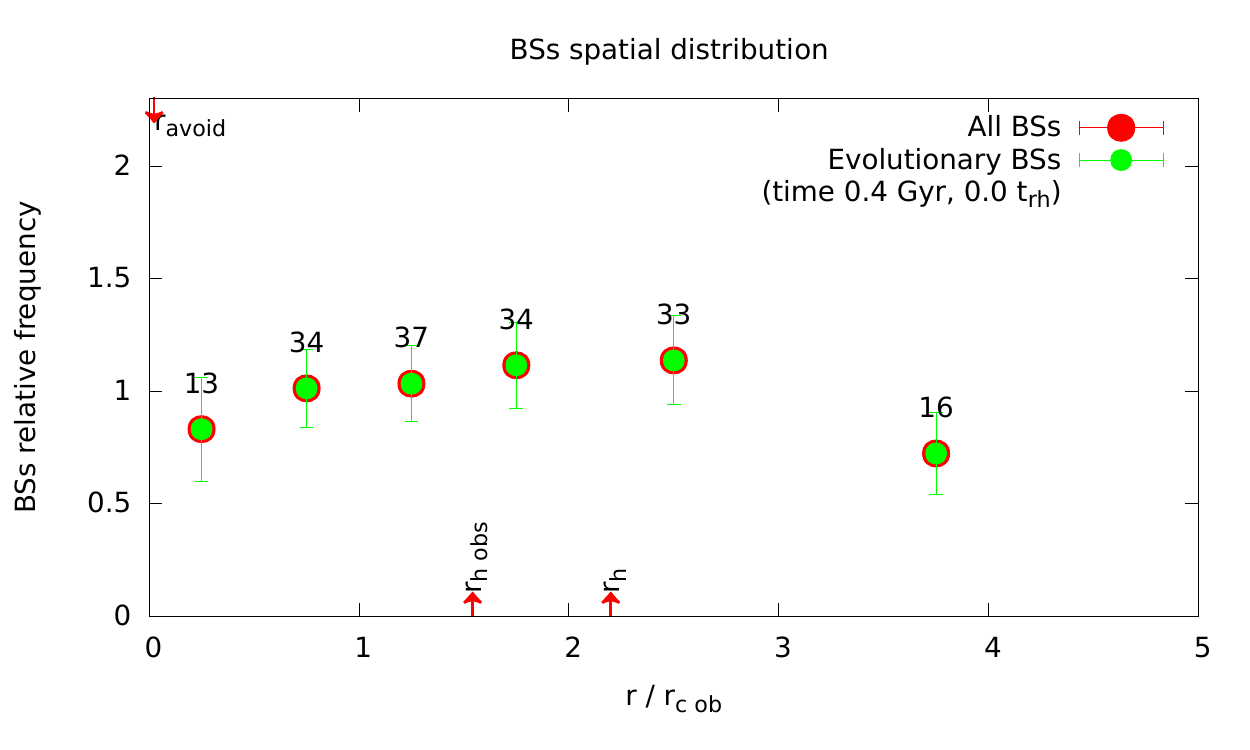}
		\includegraphics[width=5.8cm,height=3.3cm]{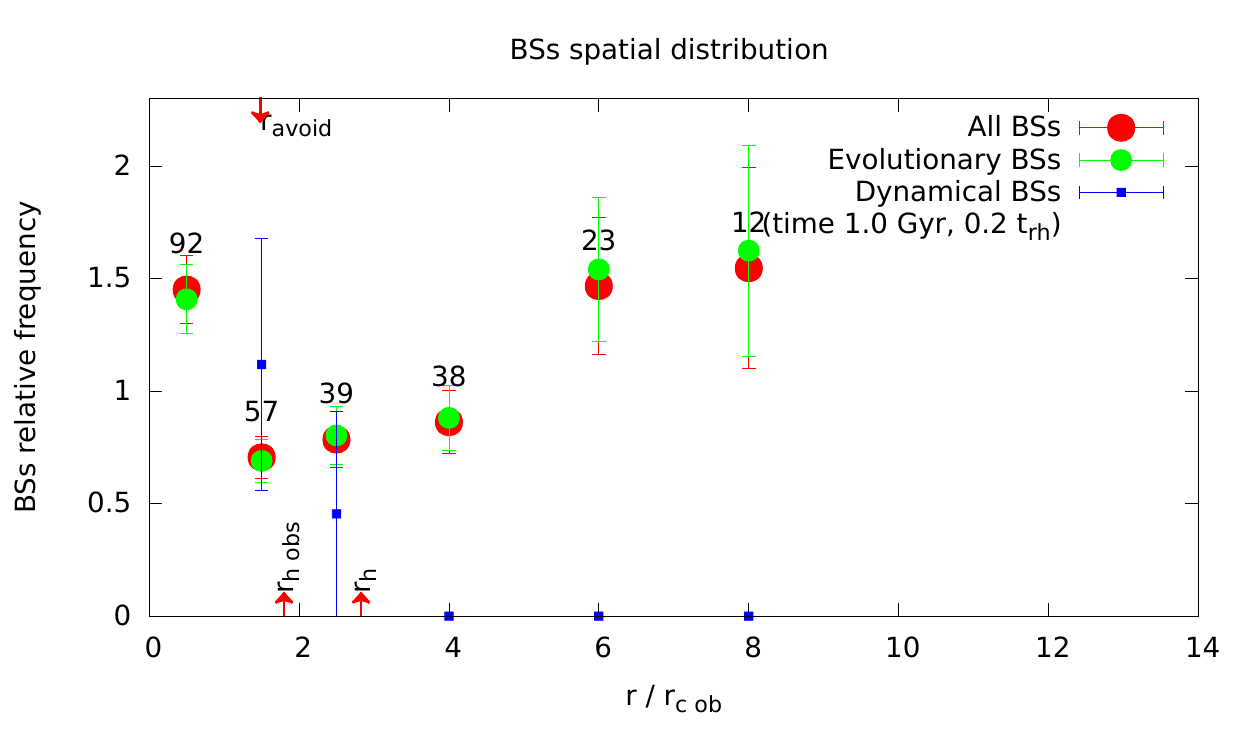}
		\includegraphics[width=5.8cm,height=3.3cm]{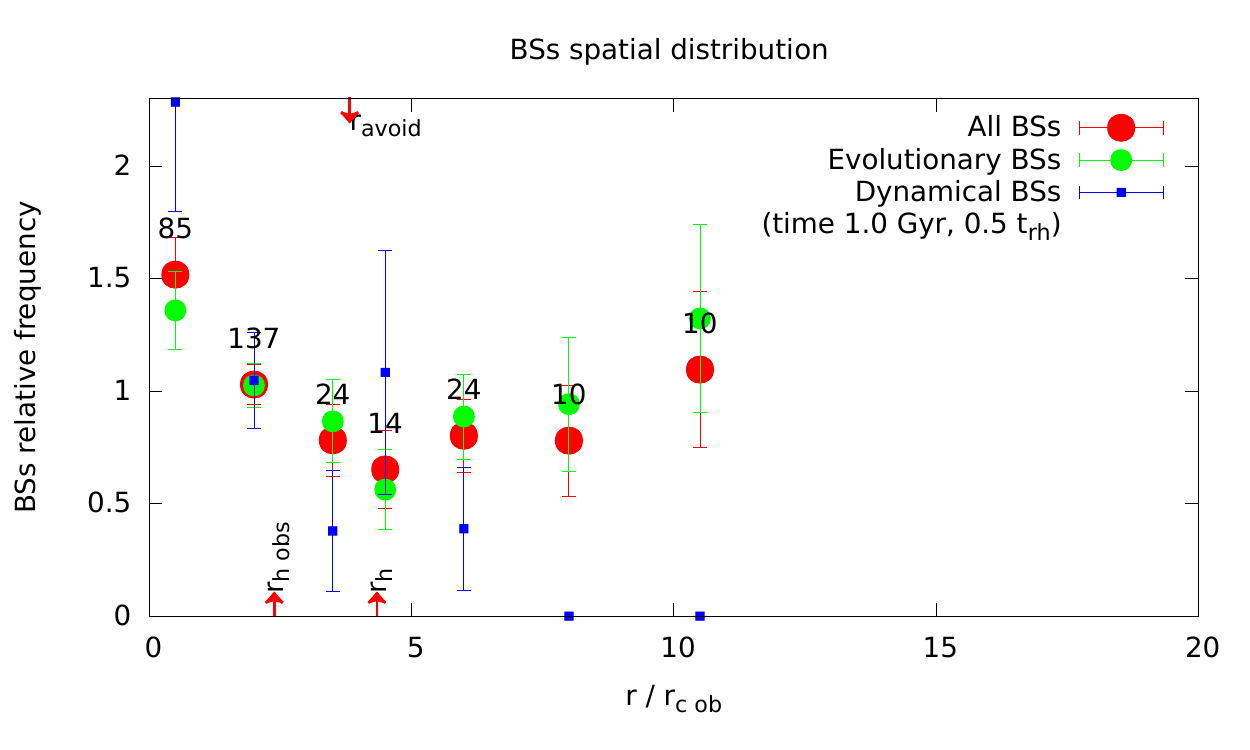}
		\includegraphics[width=5.8cm,height=3.3cm]{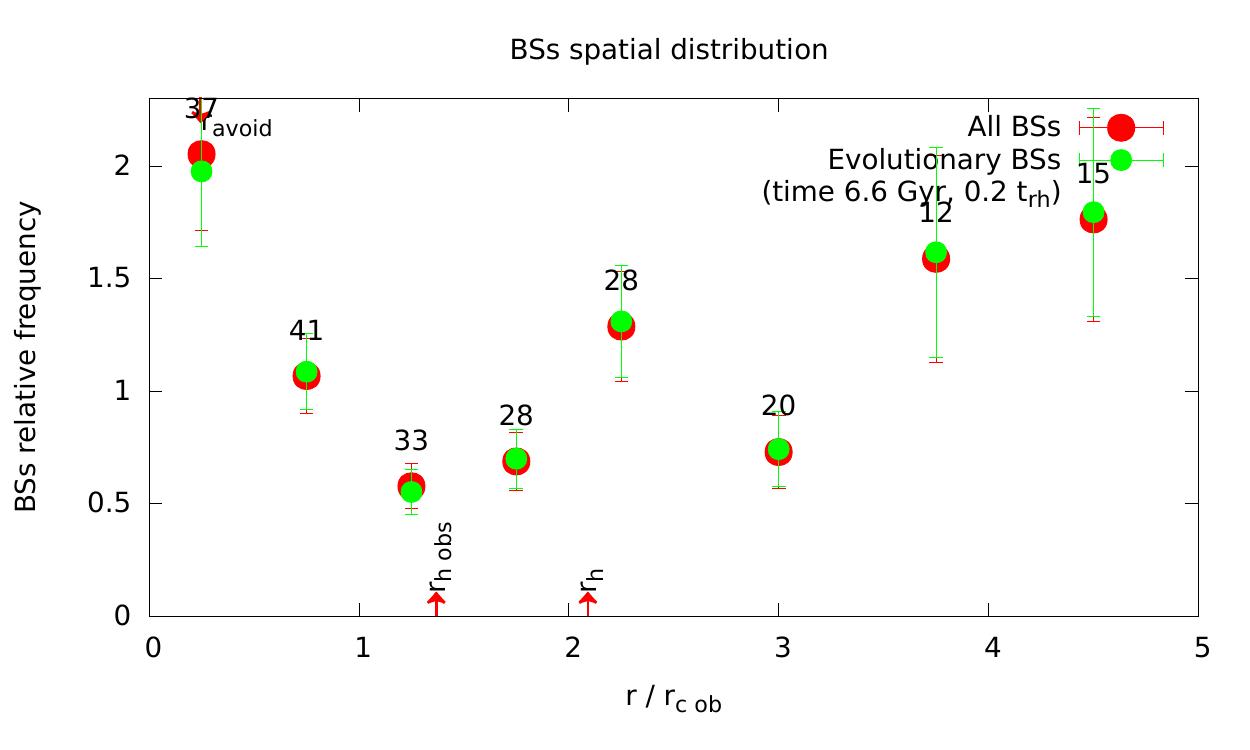}
		\includegraphics[width=5.8cm,height=3.3cm]{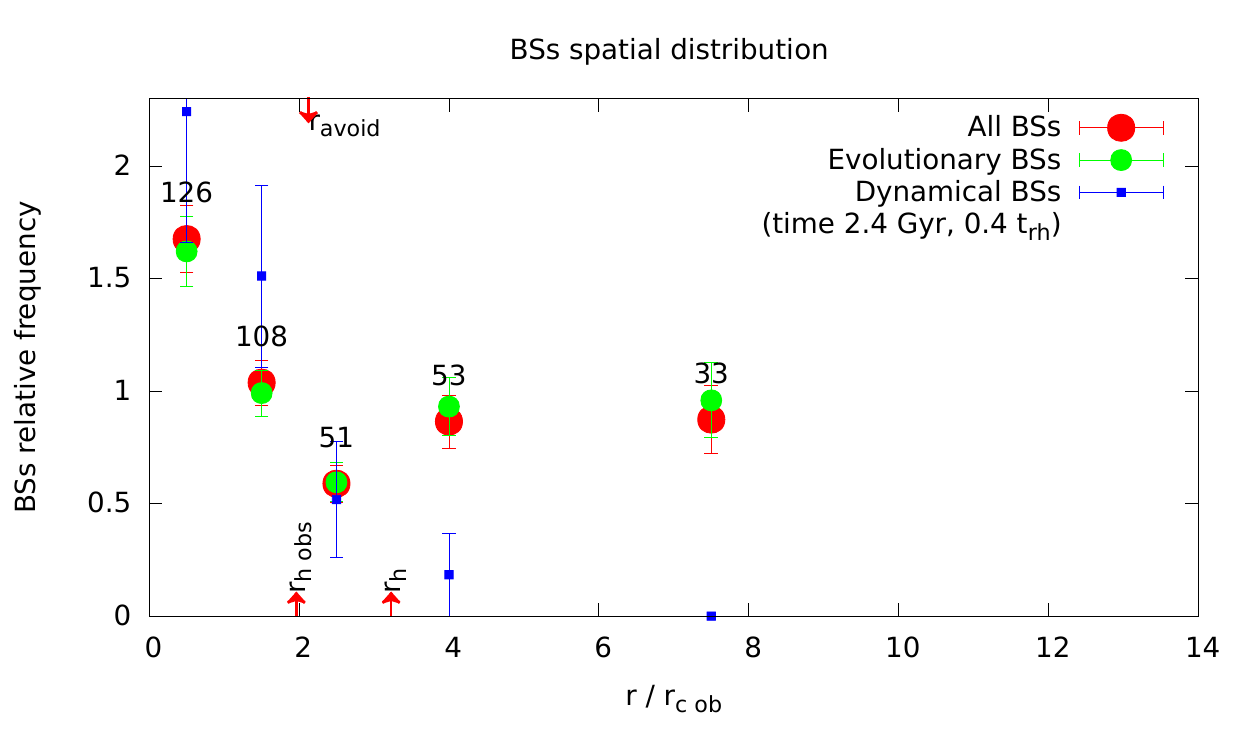}
		\includegraphics[width=5.8cm,height=3.3cm]{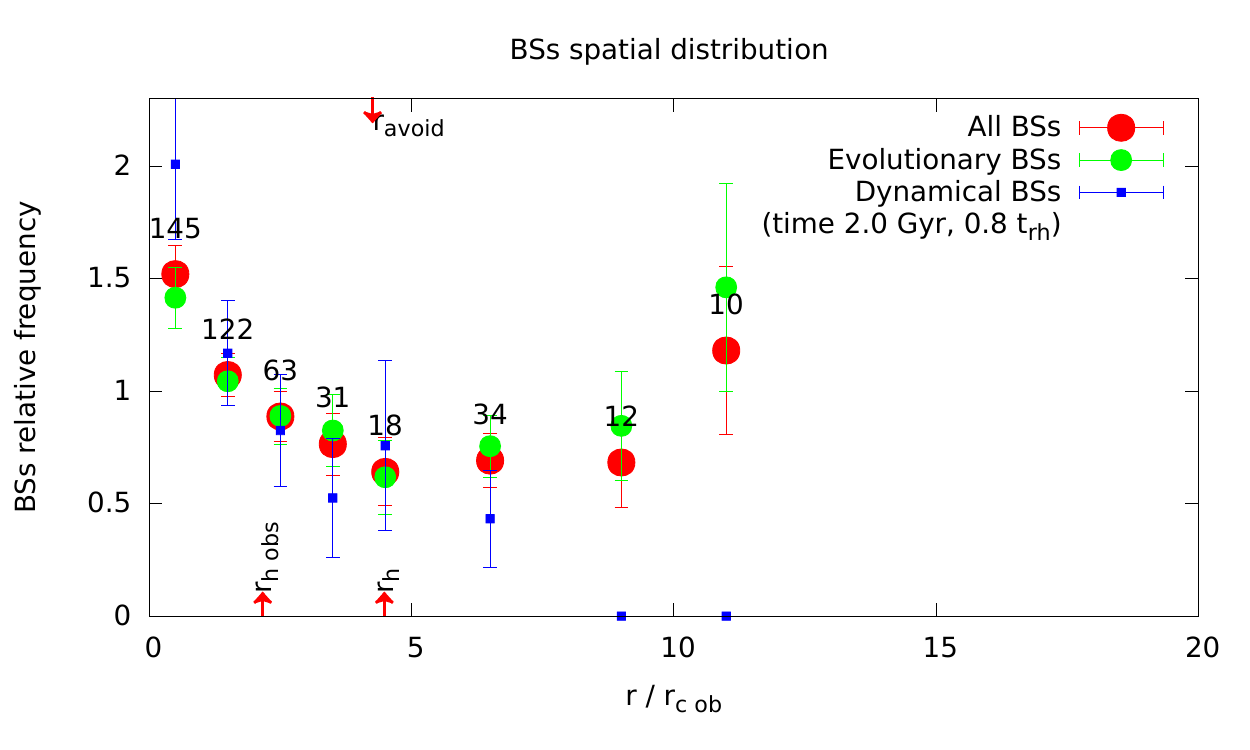}
		\includegraphics[width=5.8cm,height=3.3cm]{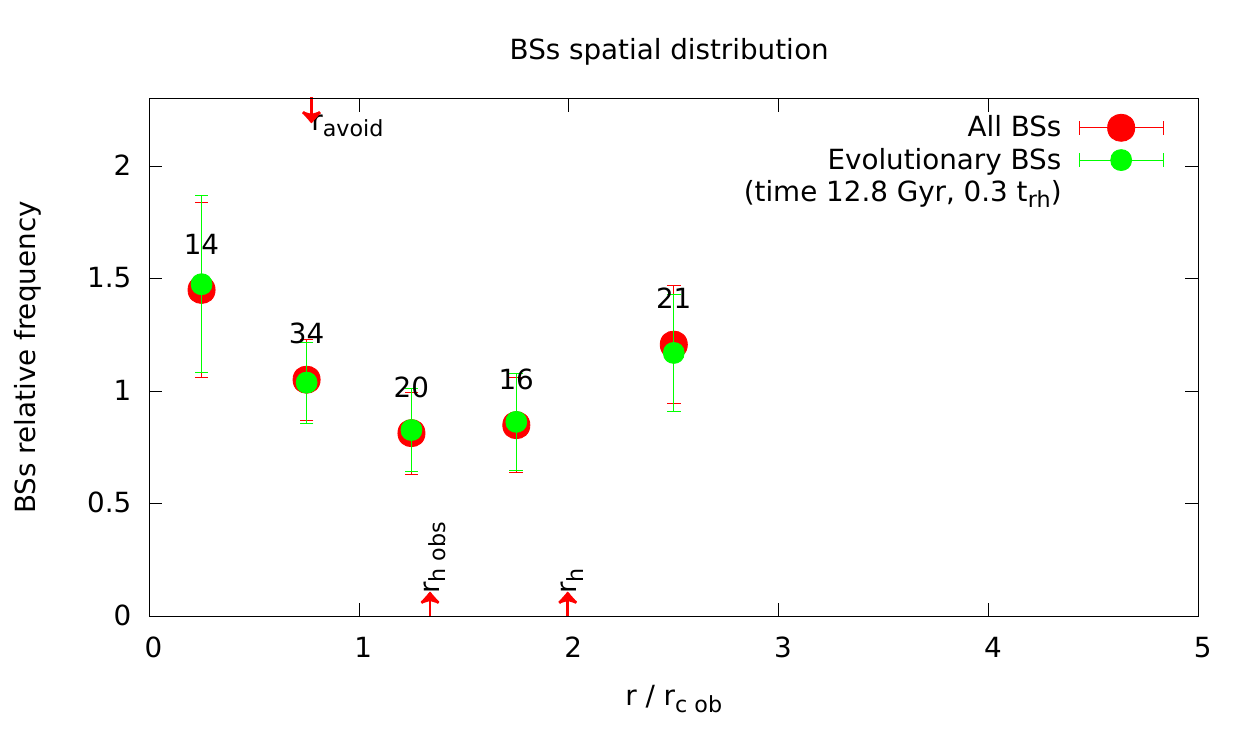}
		\includegraphics[width=5.8cm,height=3.3cm]{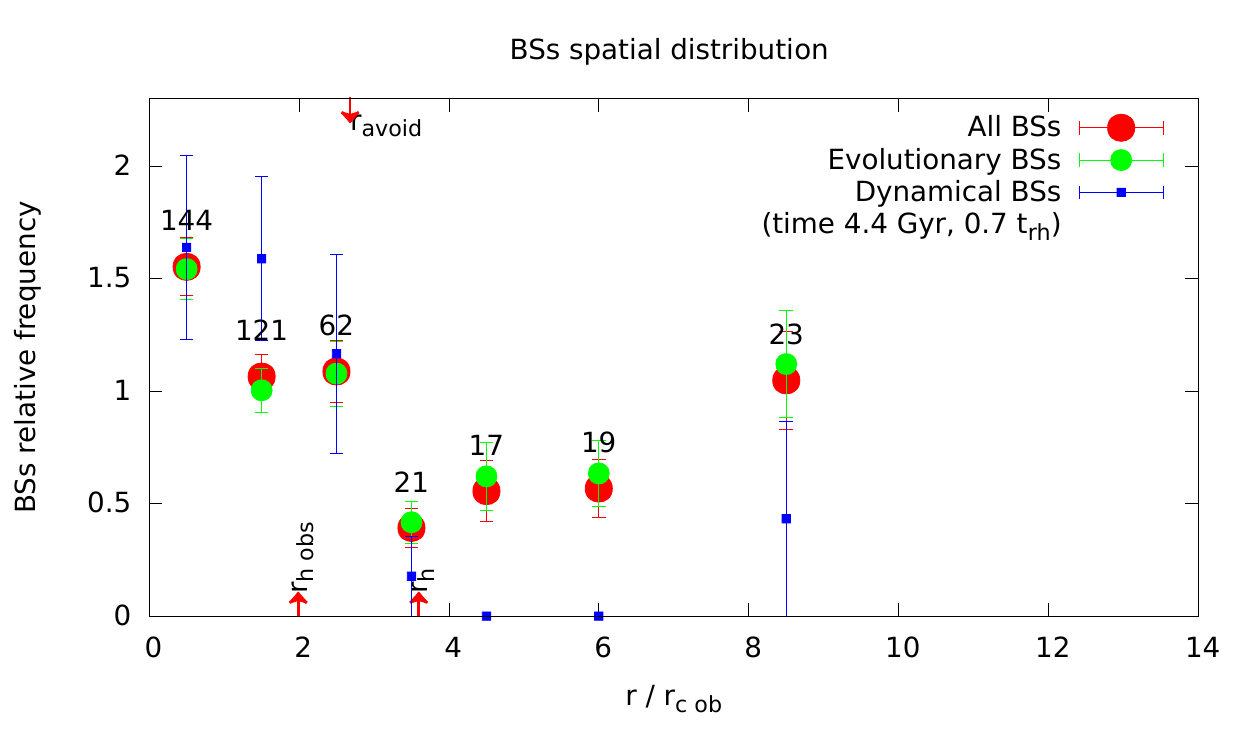}
		\includegraphics[width=5.8cm,height=3.3cm]{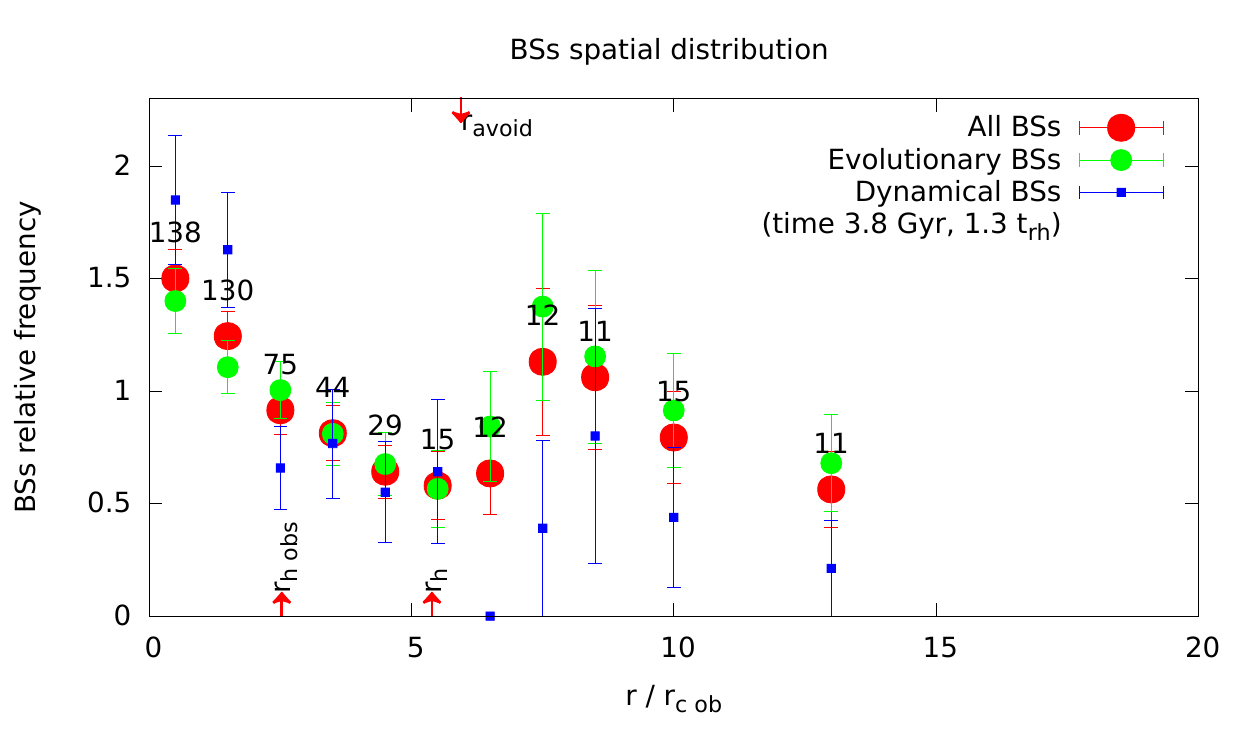}
		\includegraphics[width=5.8cm,height=3.3cm]{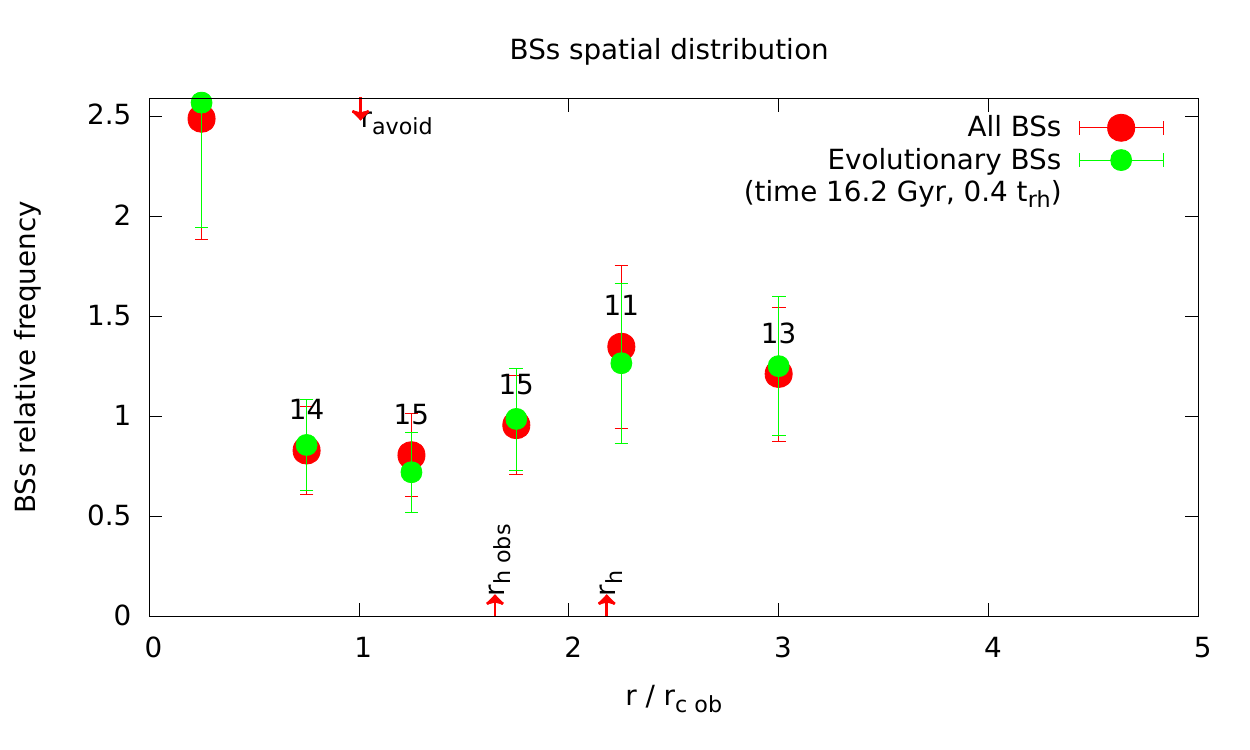}
		\includegraphics[width=5.8cm,height=3.3cm]{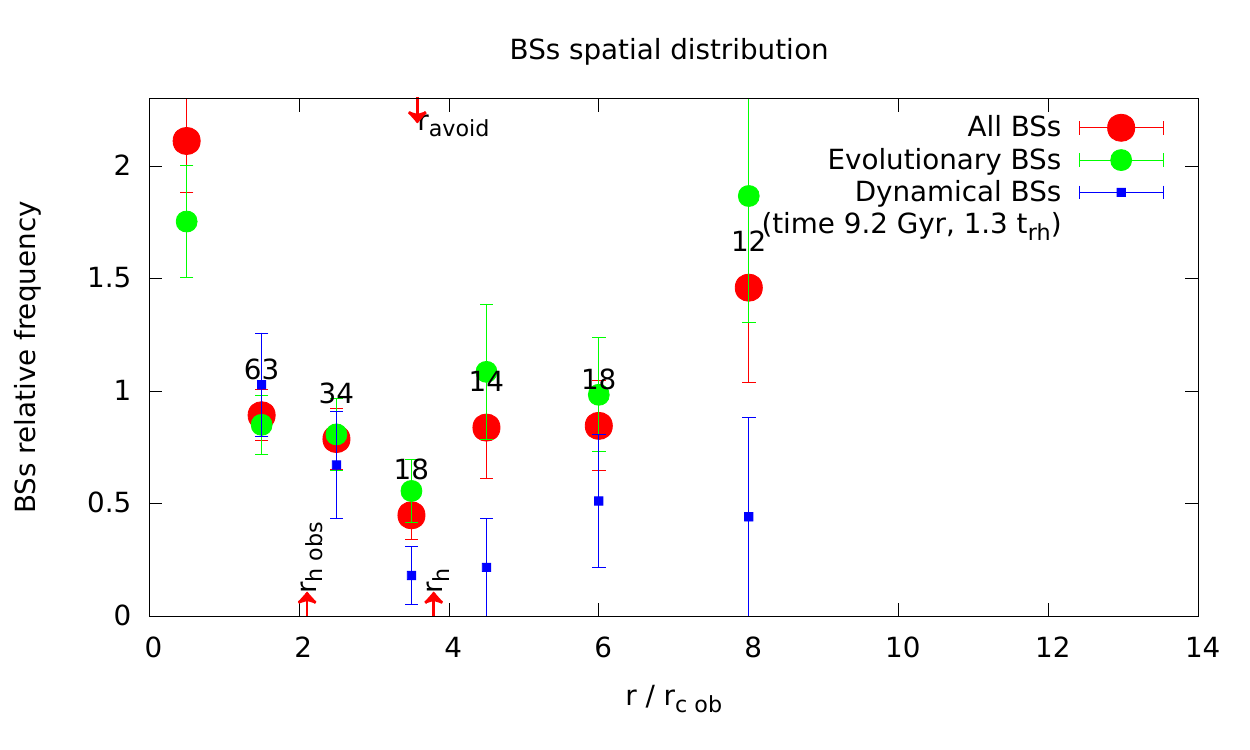}
		\includegraphics[width=5.8cm,height=3.3cm]{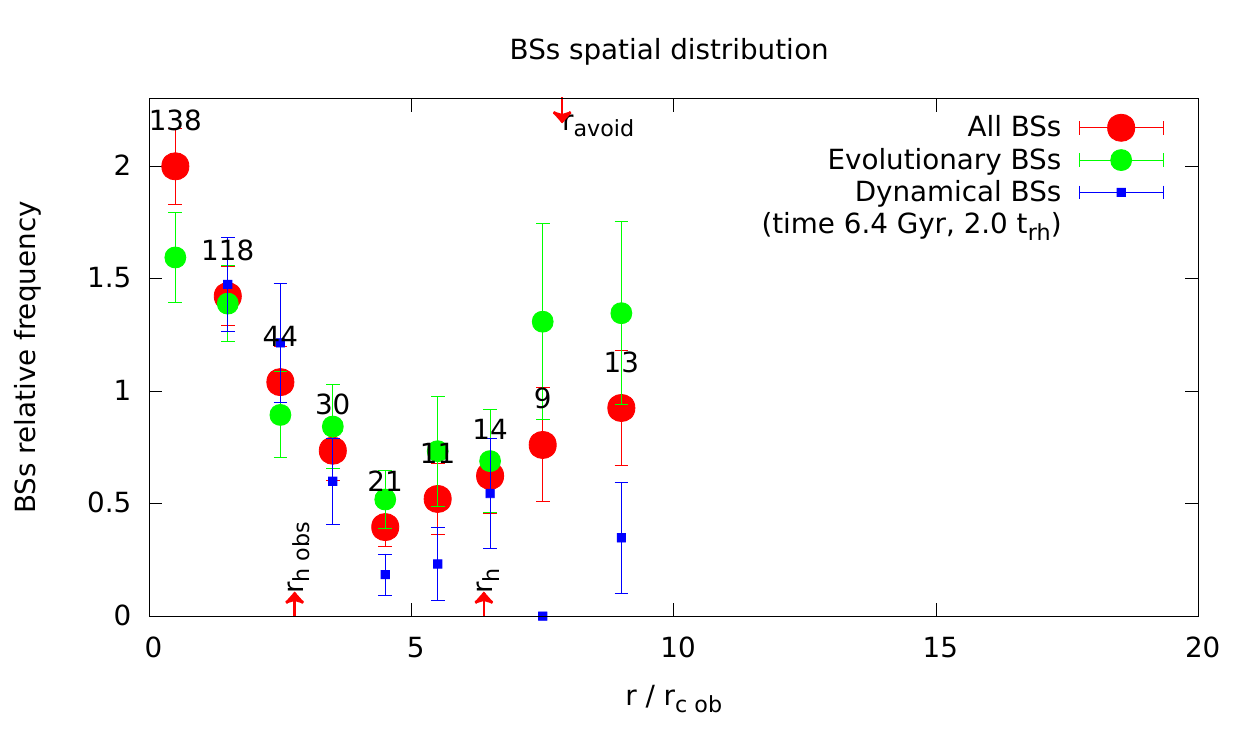}
		\includegraphics[width=5.8cm,height=3.3cm]{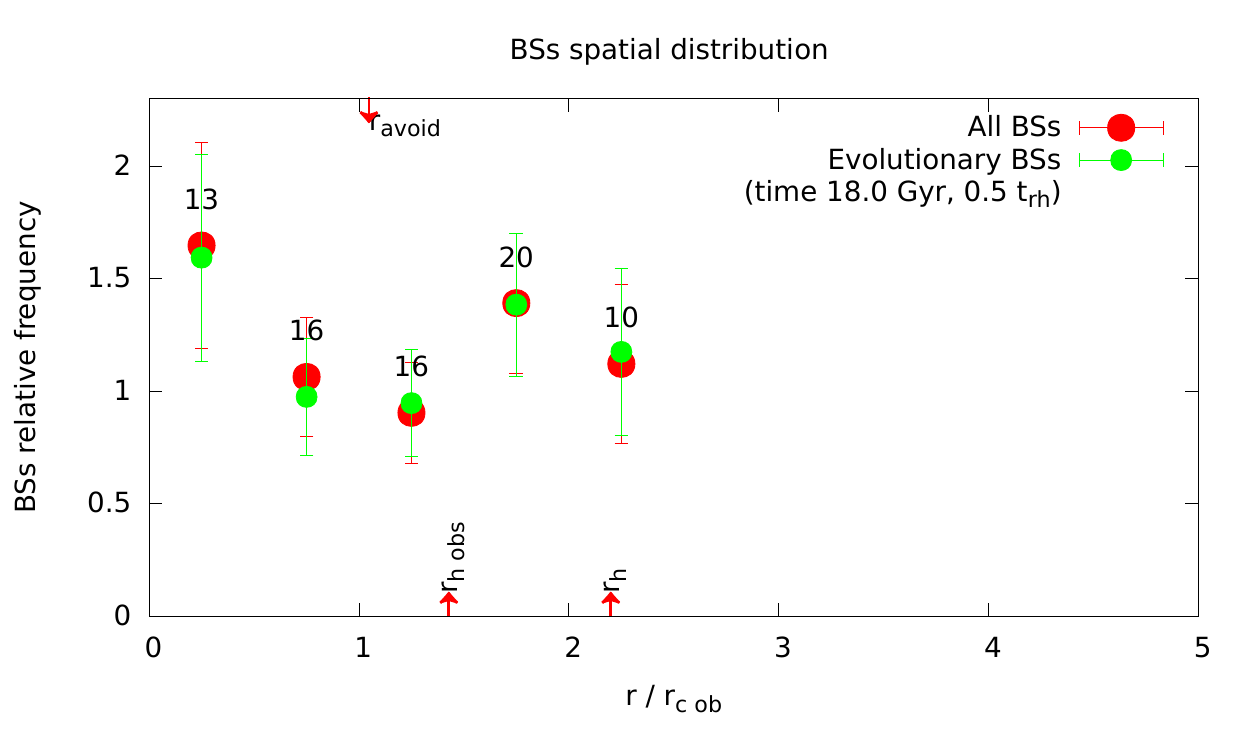}
		\includegraphics[width=5.8cm,height=3.3cm]{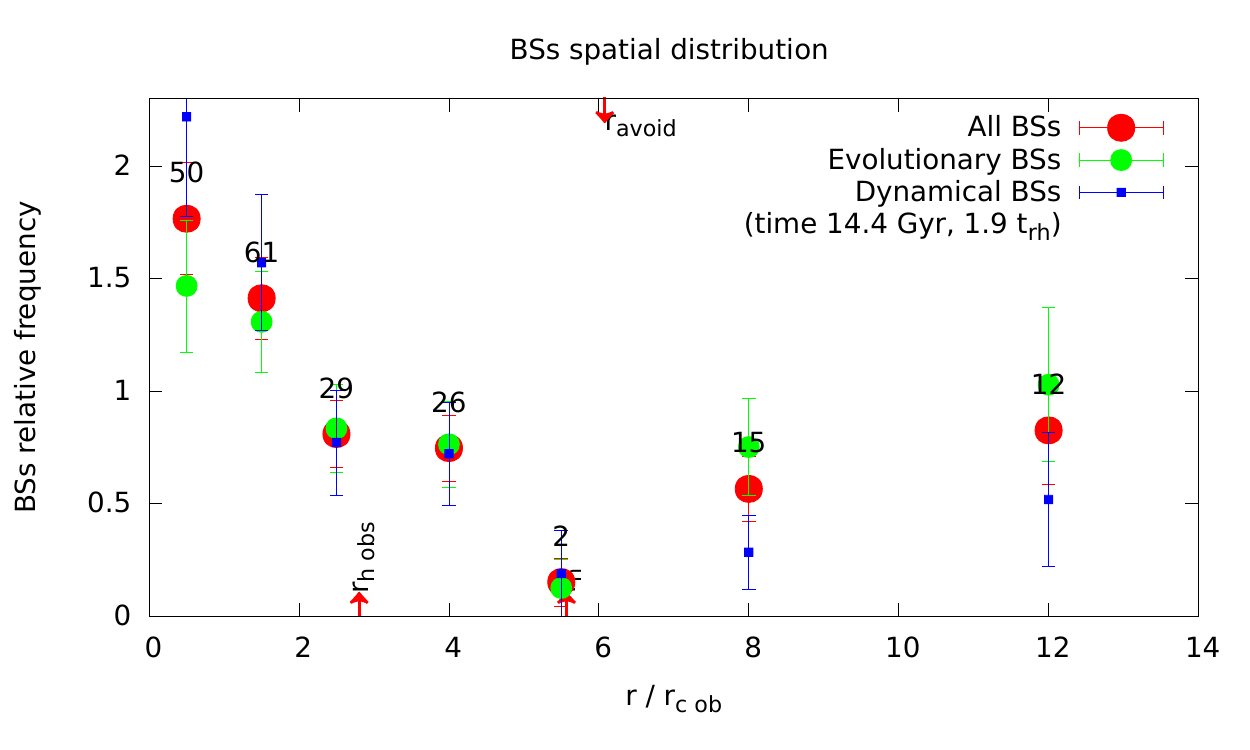}
		\includegraphics[width=5.8cm,height=3.3cm]{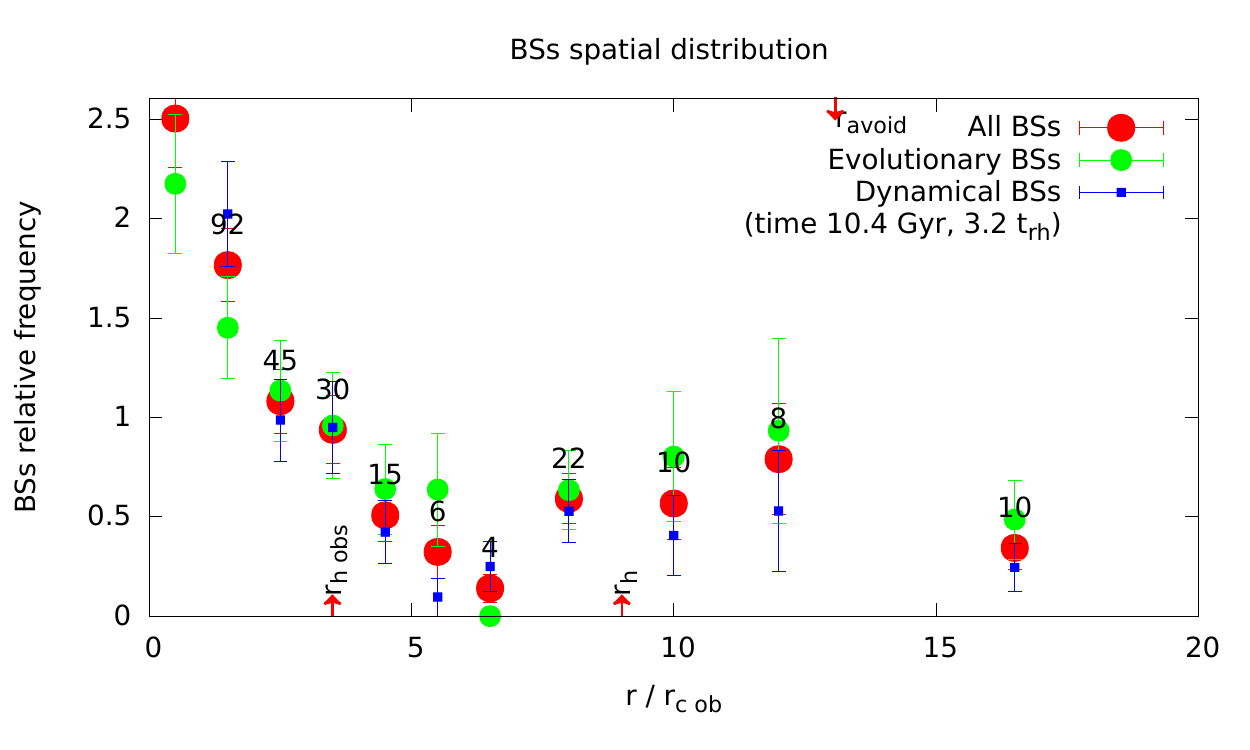}
		\includegraphics[width=5.8cm,height=3.3cm]{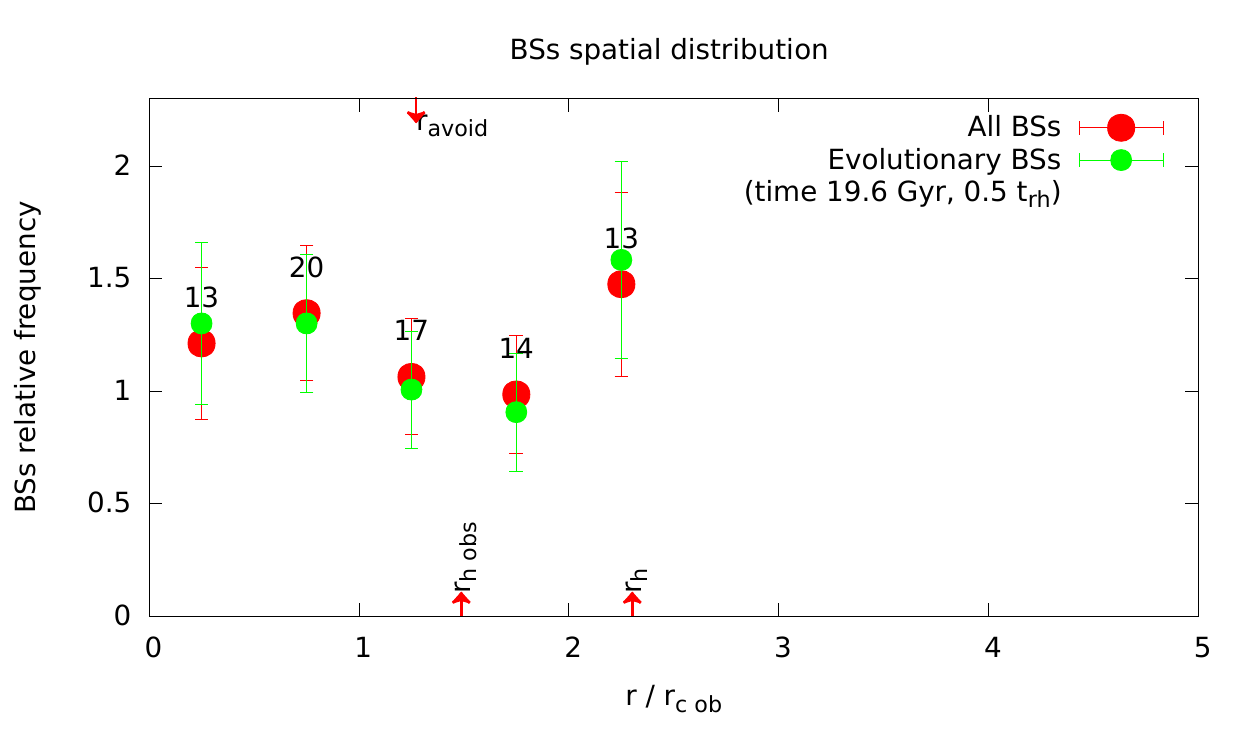}
		\includegraphics[width=5.8cm,height=3.3cm]{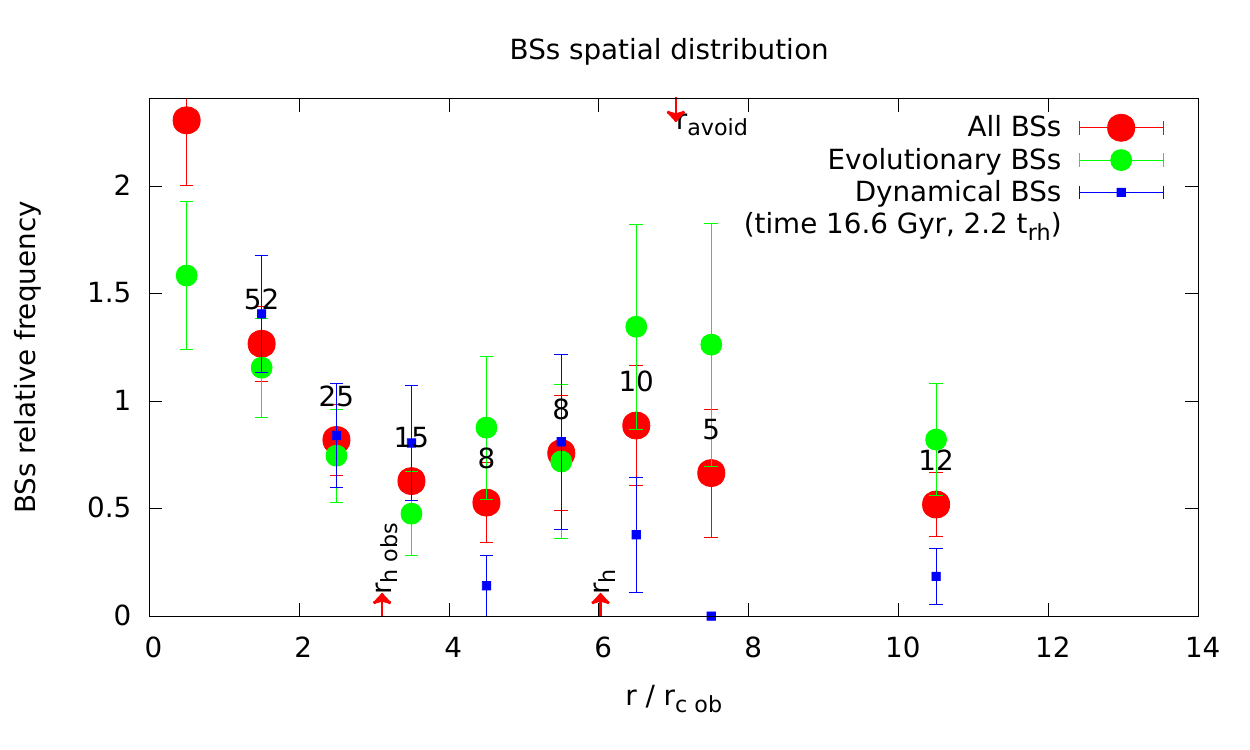}
		\includegraphics[width=5.8cm,height=3.3cm]{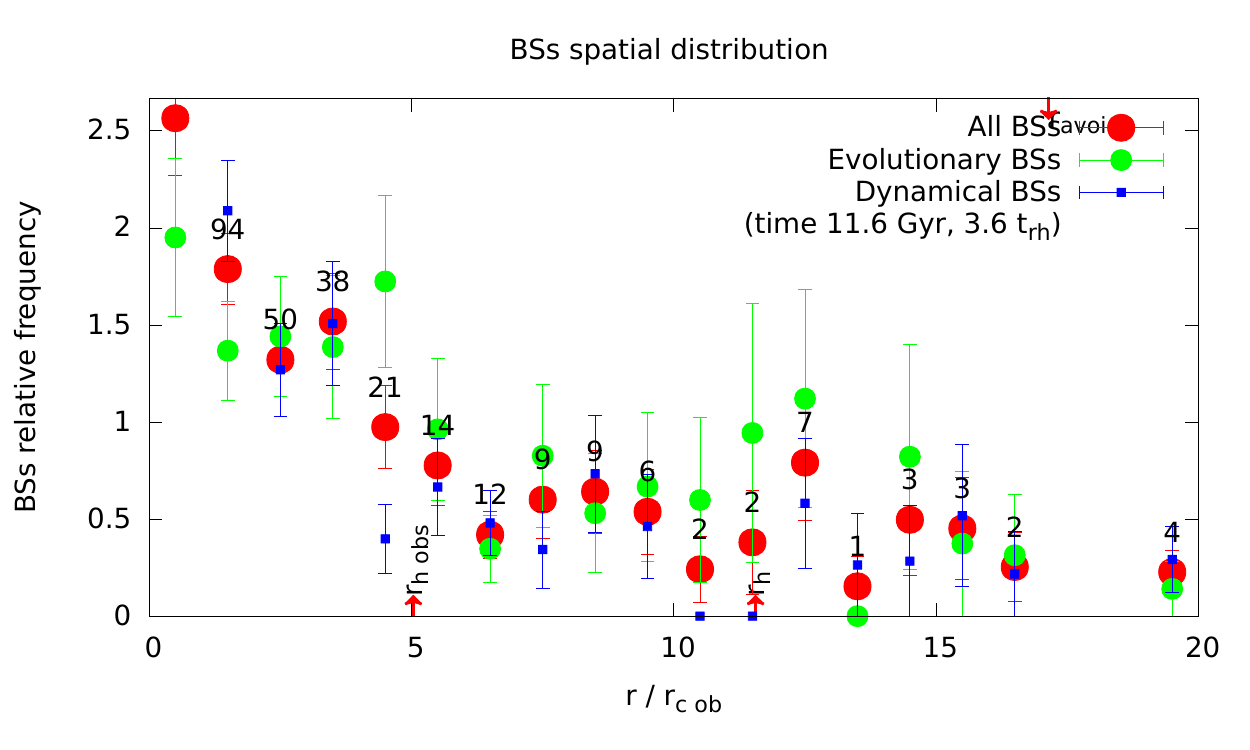}
		\caption[Signs of the bimodal spatial distribution for three models with different dynamical scales]{Figure shows the signs of bimodal spatial distribution for
\textsc{mocca-slow} (left), \textsc{mocca-medium} (middle) and
\textsc{mocca-fast} simulations (right) for a few selected times (in [Myr] and
in [$t_{rh}$], see plots' legends). Each plot contains three
characteristic radii for reference: $r_{h~obs}$, $r_h$
on the bottom X axis, and $r_{avoid}$ on the top X axis. Each plot shows three BSs
specific frequencies calculated for all BSs (red), only evolutionary (green) and
dynamical ones (blue, except the first column) together with the numbers of all BSs
per bins.
For detail discussion see Sect.~\ref{sec:Bim:RAvoidDrift}.}
\label{fig:Bim:Bimodality}
\end{figure*}

Fig.~\ref{fig:Bim:Bimodality} shows the signs of bimodal spatial distribution
for \textsc{mocca-slow} (left column), \textsc{mocca-medium} (middle) and
\textsc{mocca-fast} simulations (right) for a few selected times for which the
bimodal distribution is best visible. 
The times are specified in the plots' legends and are given in Myr and in the
units of the present $t_{rh}$. The distances on the X axis are given in the units of the
core radii ($r_c$). Each plot shows three BSs specific frequencies ($R_{BSs}$)
calculated for all BSs (red), only the evolutionary BSs (green) and the dynamical ones
(blue, except the \textsc{mocca-slow} model for which the number of dynamical
BSs is small).
Number of BSs in each bin is written on the top of red circles. The errors for
the $R_{BSs}$ are calculated as a Poisson error ($R_{BSs} / \sqrt{n_{BSs}}$).
Additionally, each plot contains three characteristic radii for a reference:
half-light radius ($r_{h~obs}$), half-mass radius ($r_h$) on the bottom X axis,
and the radius of avoidance ($r_{avoid}$) on the top X axis. The radius of avoidance
is calculated with Eq.~\ref{eq:RAvoid} and it is equal to the radius $r$ at
which the time of the dynamical friction ($t_{df}$) exceeds the age of the GC.
However, a few additional comments concerning the determination of $r_{avoid}$ are
needed here.
The radius $r_{avoid}$ strongly depends on the local parameters of the GCs. For
example, if the local density at some radius r is slightly smaller than the
average one, $t_{df}$ can suddenly exceed the age of GC. In order to avoid such
randomness, the value of $r_{avoid}$ is actually calculated as the average value
from the last 5~$r_{avoid}$ measurements. The average $r_{avoid}$ value is much
smoother and less depends on the local GC parameters. Additionally, we
decided to use $2 \times 100$ neighboring stars around a given test star to
calculate the local density and velocity dispersions (see Eq.~\ref{eq:RAvoid}). This
number has large impact on the calculated values of $r_{avoid}$ too.

The bins in Fig.~\ref{fig:Bim:Bimodality}, and other similar figures presented
later, have essentially the same widths for one simulation. Usually the
width of bins is 1.0 or 0.5~$r_{c~ob}$. However, the number of BSs at larger
distances from the center is small, and as a result, errors for later bins are
also large. Thus, we decided to join bins into larger ones to increase
the amount of BSs per bin. However, this procedure is applied only for bins larger than
the calculated $r_{avoid}$. It is consistent with the observational way of
presenting the bimodal distribution where bins further from the center are wider
too.

The signs of the bimodal distribution in Fig.~\ref{fig:Bim:Bimodality} are best
visible for the \textsc{mocca-fast} model. The bimodality for this fast evolving
model forms very quick -- just after 1~Gyr (0.5~$t_{rh}$) it is already well visible
(see the top plot on the right column of Fig.~\ref{fig:Bim:Bimodality}). The dip
gets bigger with time. After a few Gyr it gets smaller than $R_{BSs} < 0.5$ which
makes the bimodal distribution more visible. Also the first peak in the center
of GC gets bigger with time. It reaches values $R_{BSs} \sim 2.0$ after around
6~Gyr. The time needed to form some signs of the bimodal distribution is
comparable to a very low $t_{rh}$ which initially is only around 1.5~Gyr (see
Fig.~\ref{fig:Bim:TimeScales}).

A similar formation of the bimodal distribution presents
\textsc{mocca-medium} model. After 1~Gyr (the top plot in
Fig.~\ref{fig:Bim:Bimodality} in the middle column) one can see some sign of a
bimodal distribution but the central peak is only $R_{BSs} \sim 1.5$. The bimodality gets more visible with time, and after 9~Gyr (the forth plot in the middle column) is
well distinguishable. Here again the formation of the bimodal distribution is comparable
to $t_{rh}$. The bimodality starts to be well visible after 4-5~Gyr which
corresponds to $t_{rh}~\sim~6-7~Gyr$ at that time (see
Fig.~\ref{fig:Bim:TimeScales}). 

The slowest evolving model, \textsc{mocca-slow}, does not show any signs
of bimodal distribution until 6~Gyr (the second plot in
Fig.~\ref{fig:Bim:Bimodality} in the left column). Before that time, BSs spatial
distributions taking into account errors of $R_{BSs}$ are more or less flat (similarly to the time 0.4~Gyr; the top
plot in the left column). The errors for this model are the largest because of the low
number of BSs in the bins. The number of EM, and EMT BSs for \textsc{mocca-slow}
model is actually the same as for \textsc{mocca-medium} and \textsc{mocca-fast}
models (see Fig.~\ref{fig:Bim:Radii}) but \textsc{mocca-slow} is a much more
extended cluster. The BSs for \textsc{mocca-slow} are spread across the whole GC and only in the
central regions of GC amount of BSs is around or larger than 10, thus
providing a better statistics. The signs of the bimodality are not clearly visible
even after 20~Gyr. The overall lack of clear signs of the bimodality is caused
by the fact that $t_{rh}$ increases for \textsc{mocca-slow} model from the
initial 16~Gyr up to about 36~Gyr after the first few Gyr (see
Fig.~\ref{fig:Bim:TimeScales}).
This is a very slowly evolving GC. 


The radius of avoidance goes out of sync after $\sim 2 t_{rh}$ (see the middle
and right column in Fig.~\ref{fig:Bim:Bimodality}).
It was rather an unexpected result. Instead, the initial assumptions indicated
that the position of $r_{avoid}$ should more or less follow the $r_{min}$, i.e.,
the radial position at minimum $R_{BSs}$, chosen by eye.
Fig.~\ref{fig:Bim:Bimodality} shows the $r_{avoid}$ (on the top X axes) for all
three models.
For the \textsc{mocca-fast} model $r_{avoid}$ stops to follow the minimum dip
already after 6.4~Gyr (2~$t_{rh}$; the forth plot in the right column).
The $r_{avoid}$ has to increase with time because it follows the region of GC
which has to be affected by the mass segregation. It should constantly
increase with time. The same assumptions were made for $r_{min}$, namely that it has to
follow more or less the positions of $r_{avoid}$. However, the $r_{min}$ stops
to increase with time.
After around 10~Gyr $r_{avoid}$ lies even after the second peak (see the second
to the last plot in the right column). It corresponds to $> 3.0 t_{rh}$. For the
slower evolving \textsc{mocca-medium} model, $r_{avoid}$ goes out of sync with $r_{min}$ after
16~Gyr (the last plot in the middle column).

As a remark, it is necessary to say that for a few snapshots of
\textsc{mocca-fast} model, even after 6~Gyr, the plots with the BSs relative
frequency ($R_{BSs}$) show agreement between $r_{avoid}$ and $r_{min}$. But in
majority of cases $r_{avoid}$ is clearly out of sync with $r_{min}$. These plots
are not shown in the Fig.~\ref{fig:Bim:Bimodality}.
The randomness of $r_{avoid}$ is discussed in
Sect.~\ref{sec:Bim:Stochasticity}.

The bimodal spatial distribution changes to unimodal after several $t_{rh}$.
After that time the signs of bimodal distribution appear very rarely. An example
of such feature one can see for the \textsc{mocca-fast} model. Thel last plot in
Fig.~\ref{fig:Bim:Bimodality} in the right column shows the BSs distribution for
the time 11.6~Gyr (3.6~$t_rh$). Only the central peak is present, then
$R_{BSs}$ drops continuously throughout the whole cluster. In the
\textsc{mocca-fast} model after $\sim~12$~Gyr BSs are already mass segregated.
Almost all BSs at the time of 12~Gyr had already enough time to sink to the center of
GC (see mass-segregation time scales in Fig.~\ref{fig:Bim:TimeScales}). After
12~Gyr there were a few snapshots for which the bimodal distribution was also
visible. However, in majority of cases the distribution was already unimodal.
For details on the transientness of the signs of the bimodality see
Sect.~\ref{sec:Bim:Stochasticity}. The same feature of transition to unimodal
distribution was observed also for other models for which $t_{rh}$ were relatively small (not shown in the paper,
though).

Figure~\ref{fig:Bim:Bimodality} shows the BSs specific frequencies for all BSs
($R_{BSs}$, red circles) but also separately for only the evolutionary
($R_{BSs}^{evol}$, green), and the dynamical ones ($R_{BSs}^{dyn}$, blue). There
is one feature which is consistent across both \textsc{mocca-medium} and
\textsc{mocca-fast} models. For the bins which are before the apparent minimum
of the bimodal distributions the values of $R_{BSs}^{evol}$ are
consistently below the values of $R_{BSs}$ (red).
In turn, the values of $R_{BSs}^{dyn}$ are for the same bins above them.
For the bins outside the apparent minimum of the bimodal distributions the
situation is reversed. The values of $R_{BSs}^{dyn}$ are consistently below the
values of $R_{BSs}$, while the values $R_{BSs}^{evol}$ are consistently above the
values of $R_{BSs}$. This situation is a consequence of a several factors. First, the number
of dynamical BSs is the highest in the center of a GC because of the higher
density there and thus also the higher probabilities for strong dynamical
interactions. Second, the dynamical BSs are on average more massive than
the evolutionary ones \citep[Fig.~6]{Hypki2013MNRAS.429.1221H}.
This causes that the mass segregation times for them are slightly smaller and
they can sink faster to the center. Some of the dynamical BSs are ejected due
to strong dynamical interactions to larger orbits. Thus, some amount of
them can be found far from the center. The last factor is the fact that the evolutionary BSs are being
created from unperturbed binaries, so
 they are created more or less in any part of the system. Thus, the
values $R_{BSs}^{evol}$ follow quite closely the values of $R_{BSs}$. The values
of $R_{BSs}$ are changed mainly due to the population of dynamical BSs.

\subsection{Transientness of the bimodal spatial distribution}
\label{sec:Bim:Stochasticity}

\begin{figure*}
		\includegraphics[width=9cm,height=4.1cm]{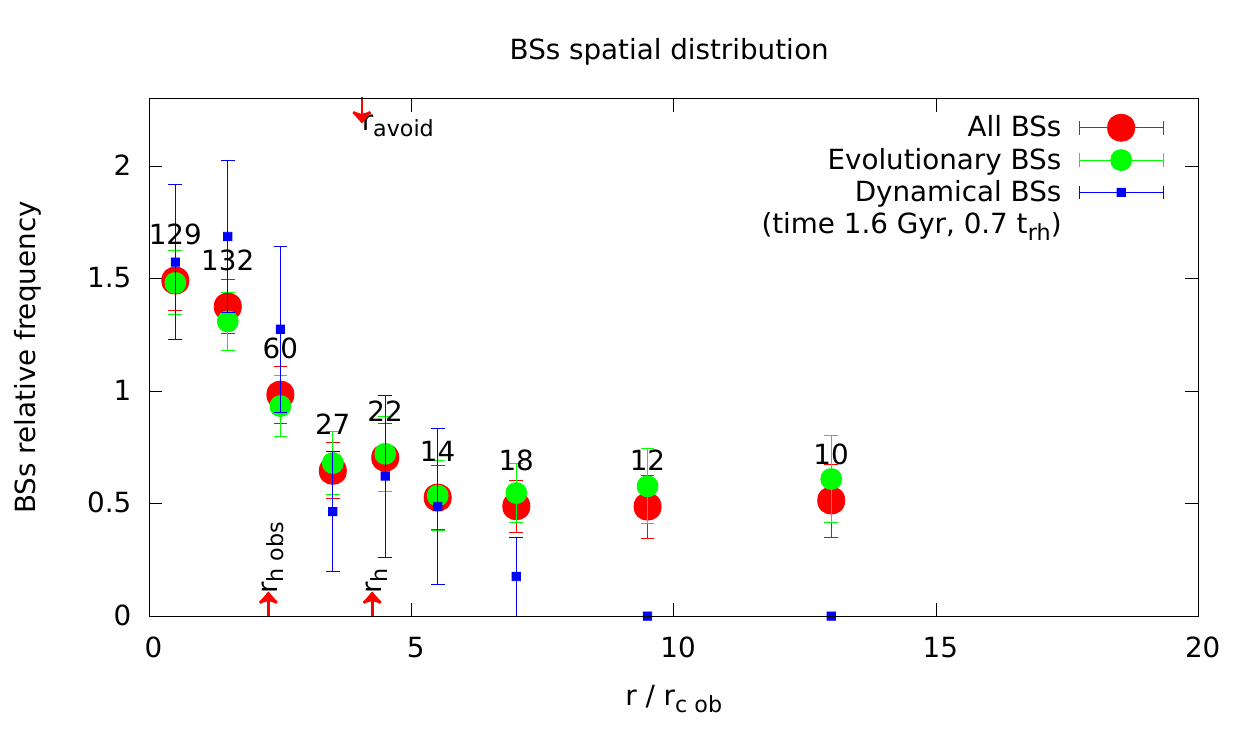}
		\includegraphics[width=9cm,height=4.1cm]{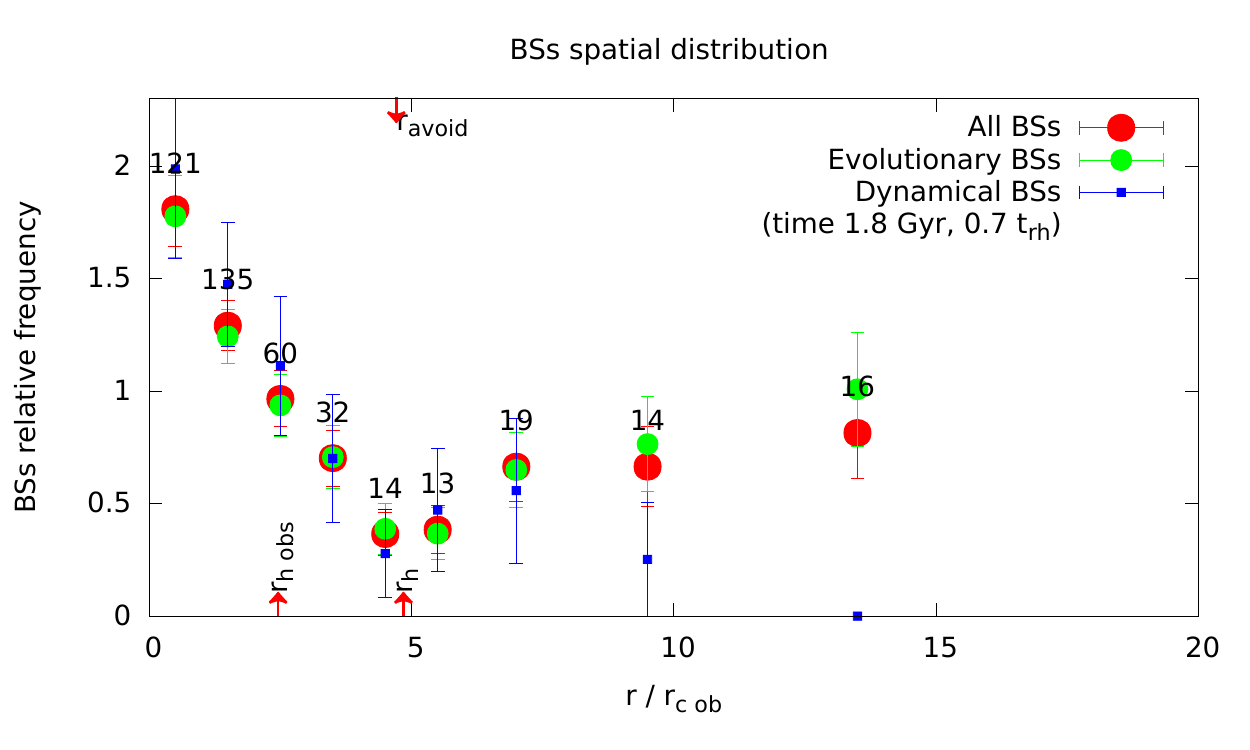}
		\includegraphics[width=9cm,height=4.1cm]{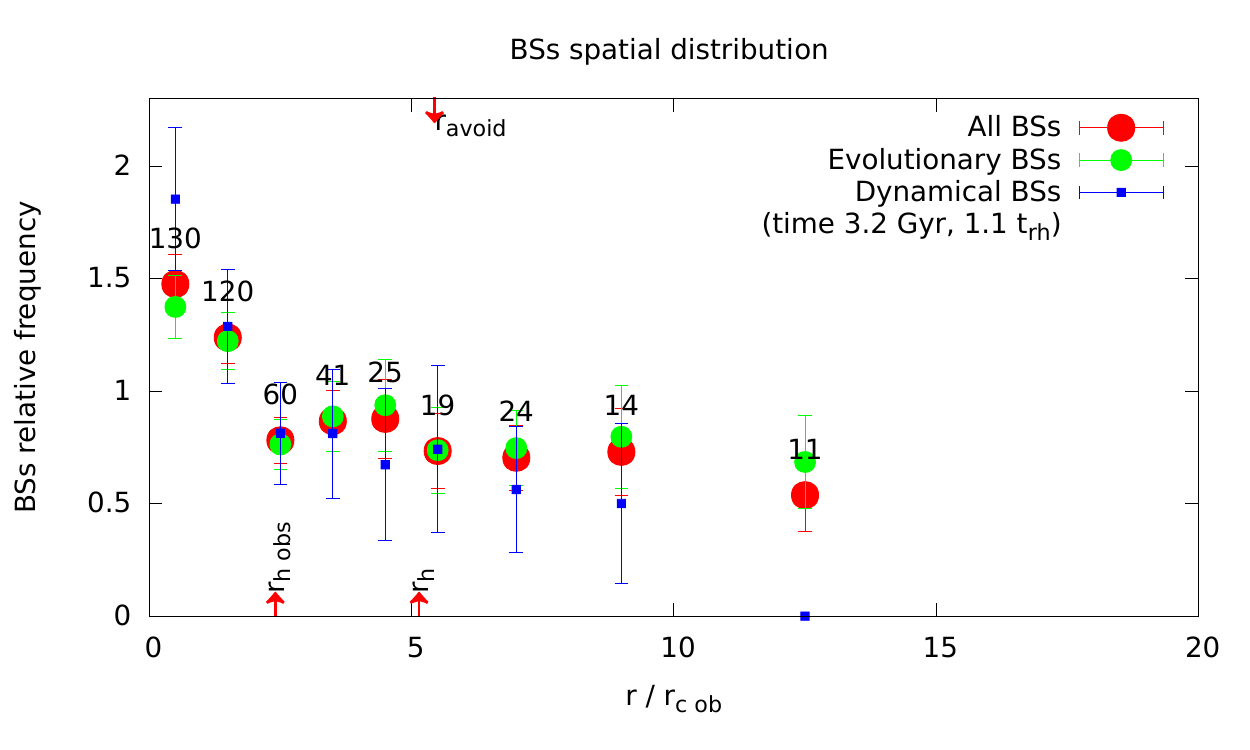}
		\includegraphics[width=9cm,height=4.1cm]{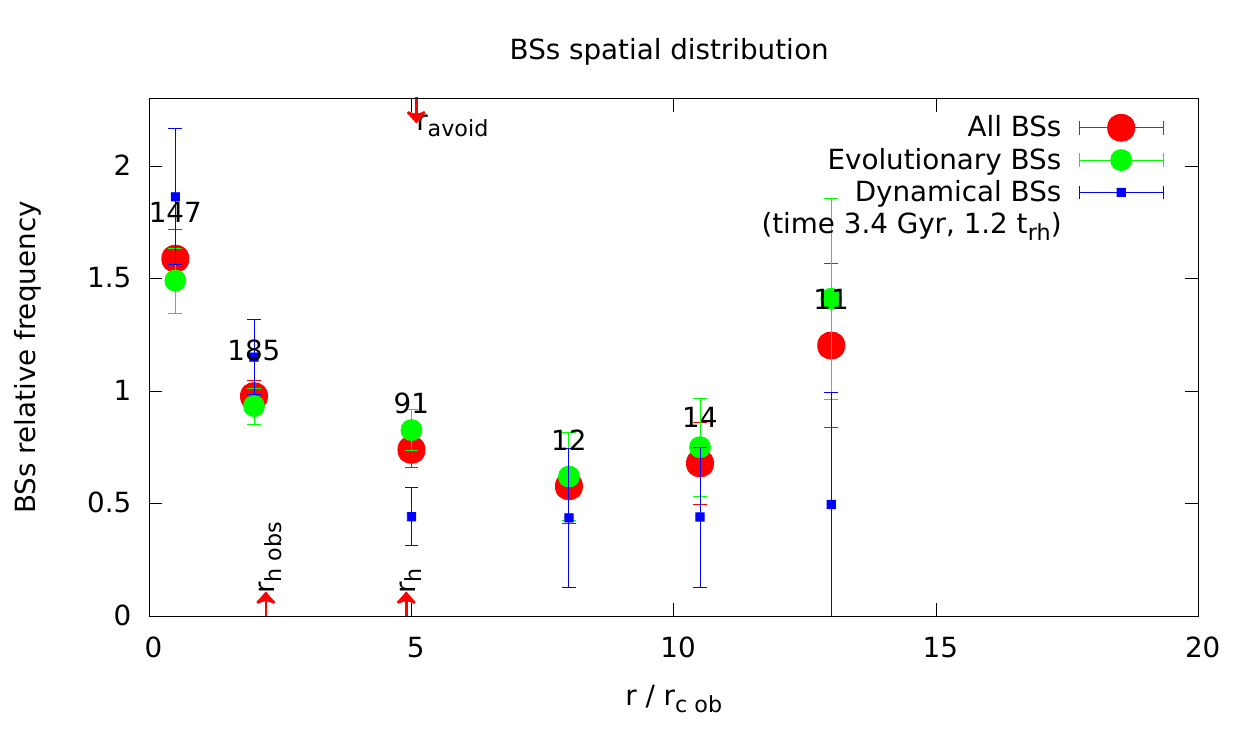}
		\includegraphics[width=9cm,height=4.1cm]{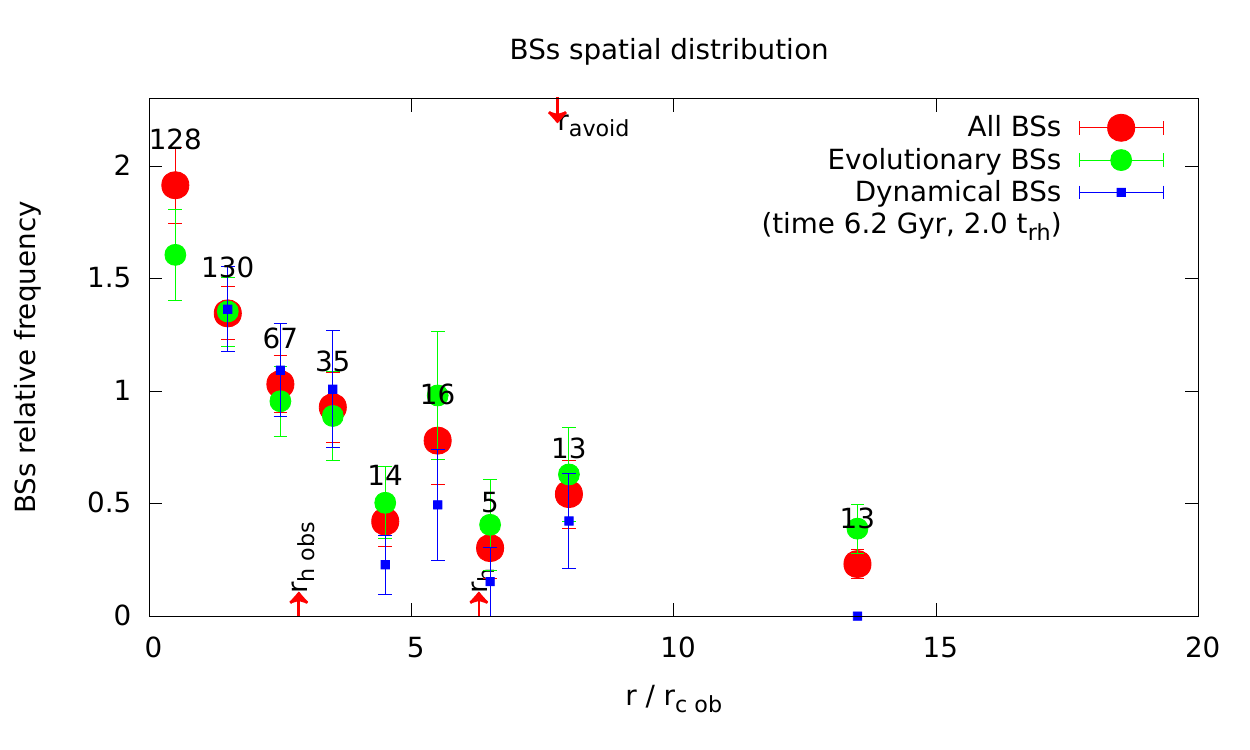}
		\includegraphics[width=9cm,height=4.1cm]{./bimodality/rbar55/bimodality-6414.pdf}
		\includegraphics[width=9cm,height=4.1cm]{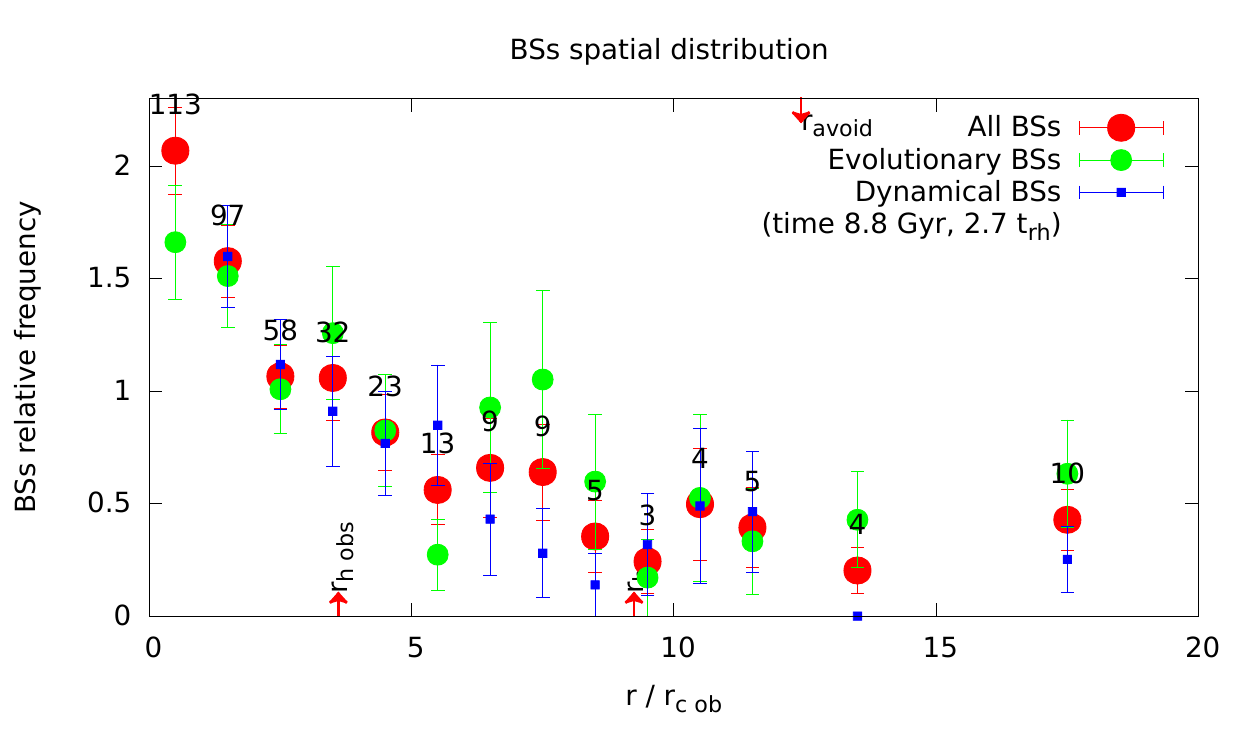}
		\includegraphics[width=9cm,height=4.1cm]{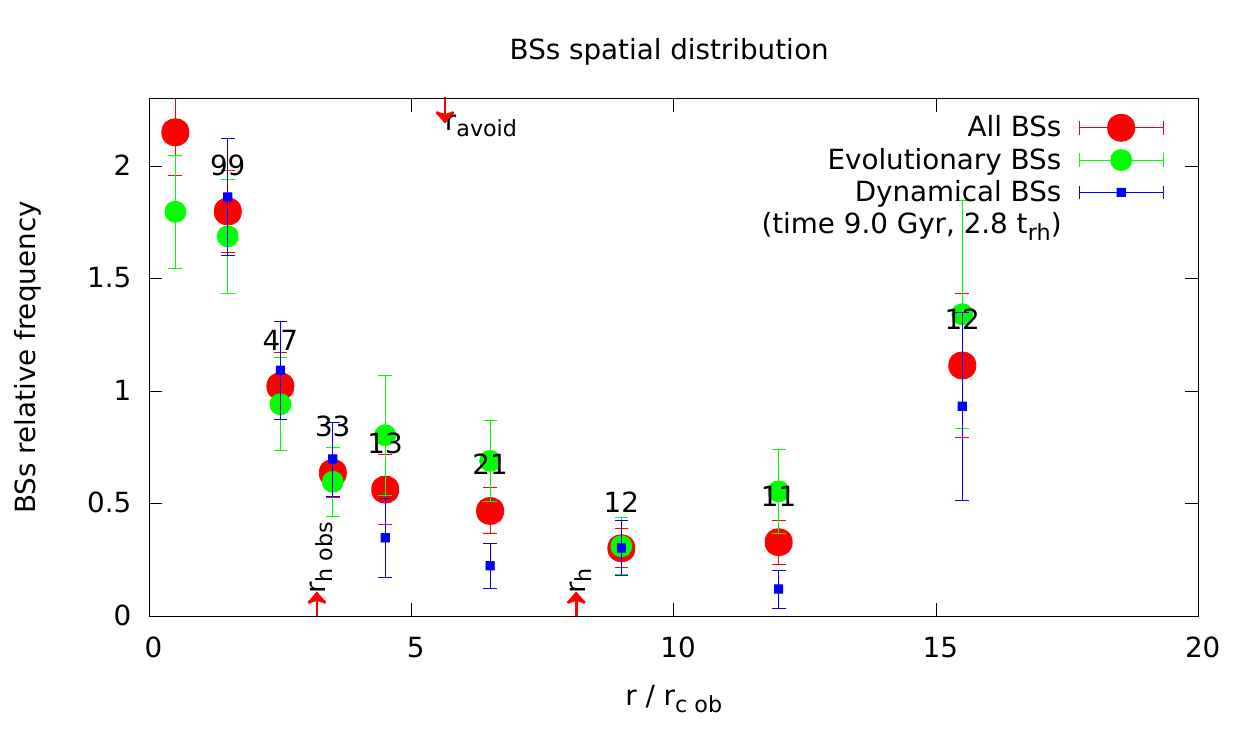}
\caption[Examples of the transientness of the signs of the
bimodal spatial distribution]{Figure shows a few examples of the transientness of the signs of the
bimodal spatial distribution taken only from the \textsc{mocca-fast} simulation.
Description of the plots are the same as for Fig.~\ref{fig:Bim:Bimodality}. The
examples show that between only two snapshots ($\Delta~t = 200$~Myr, from left
to right column) the clear sign of the bimodality can change to unimodal and vice versa. For detailed discussion see
Sect.~\ref{sec:Bim:Stochasticity}.}
\label{fig:Bim:Stochastic}
\end{figure*}

The bimodal spatial distribution is very transient. It can be clearly
visible in one snapshot and vanish just in the next one 
(snapshot are written every 200~Myr). Fig.~\ref{fig:Bim:Stochastic} shows a few
examples of such transitions between clear signs of the bimodal distribution
and the unimodal distribution and vice versa. This change from one state into
another can happen actually at any stage of GC evolution, even at later stages
when GC is dynamically old.
However, after 2-3~$t_{rh}$ the transitions from the unimodal distribution into bimodal are
clearly less frequent (see Sect.~\ref{sec:Bim:RAvoidDrift}). 

The transientness of the bimodal distributions is a consequence of the fact that
BSs often change their bins. This is also the reason why the values of the
BSs relative frequencies vary so much, for the same bins, between two consequent
snapshots in time (see Fig.~\ref{fig:Bim:AllTimesteps}). The BSs change their
bins due to a combination of a few factors. BSs have
small mass-segregation times (see Fig.~\ref{fig:Bim:TimeScales}), thus, they can sink to the center faster
than RGB or MS stars. Additionally, the orbits of stars around the GC's center
can be elongated (high eccentricity). The time which BSs spend close to their 
apocenters is larger than in pericenter, but still the high eccentricity of
the orbits mean that the radial distances (thus also the bins) for BSs can vary
significantly between snapshots. Additionally, the $r_c$, which is used to
compute BSs relative frequencies, is small in comparison to $r_h$. It means that
for the bins $\gtrsim r_h$ the number of BSs per bin can differ substantially.
This is also the reason why it is so important to join bins into larger ones for these
bins to see some signs of the bimodal distributions.

\begin{figure*}
		\includegraphics[width=9cm,height=4.1cm]{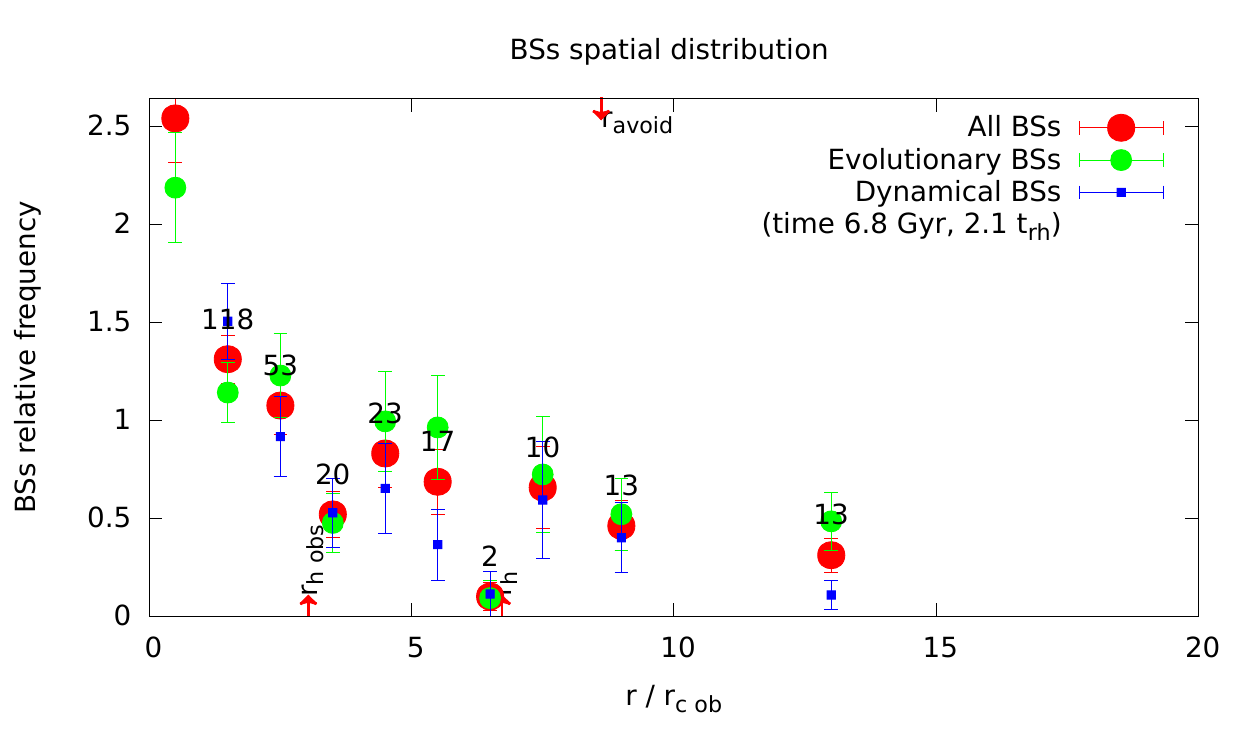}
		\includegraphics[width=9cm,height=4.1cm]{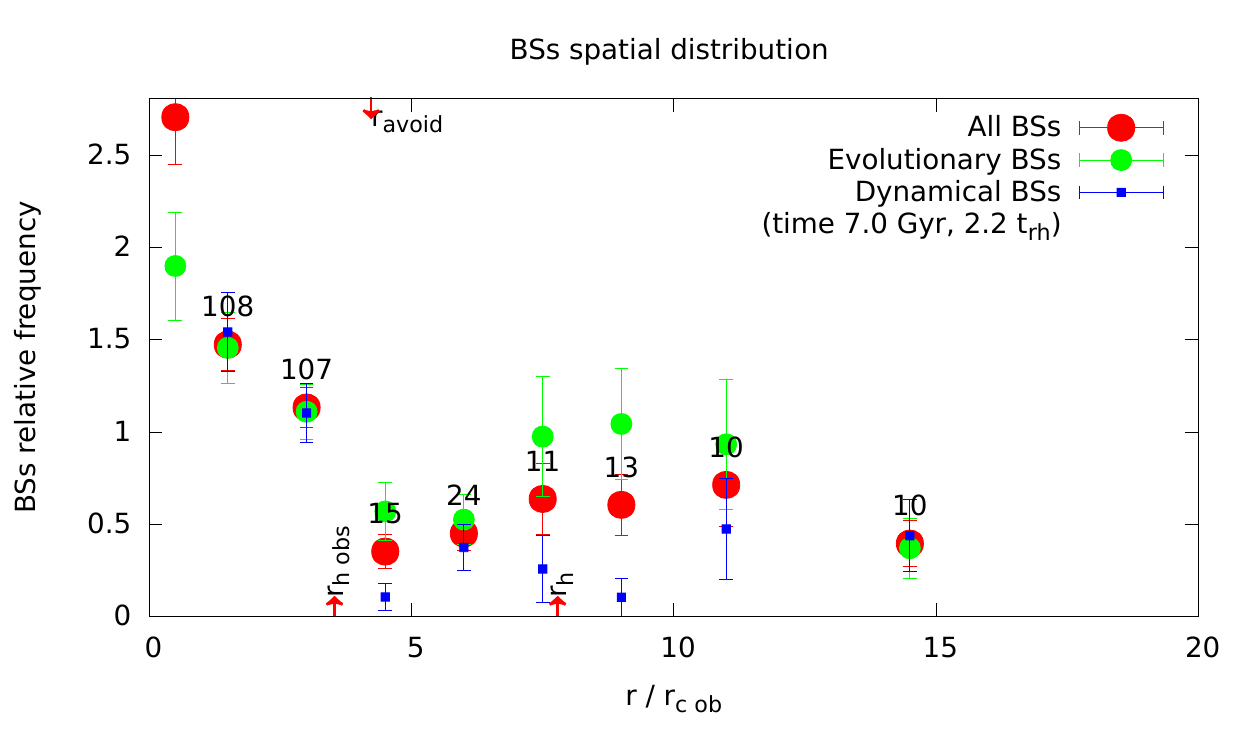}
		\includegraphics[width=9cm,height=4.1cm]{./bimodality/rbar55/bimodality-8802.pdf}
		\includegraphics[width=9cm,height=4.1cm]{./bimodality/rbar55/bimodality-9016.pdf}
		\includegraphics[width=9cm,height=4.1cm]{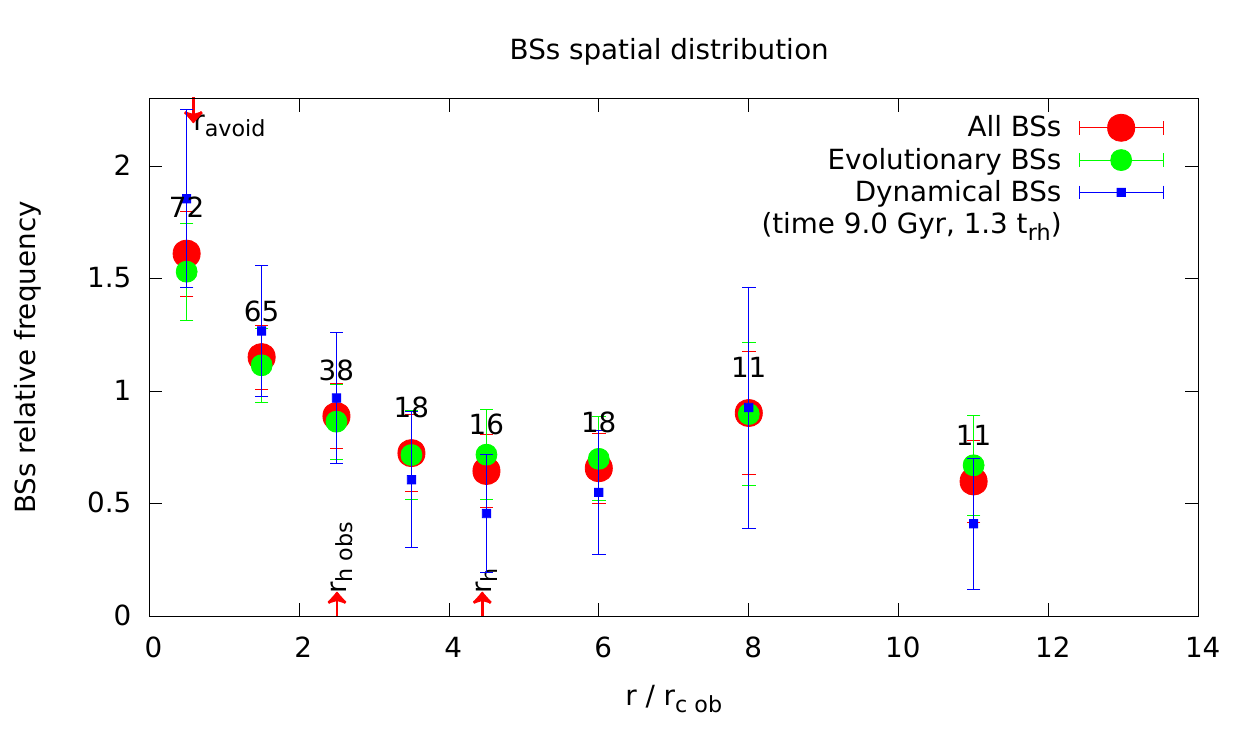}
		\includegraphics[width=9cm,height=4.1cm]{./bimodality/rplum20/bimodality-9200.pdf}
\caption[Examples of a randomness of the radius of avoidance]{Figure shows three examples on how significantly the radius of avoidance
($r_{avoid}$) can change its value between two consequent snapshots in time ($\Delta~t =
200$~Myr, from left to right column). The first two rows are taken from
\textsc{mocca-fast} and the third one from \textsc{mocca-medium} simulation. The description of the plots is the same as
for Fig.~\ref{fig:Bim:Bimodality}. For details 
see Sect.~\ref{sec:Bim:Stochasticity}.
}
\label{fig:Bim:RAvoidNoise}
\end{figure*}

Radius of avoidance ($r_{avoid}$) is rather chaotic and very difficult to
compute too. The procedure of its calculation was explained in
Sect.~\ref{sec:Bim:RAvoidDrift}. Fig.~\ref{fig:Bim:RAvoidNoise} shows three examples on
how significantly $r_{avoid}$ can change its value between just two consequent
snapshots in time ($\Delta~t = 200$~Myr). Figure shows how unstable the
$r_{avoid}$ is for computation of the ,,dynamical clocks'' of GCs. The randomness of the $r_{avoid}$ is a result of its large dependence on the local parameters of GCs. 

\begin{figure*}
		\includegraphics[width=9cm,height=4.1cm]{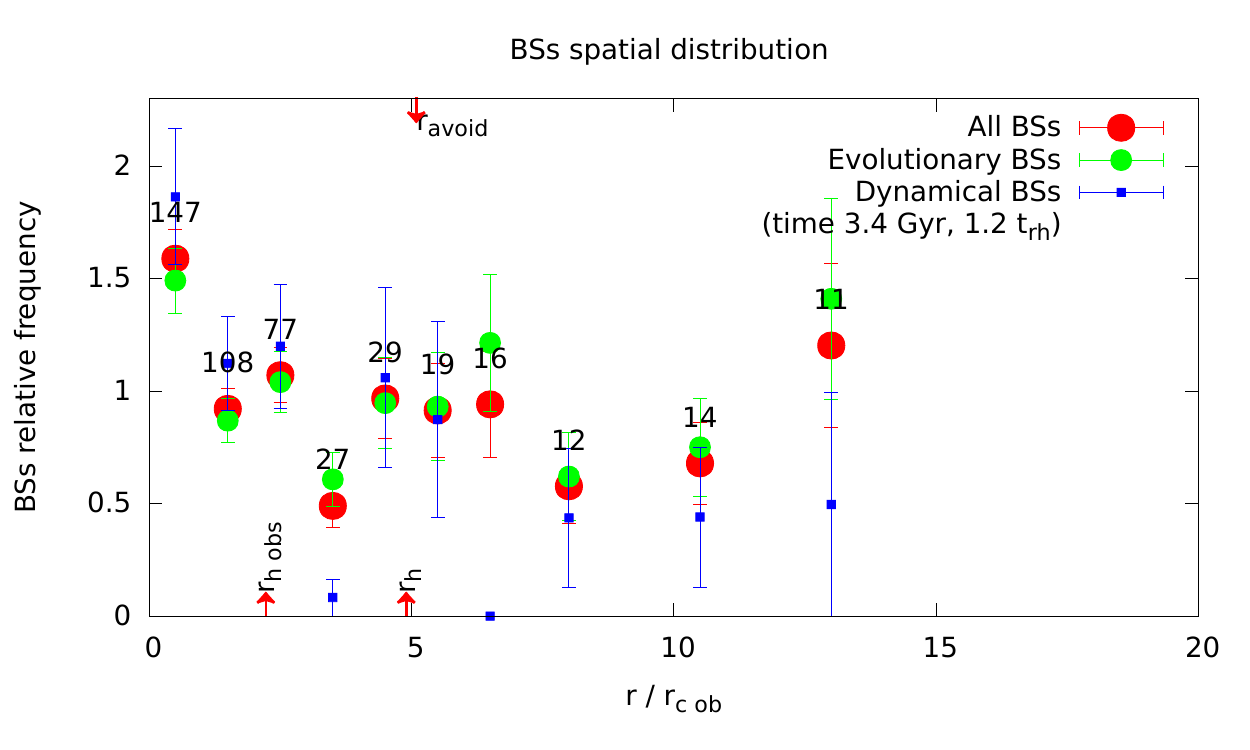}
		\includegraphics[width=9cm,height=4.1cm]{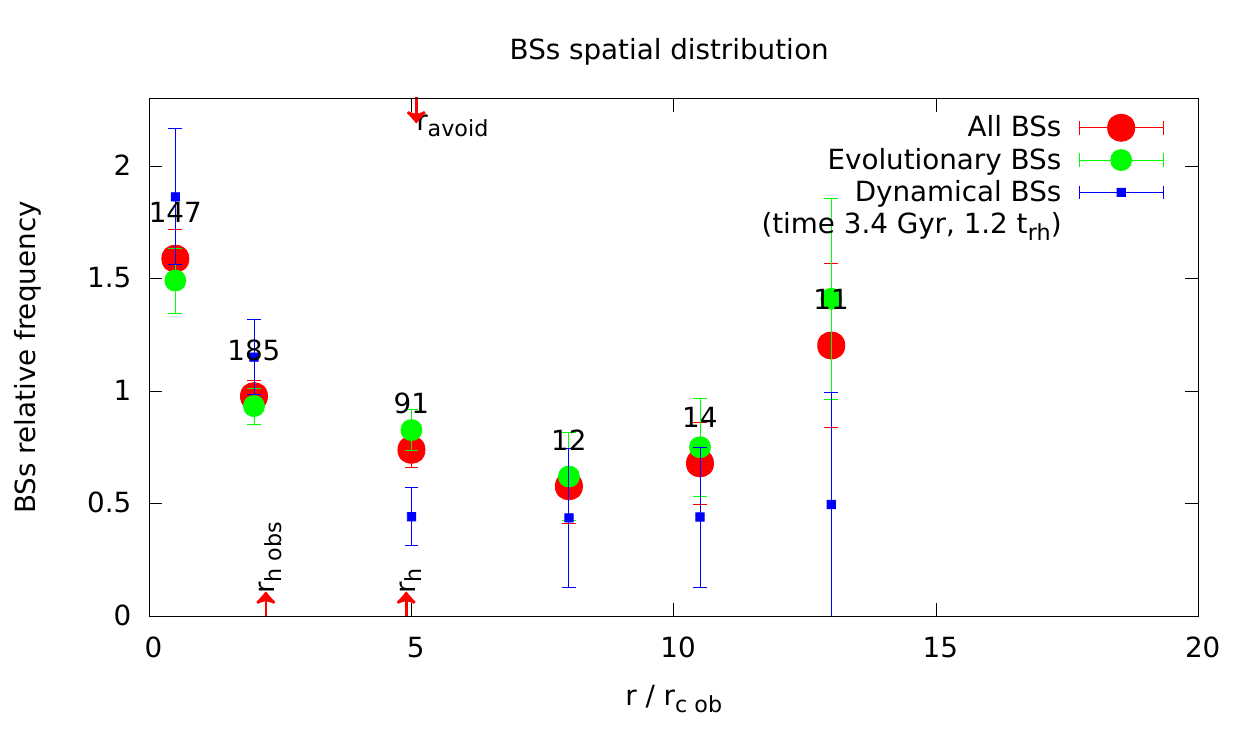}
		\includegraphics[width=9cm,height=4.1cm]{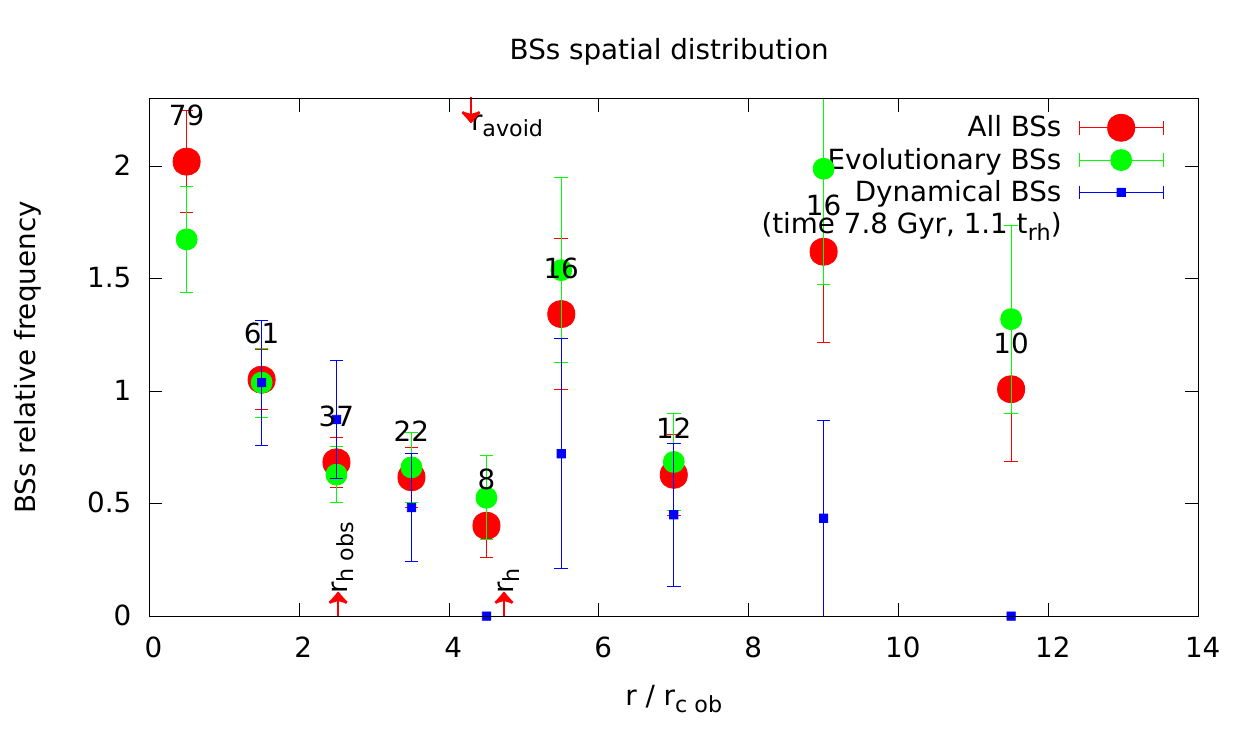}
		\includegraphics[width=9cm,height=4.1cm]{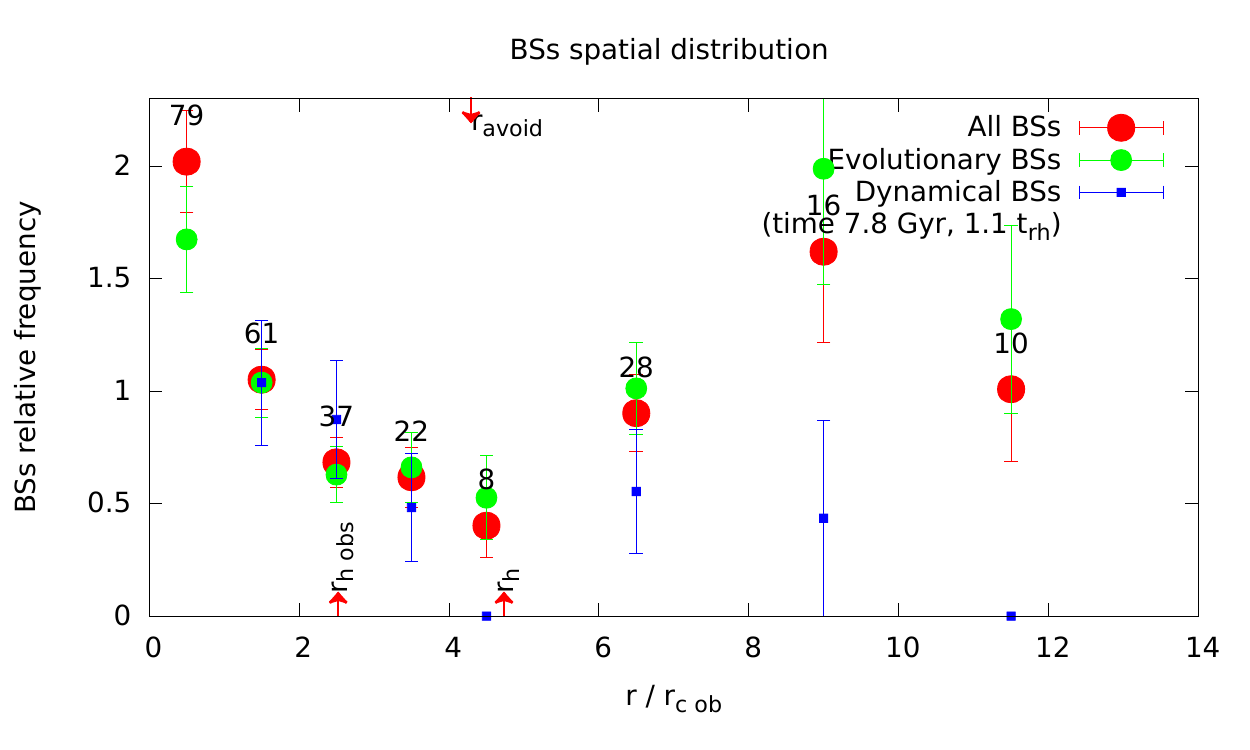}
		\caption[Examples on how careful binning can change the signs of the bimodal
		spatial distributions]{Figure shows two examples on how, only by combining two
		neighbouring bins into larger ones, one can considerably improve the
		apparent visibility of the bimodal distribution. These examples are taken from
		the \textsc{mocca-fast} simulation.
The description of the plots is the same as for Fig.~\ref{fig:Bim:Bimodality}. For details 
see Sect.~\ref{sec:Bim:Stochasticity}.}
\label{fig:Bim:BinningMagic}
\end{figure*}

The way of binning has a large impact on the apparent visibility of the bimodal
distributions. Too small or too wide bins can obviously hide any signal in any
data.
However, for the signs of bimodal distribution the procedure of binning
histograms seems to be especially important. Fig.~\ref{fig:Bim:BinningMagic}
shows two examples where just by combining two neighbouring bins into larger
ones one can considerably improve the apparent visibility of the bimodal
distribution.
These examples shows only that the bimodality itself is very noisy. These
examples are intended only to make readers sensitive to this problem. By a
careful binning one can increase the overall visibility of the bimodal
distribution and simultaneously create artificial impression of signals which
might not really be present in the data.
For example, the widths of bins in \citet{Ferraro2012Natur.492..393F} are
different and look quite random. In this paper the way of binning is kept to be
quite simple, and what is most important, consistent for all models (see
Sect.~\ref{sec:Bim:RAvoidDrift}) in order to avoid creation of any artificial
signals.

\section{Summary and discussion} 
\label{sec:Bim:DynamicalClock}
\label{sec:Summary}


The goal of this work was to check the evolution of the spatial
positions of BSs in GCs. Particularly interesting is the formation of the
bimodal distribution of BSs which was already observed in several GCs. This
phenomena is also important from the point of view of the dynamical
processes which take place in GCs. The formation of the bimodal distribution
of BSs is believed to be a result of the mass segregation -- the main
manifestation of the relaxation processes. In turn, the relaxation describes the
age of GCs. Thus, the bimodal distribution might help to determine which GCs are
dynamically old and which dynamically young.

The clear sign of the bimodal distribution forms during
time comparable with the half-mass relaxation time ($t_{rh}$). Thus, it forms
earlier for faster evolving GCs, i.e., these GCs which are denser or have smaller tidal
radii (see Sect.~\ref{sec:Bim:RAvoidDrift}). To check the formation of the
bimodal distributions three \textsc{mocca} simulations with different dynamical
ages were selected. Only for the slowest evolving model, \textsc{mocca-slow},
the signs of the bimodal distribution were not as clear as for the other
models (see Sect.~\ref{sec:Bim:RAvoidDrift}). For the \textsc{mocca-slow} model
the half-mass relaxation time was simply too large and BSs did not have enough time to
sink to the center and form a clear central peak.

The bimodal distribution is very transient. It can appear at some point and then
vanish after relatively short time ($\sim 200$~Myr, i.e., duration between two
subsequent snapshots in time in the \textsc{mocca} simulations). After the next
snapshot it may appear again. The \textsc{mocca-fast} model (see
Sect.~\ref{sec:Bim:RAvoidDrift}) is the fastest evolving model
from the selected ones. For this model a clear sign of the bimodal
distribution forms  just after 1~Gyr (see Fig.~\ref{fig:Bim:Bimodality}). The
half-mass relaxation time changes from $t_{rh}^{T = 0} \sim 1.0$~[Gyr] to
$t_{rh}^{T = 20} \sim 3.0$~[Gyr]. The number of clear signs of the bimodal
distribution was observed only in 13 out of 53 snapshots between time
1~Gyr and 11.6~Gyr (interval between the first and the last plot for this model
in Fig.~\ref{fig:Bim:Bimodality}). It gives 25\% chance to see a bimodality for
this model. For the model \textsc{mocca-medium}, which evolves a bit slower
($t_{rh}^{T = 0} \sim 3$~Gyr, $t_{rh}^{T = 20} \sim 7.5$~Gyr), the chance drops
to 22\% (between 1 and 16.6~Gyr). In other words, it is about 4~times more
probable not to see the signs of the bimodal distribution than to actually see
it. 

\begin{figure}
  \includegraphics[width=\columnwidth]{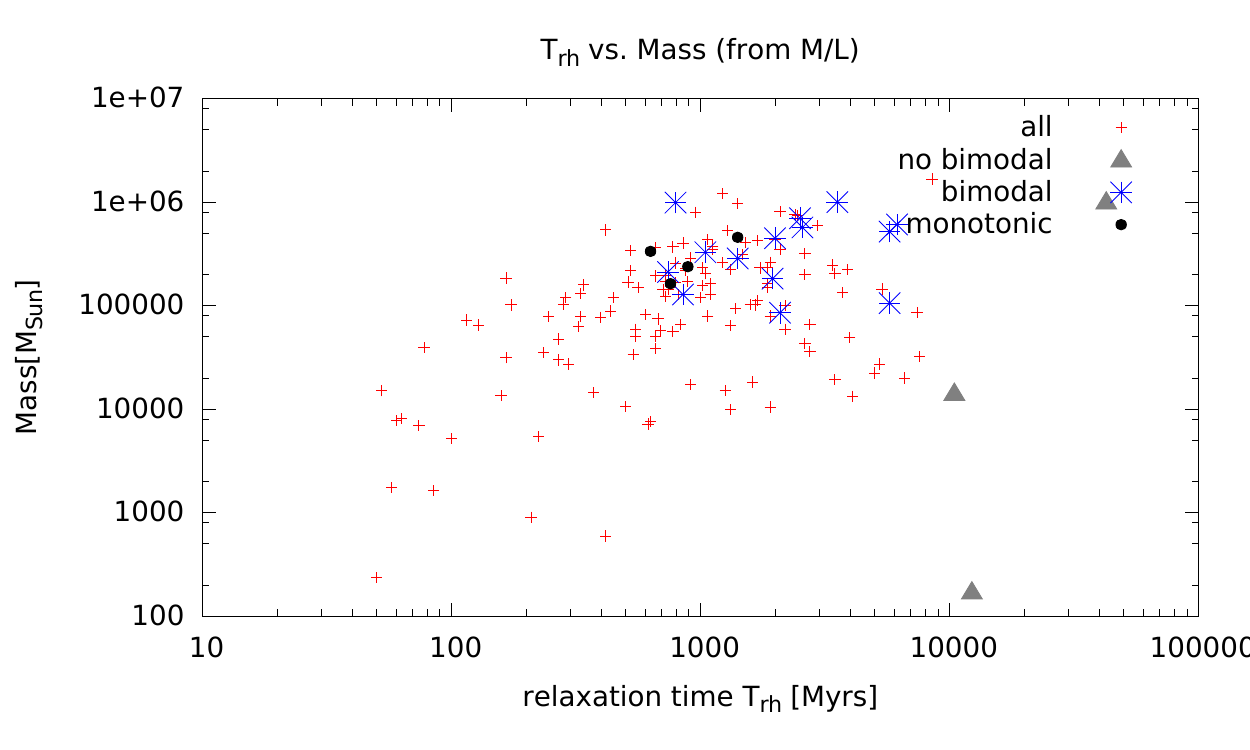}
  \caption[Globular clusters with observed bimodal, unimodal, and flat
  distributions showed on top of all GCs from the Harris catalogue]{Globular
  clusters with observed bimodal, unimodal, and flat distributions showed on 
  top of all GCs from the Harris catalogue \citep{Harris1996AJ....112.1487H}. On
  the X axis there is the half-mass relaxation time of a GC, while on the Y axis
  the mass from the M/L relation. The GCs marked with known bimodal distribution
  are: NGC104, NGC288, NGC5024, NGC5272, NGC5466, NGC5824, NGC5904, NGC6121,
  NGC6205, NGC6229, NGC6341, NGC6388, NGC6752, NGC6809, NGC7089, with known
  unimodal distribution: NGC1904, NGC6093, NGC6864, NGC7099 and with known flat
  distribution: NGC2419, NGC5139, Pal14.
  }
  \label{fig:Bim:Harris}
\end{figure}

The transientness of the signs of bimodal distributions could have important
implications for the observations of the real GCs. Fig.~\ref{fig:Bim:Harris}
shows the GCs with observed bimodal, unimodal and flat distributions marked on
the top of all GCs from the Harris catalogue \citep{Harris1996AJ....112.1487H}.
On the X axis there is a half-mass relaxation time ($t_{rh}$) of a GC and on the Y axis -- the
mass of GC estimated from the M/L relation. Three GCs with flat
distributions have large $t_{rh} \gtrsim 10$~Gyr. The bimodal distribution has not been
formed for them yet. It is in agreement with the \textsc{mocca-slow} model
(see Fig.~\ref{fig:Bim:Bimodality}).

The GCs with unimodal and bimodal distributions of BSs intersect each other in
Fig.~\ref{fig:Bim:Harris}. The intersection corresponds to
GCs with $t_{rh} \sim 1$~Gyr. The GCs located there should have enough time to
form the signs of the bimodal distribution, which they do. However, because of
the fact that the bimodality seems to be very transient, the GCs with unimodal
distributions occupy the same regions. It is possible that the GCs located there
have similar dynamical ages but the signs of the bimodal distributions switches
cyclically with the unimodal distributions. This, in turn, implies that the
unimodal distributions are not necessarily characteristic for dynamically old
GCs which are already well segregated. Because of the transientness, the signs of the
bimodal distribution can reappear after a few hundreds of Myr.

The ,,dynamical clock'' proposed by \citet{Ferraro2012Natur.492..393F} allows to
relate the position of the $r_{avoid}$ to the age of GC. However, the
simulations from Sect.~\ref{sec:Bim:RAvoidDrift} suggest that the ,,dynamical
clock'' works only for the first $\sim 2 t_{rh}$. After that time $r_{avoid}$
goes out of sync with $r_{min}$ (the position of the dip in the bimodal
distribution chosen by eye). The position of $r_{avoid}$ keeps increasing its
value, while $r_{min}$ stays closer to the center of GC (as the bimodal
distribution, see Fig.~\ref{fig:Bim:Bimodality}). An example of the real GC
which is consistent with this finding is NGC~6388 (see
\citet[Fig.~12]{Lanzoni2007ApJ...663..267L},
\citet{Dalessandro2008ApJ...677.1069D}). The figure presents a clear bimodal
distribution of the GC, but the radius $r_{avoid}$ does not correspond to the
visible minimum (the place with the lowest number of the relative BSs
frequency). The position of $r_{avoid}$ is about $15 r_c$, whereas the position
of the minimum is about 3 times smaller ($5 r_c$).
\citet{Dalessandro2008ApJ...677.1069D} suggests that the dynamical friction,
responsible for creating so-called ,,zone of avoidance'' (regions around
$r_{avoid}$), is not as efficient as previously thought. However, on the basis
of this paper, another explanation is argued. The radius $r_{avoid}$ goes out of
sync with $r_{min}$ for models if the age of GC exceeds $\sim 2 t_{rh}$. If
$r_{avoid}$ does not correspond to $r_{min}$, it suggests that NGC~6388 is a
dynamically old GC -- not dynamically younger as it is stated by
\citet{Dalessandro2008ApJ...677.1069D}. These authors discuss also the
possibility of the existence of IMBH in the center of NGC~6388 but it is not
clear how it would affect GC at radii of about $5 - 15 r_c$.
However, the \textsc{mocca} simulations which do not show agreement between
$r_{avoid}$ and $r_{min}$ after $\sim 2 t_{rh}$, do not have any IMBHs in their
centers. Is is possible that there are GCs with IMBHs in their centers and that
for them these two radii would be out of sync too. Nevertheless, the paper shows
that there is no need for IMBHs to explain such a feature.

Radius $r_{avoid}$ is found to be very hard to compute. It strongly depends on
the local parameters of GCs. Thus, its value can differ significantly between
two consequent snapshots in time (200~Myr for all models in
Sect.~\ref{chap:Bimodality}). It rises additional difficulties for the
,,dynamical clock'' which relates the values of $r_{avoid}$ with the ages of
GCs. The mass segregation is in fact the mechanisms which stands behind the
formation of the bimodal distributions (see Sect.~\ref{sec:Bim:RAvoidDrift}).
However, the calculations of $r_{avoid}$ in Sect.~\ref{sec:Bim:Stochasticity}
shows and stresses that $r_{avoid}$ is actually a very challenging quantity
to compute. Their values should be taken with caution.

It is argued that the ,,dynamical clock'' is not as promising tool for dating
the dynamical ages of GCs as previously thought. The main challenges constitute
the transientness of the signs of the bimodal distribution, the fact that
$r_{avoid}$ goes out of sync with the apparent minimum, and the strong dependence
of $r_{avoid}$ on the local GC's parameters.

\section*{Acknowledgment}

We would like to thank Douglas C. Heggie for N-body simulations (referred in the text as
\textsc{DH} simulations) and for very
valuable discussions concerning blue straggler spatial distributions in star
clusters. Results from these simulations added much confidence to the validity of the results
obtained with the \textsc{mocca} code.

The project was supported partially by Polish National Science Center grants
DEC-2011/01/N/ST9/06000 and DEC-2012/07/B/ST9/04412.





\bibliographystyle{mn2e} 
\bibliography{biblio}



\appendix



\bsp	
\label{lastpage}
\end{document}